\documentclass{JHEP3} 

\usepackage{epsfig}
\usepackage{amsfonts}
\usepackage{amssymb,amsmath}

\newcommand{\Tr}[1]{\:{\rm Tr}\,#1}

\newcommand{\mbf}[1]{{\boldsymbol {#1} }}
\newcommand{\complex}{{\mathbb C}} 
\newcommand{\zed}{{\mathbb Z}} 
\newcommand{\nat}{{\mathbb N}} 
\newcommand{\real}{{\mathbb R}} 
\newcommand{\rat}{{\mathbb Q}} 
\newcommand{\mat}{{\mathbb M}} 
\newcommand{\torus}{{\mathbb T}}
\def\e{{\,\rm e}\,}
\newcommand{\wt}{\widetilde}

\newcommand{\id}{{1\!\!1}}
\newcommand{\alg}{{\cal A}}
\newcommand{\Fock}{{\cal F}}
\newcommand{\proj}{{\sf P}}
\newcommand{\shift}{{\sf S}}

\newcommand{\vq}{{\mbf q}}

\newcommand{\vnu}{{\mbf\nu}}
\newcommand{\K}{{\rm K}}
\def\ii{{\,{\rm i}\,}}
\def\dd{{\rm d}}
\def\Li{{\rm Li}}
\newcommand{\NO}{\,\mbox{$\circ\atop\circ$}\,} 

\def\beq{\begin{equation}}
\def\eeq{\end{equation}}
\def\bea{\begin{eqnarray}}
\def\eea{\end{eqnarray}}
\def\bd{\begin{displaymath}}
\def\ed{\end{displaymath}}

\newcommand{\be}{\begin{equation}}
\newcommand{\ee}{\end{equation}}

\newcommand{\h}{{\cal E}}
\newcommand\fverb{\setbox\pippobox=\hbox\bgroup\verb}
\newcommand\fverbdo{\egroup\medskip\noindent%
                        \fbox{\unhbox\pippobox}\ }
\newcommand\fverbit{\egroup\item[\fbox{\unhbox\pippobox}]}
\newbox\pippobox

\newtheorem{theorem}{Theorem}
\newtheorem{proposition}{Proposition}

\allowdisplaybreaks

\title{Instantons, Fluxons and Open Gauge String Theory}

\author{ Luca Griguolo \\
Dipartimento di  Fisica, Universit\`a  di Parma,
INFN Gruppo Collegato di Parma\\
Parco Area delle Scienze 7/A, 43100 Parma, Italy\\
E-mail: \email{griguolo@fis.unipr.it}}
\author{Domenico Seminara\\
Dipartimento di Fisica, Polo Scientifico Universit\`a di Firenze,
INFN Sezione di Firenze\\
Via  G. Sansone 1, 50019 Sesto Fiorentino, Italy\\
Email: \email{seminara@fi.infn.it}}
\author {Richard J. Szabo\\
Department of Mathematics, Heriot-Watt University\\
Scott Russell Building, Riccarton, Edinburgh EH14 4AS, UK\\
Email: \email{R.J.Szabo@ma.hw.ac.uk}}

\received{\today}               

\accepted{\today}               
\preprint{ {\tt HWM-04-22} \ \ {\tt EMPG-04-11}
\\ \hepth{0411277} \\ November 2004} 

\date{data}
\abstract{We use the exact instanton expansion to illustrate
various string characteristics of noncommutative gauge theory in
two dimensions. We analyse the spectrum of the model and present
some evidence in favour of Hagedorn and fractal behaviours. The
decompactification limit of noncommutative torus instantons is
shown to map in a very precise way, at both the classical and
quantum level, onto fluxon solutions on the noncommutative plane.
The weak-coupling singularities of the usual Gross-Taylor string
partition function for QCD on the torus are studied in the
instanton representation and its double scaling limit, appropriate
for the mapping onto noncommutative gauge theory, is shown to be a
generating function for the volumes of the principal moduli spaces
of holomorphic differentials. The noncommutative deformation of
this moduli space geometry is described and appropriate
open string interpretations are proposed in terms of the fluxon
expansion.}


\begin{document}

\section{Introduction and Summary\label{Intro}}

One of the most challenging problems in modern theoretical
high-energy physics is to derive an effective string theory
description for the quantum dynamics of strong interactions. Many interesting
approaches were triggered by the observation that the Feynman diagram
expansion of non-abelian gauge theories, in the limit of a large
number of colours $N$, is naturally organized as in the genus
expansion of a string theory~\cite{hooft}. Although four-dimensional
QCD has always resisted any precise string interpretation, there exist
a number of examples in which the project has been effectively worked
out, such as ${\cal N}=4$ supersymmetric Yang-Mills theory in four
dimensions~\cite{ads}, two-dimensional QCD~\cite{gross1}--\cite{gt2}, and
Chern-Simons gauge theory on ${\mathbb S}^3$~\cite{VG}. In all of these
models the string description appears in the large $N$ limit,
confirming the expectation that an effective dynamics of extended
objects emerges when the rank of the underlying gauge group goes
to infinity.

Noncommutative gauge theories on Moyal-deformed spacetimes
(see~\cite{ksrev}--\cite{szrev} for reviews) describe the low-energy
dynamics of D-branes in a constant $B$-field background~\cite{sw}, and
their local symmetry group is a particular realization of the infinite unitary
group $U(\infty)$ that includes spacetime translations and more
generally symplectomorphisms of the flat spacetime~\cite{lsz1}. These
theories encode from the very beginning the concept of large $N$
in a well-defined mathematical framework, and therefore seem to be a
promising arena in which to tackle the difficult problem of deriving a
gauge string dynamics. There are various indications that this
framework could be easier than in the ordinary (commutative)
case. Noncommutative field theories are in general naturally induced
in string theory and they possess many unconventional stringy
properties themselves reflecting the non-locality that their
interactions retain~\cite{dnrev,szrev}. A first important observation is
that, in the limit of large noncommutativity parameter $\theta$, the
perturbative expansion nicely organises itself into planar and
non-planar Feynman diagrams in exactly the same way that the
large $N$ expansion of multicolour field theories does. Moreover,
they can be represented and analysed exactly as matrix models,
indicating a potential connection with non-critical strings. Their
fundamental degrees of freedom are electric dipoles, extended
rigid rods whose lengths are proportional to their momenta, and
their interactions are therefore governed by string-like
mechanisms. Finally, some of these theories admit novel soliton
and instanton solutions which have no counterparts in their
commutative cousins and can be naturally interpreted as D-branes.

These aspects become particularly interesting in two dimensions,
because ordinary Yang-Mills theory on a Riemann surface has a very
precise interpretation as a string theory~\cite{gross1}--\cite{gt2}, while the
exact solution of gauge theory on the noncommutative torus
$\torus_\theta^2$ has been recently presented through an expansion in
noncommutative instantons in~\cite{pasz1,pasz2}. An explicit
realization of two-dimensional noncommutative gauge theory as a matrix
model has also been constructed in~\cite{pasz3,gs1} by exploiting the
general relation~\cite{amns1} between its lattice regularization and
the twisted Eguchi-Kawai (TEK) model~\cite{ek1}--\cite{ek3}. According
to the general paradigm of reduced models, ordinary QCD$_2$ at large
$N$ should be recovered from the TEK model as well. The exact solution
of gauge theory on the fuzzy sphere, and its connection to instantons,
has also been described via a matrix model representation
in~\cite{stein1}. Moreover, concrete computations of observables can
be performed in a rather explicit way in these settings. For
these reasons a formulation of two-dimensional noncommutative gauge
theory as a string theory appears to be a fruitful line of attack and
some interesting relations between large $N$ gauge dynamics, instantons and
strings can be expected to emerge.

In this paper we present the results of an intensive investigation
of two-dimensional Yang-Mills theory on the torus at large $N$ and its
noncommutative cousin, with particular emphasis on its string
and instanton aspects. Our main technical tool for connecting commutative
and noncommutative theories is Morita equivalence~\cite{szrev}, which
among other things implies that when the noncommutativity parameter
$\theta=n/N$ is a rational number then Yang-Mills theory
on $\torus^2_\theta$ is exactly equivalent to ordinary gauge
theory on $\torus^2=\torus^2_{\theta=0}$ with gauge group $SU(N)$
in a specific sector of magnetic flux. Its fruitfulness lies in the fact that
a generic irrational noncommutative gauge theory can be obtained
as the limit of commutative ones as $n,N\to\infty$ with
$\theta=n/N$ fixed~\cite{HI1}--\cite{lls}. This suggests that great
insight into the string representations of noncommutative gauge
theories could be obtained from their commutative
counterparts. However, the limit required is ${\it not}$ the usual
't~Hooft limit, as Morita equivalence implies that the area of the
torus must also be scaled with $N$, forcing a double scaling
limit. This is the same sort of large $N$ limit required of the TEK
model in order to reproduce continuum noncommutative Yang-Mills
theory~\cite{amns1}. We have adapted this technique to describe
gauge theory on the noncommutative plane, performing accordingly
the decompactification limit of the torus theory~\cite{gsv1}. The
partition function becomes dominated by contributions from fluxons in
this case, the classical solutions on the noncommutative plane which
have a direct description in terms of
D-branes~\cite{gn}--\cite{gn2}. Remarkably, we are able to describe
the correct mathematics underlying the emergence of fluxons from noncommutative
torus instantons. But the most surprising results are obtained
by applying the double scaling limit, implied by the Morita
equivalence, to the complete commutative $U(N)$ gauge theory on
$\torus^2$. We will find that the partition function is dominated, in
this limit, by string states of infinite winding number and it
provides a generating function for the volumes of the principal moduli
spaces of holomorphic differentials on Riemann
surfaces~\cite{gss1}. Our approach, based on a simple saddle-point
equation, represents a very quick and efficient way to extract exact
formulas for these volumes which avoids the cumbersome combinatorial
techniques usually employed in the mathematics literature. The
appearence of a rather peculiar moduli space suggests the presence
of some particular topological string theory behind this limit, and
its generalization to the flux case, relevant for the noncommutative
gauge theory, seems to lead to even more exotic geometrical
structures.

This paper is organised as follows. We start in
Section~\ref{NCInsts} by reviewing the structure of instantons on
the noncommutative torus and by presenting the partition function
for gauge theory on a fixed Heisenberg module, written as a sum
over the noncommutative instantons. In Section~\ref{InstStrings}
we discuss a number of stringy features exhibited by the
noncommutative instantons on $\torus^2$. We
illustrate the exponential behaviour in energy of the asymptotic
distribution of instanton solutions and the possibility of a
Hagedorn type phase transition in the theory is briefly addressed.
We also present some arguments for a fractal
description of the instanton spectrum. In Section~\ref{Fluxon} we
turn our attention to gauge theory on the noncommutative plane
which can be analysed in detail. Section~\ref{Decomp} describes
the decompactification limit of the torus partition function and
how it can be written as a semi-classical expansion around
fluxons, Section~\ref{DecompFields} rigorously proves using the
language of projective modules that torus instantons map in a very
precise way onto fluxons, Section~\ref{FluxonProps} illustrates
how fluxon characteristics are naturally inherited from the torus
instantons, and Section~\ref{TFPF} presents the resummation of the
semi-classical expansion of the partition function on the
noncommutative plane. The Gross-Taylor string expansion of ${\it
commutative}$ $U(N)$ Yang-Mills theory on the torus is revisited
in Section~\ref{SLISE} and its manifestation in the instanton
series is analysed. Section~\ref{TG-TET} describes the singular
small area behaviour of the genus expansion, Section~\ref{IEUNGT}
derives the dual instanton expansion, Section~\ref{SPA} derives a
saddle-point equation neglecting higher instanton contributions
that controls the weak-coupling limit of the string expansion, and
Section~\ref{CFP} explores in detail the quasi-modular property of
the full string partition function in terms of the instanton
contributions at weak-coupling. In Section~\ref{DSL} we analyze
the $U(N)$ gauge theory in the double scaling limit.
Section~\ref{DSLsaddle} gives the general solution of the
saddle-point equation and the structure of the exact free energy,
Section~\ref{ASHN} relates the terms in the saddle-point expansion
to the asymptotic growth of simple Hurwitz numbers,
Section~\ref{MSHD} explores the geometrical structure behind this
limit in terms of the principal moduli spaces of holomorphic
differentials, and Section~\ref{OSI} discusses the open string
interpretation of these results. Finally, in Section~\ref{ISRNCGT}
we come back to the noncommutative setting and analyse the double
scaling limit of $SU(N)/\zed_N$ gauge theory on $\torus^2$.
Section~\ref{IESUNZNGT} derives the generalization of the
instanton expansion in this case, Section~\ref{NCDSL} performs the
double scaling limit illustrating how the chiral fluxon expansion
emerges, Section~\ref{PPT} shows that the fluxon expansion always
dominates the theory thus verifying the absence of phase
transitions, and Section~\ref{NCOSI} proposes an open string
interpretation of these results. Appendix~\ref{appA} contains some
additional technical details, while Appendix~\ref{FFR} presents a
free fermion representation of the double scaling limit.

\section{Noncommutative Instantons in Two Dimensions\label{NCInsts}}

In this section we will briefly describe the structure of
instantons on the noncommutative torus. They are defined as
solutions of the noncommutative Yang-Mills equations on a square
torus $\torus^2$ of area $A$ which are not gauge transformations
of the trivial connection. For this, we will first recall the
notion of {\it Heisenberg modules}, as they classify all possible
types of gauge theories on the noncommutative torus. Further
details may be found in~\cite{pasz1}.

Let $\alg={\cal S}(\torus_\theta^2)$ be
the algebra of Schwartz functions on the noncommutative torus of
deformation parameter $\theta\in\real$, and let
$\mat_n(\alg)$ denote the algebra of $n\times n$ matrices with entries
in $\alg$. For any $n\in\nat$ and any projector $\proj\in\mat_n(\alg)$ with
$\proj^2=\proj=\proj^\dag$, the vector space $\h=\proj\alg^n$ is a
finitely-generated projective module over the algebra
$\alg$. Yang-Mills theory on the noncommutative torus is constructed
on a fixed stable equivalence class of projective modules. Since the
K-theory of the noncommutative torus is
$\K^0(\torus_\theta^2)=\zed^2\cong\zed+\zed\,\theta$ (for $\theta$
irrational), any projector $\proj=\proj_{p,q}$ is characterized by a pair of
integers $(p,q)$. Stable modules
\beq
\h_{p,q}=\proj_{p,q}\alg^n
\label{stablemodule}\eeq
live in the restriction to the positive cone of $\K^0(\torus_\theta^2)$
defined by positivity of the corresponding Murray-von~Neumann
dimension
\begin{equation}
\dim\h_{p,q}:=\Tr\,\proj_{p,q}=p-q\,\theta>0 \ .
\label{dimhpq}\end{equation}
Such modules $\h_{p,q}$ are called Heisenberg modules, and on them the
Yang-Mills action can be defined. The integer $q$
is the Chern number of the corresponding gauge bundle and it can be
computed explicitly in terms of the curvature of the
module~\cite{pasz1}. The rank of $\h_{p,q}$ is the positive integer
\beq
{\rm rank}\,\h_{p,q}:={\rm gcd}(p,q) \ .
\label{rankE}\eeq
Any finitely generated projective
module over the noncommutative torus is of the form $(\h_{p,q})^l\oplus\alg^m$
for some $(p,q)\in\zed^2$ and $l,m\in\nat_0$. We will always assume
in the following that $q\neq0$ and that all modules have been
implicitly completed into separable Hilbert spaces using the natural
inner products on the vector spaces involved.

The characteristic feature of any Heisenberg module $\h_{p,q}$ is that
it always admits an anti-Hermitian connection $\nabla$ of
constant curvature
\beq
F_{p,q}:=\left[\nabla_1\,,\,\nabla_2\right]=
\frac{2\pi\ii}A\,\frac q{p-q\,\theta}~\proj_{p,q} \ .
\label{Fpqdef}\eeq
The curvature (\ref{Fpqdef}) is proportional to the identity element
of the endomorphism algebra of (\ref{stablemodule}), because the
projector $\proj_{p,q}$ clearly acts as the identity operator on the
module,
$\proj_{p,q}|\psi\rangle=|\psi\rangle~~\forall|\psi\rangle\in\h_{p,q}$.
{}From this fact one may exhibit $\h_{p,q}$ explicitly as the separable
Hilbert space
\beq
\h_{p,q}={\cal F}\otimes{\cal W}_{p,q} \ .
\label{hpqexpl}\eeq
The space ${\cal F}\cong L^2(\real)$ is the Schr\"odinger
representation of the Heisenberg commutation relation (\ref{Fpqdef}),
and it is the unique irreducible module over the Heisenberg Lie
algebra. The finite dimensional vector space ${\cal
W}_{p,q}\cong\complex^{|q|}$ is the $|q|\times|q|$ unitary
representation of the Weyl-'t~Hooft algebra
\beq
\Gamma_1\,\Gamma_2=\e^{2\pi\ii p/q}~\Gamma_2\,\Gamma_1 \ ,
\label{WtHalg}\eeq
whose unique irreducible module has dimension $|q|/{\rm
  rank}\,\h_{p,q}$. The unitary generators $U_i$, $i=1,2$ of the
noncommutative torus, defined by the commutation relation
\beq
U_1\,U_2=\e^{2\pi\ii\theta}~U_2\,U_1 \ ,
\label{U12commrel}\eeq
can then be represented on (\ref{hpqexpl}) as
\beq
U_i=\e^{\sqrt A\,(p-q\,\theta)\,\nabla_i/q}\otimes\Gamma_i \ , ~~
i=1,2 \ .
\label{Uirep}\eeq

Within this framework it is also possible to classify all critical
points of the noncommutative Yang-Mills action on a fixed Heisenberg
module $\h_{p,q}$~\cite{pasz1}. The constant curvature connection
$\nabla$, obeying (\ref{Fpqdef}), solves the equations of motion
and yields the absolute minimum value of the action on $\h_{p,q}$. It
can also be used to construct all solutions of the classical equations
of motion. The main observation is that, in a neighbourhood of a
solution of the Yang-Mills equations, the module $\h_{p,q}$ may be
regarded as a direct sum of submodules, and a critical point is always
gauge equivalent to the direct sum of the corresponding absolute
minima provided by the constant curvature connections on the
submodules~\cite{pasz1}. The classification of classical solutions is
equivalent to finding all possible decompositions of the original
Heisenberg module $\h_{p,q}$ into submodules $\h_{p_k,q_k}$ of the
form
\begin{equation}
\h_{p,q} = \bigoplus_{k}\h_{p_k, q_k} \ .
\label{deco}
\end{equation}
For any given value of $\theta$, a classical solution
of Yang-Mills theory on $\h_{p,q}$ is therefore completely characterized by the
sets of  pairs of integers $\underline{(p,q)}:=\{(p_k,q_k)\}$ satisfying the
three constraints
\begin{eqnarray}
p_k-q_k\theta&>&0 \ , \nonumber\\
\sum_{k}(p_k-q_k\theta)&=&p-q\,\theta \ , \nonumber\\
\sum_{k}q_k&=&q \ .
\label{costri}
\end{eqnarray}
The collection $\underline{(p,q)}$ is called a {\it partition} of the
topological numbers $(p,q)$ of the Heisenberg module $\h_{p,q}$. The
third equation is the requirement that the Chern number of the
module be equal to the total magnetic flux of the direct sum decomposition.
When the noncommutativity parameter $\theta$ is irrational this
constraint is an automatic consequence of the second one, while in the
rational case it is understood as the requirement of K-theory charge
conservation.

The Yang-Mills action, evaluated on a solution of the classical
equations of motion, is computed in terms of the
noncommutativity parameter $\theta$ and of a partition
$\underline{(p,q)}$ as
\begin{equation}
S\left(\,\underline{(p,q)}\,;\,\theta
\right)=\displaystyle{\frac{2\pi^2}{g^2A}\,\sum_{k}
\frac{q_k^2}{p_k-q_k\theta}} \ ,
\label{SYMpart}\end{equation}
where $g$ is the Yang-Mills coupling constant and the sum runs through
all components $(p_k,q_k)$ of $\underline{(p,q)}$. At the quantum
level, the significance of these noncommutative instantons is that the
partition function and observables of noncommutative gauge theory in
two-dimensions are given exactly by their semi-classical
approximations and can be represented as sums over classical
solutions~\cite{pasz1,pasz2}. This leads to explicit weak coupling
expansions over partitions. In particular, the partition function on a
fixed Heisenberg module $\h_{p,q}$ can be written as
\beq
Z_{p,q}\left(g^2A\,,\,\theta\right)=\sum_{\stackrel{\scriptstyle{\rm
partitions}}{\scriptstyle\underline{(p,q)}}}\,
\frac{(-1)^{|\,\underline{(p,q)}\,|}}{\prod_a\nu_a!}\,
\prod_{k=1}^{|\,\underline{(p,q)}\,|}
\sqrt{\frac{2\pi^2}{g^2A\,(p_k-q_k\theta)^3}}~
\e^{-S(\,\underline{(p,q)}\,;\,\theta)} \ ,
\label{Zpqpart}\eeq
where the integer $\nu_a>0$ is the number of partition components
$(p_k,q_k)\in\underline{(p,q)}$ which have the $a^{\rm th}$ least dimension
$p_a-q_a\theta$, while $|\,\underline{(p,q)}\,|:=\nu_1+\nu_2+\dots$ is
the total number of components in the partition. This definition requires us to
order components of partitions according to their increasing
Murray-von~Neumann dimensions $p_k-q_k\theta$ and regard any two
partitions as equivalent if they differ only in their partial orderings. In
other words, the sum in (\ref{Zpqpart}) is implicitly assumed to run
over {\it unordered partitions}. By resumming (\ref{Zpqpart}) over
gauge inequivalent partitions, we may express it as a sum over
unstable instantons~\cite{pasz1}. The exponential prefactors then represent the
exact corrections due to quantum fluctuations about each instanton.

\section{The Spectrum of Noncommutative Instantons\label{InstStrings}}

We will now proceed to demonstrate that the noncommutative
instantons on the torus exhibit a number of stringy features which
will direct the way towards the string interpretation of the gauge
theory partition function (\ref{Zpqpart}). The simplest way to
understand this stringiness is through the strong coupling
expansion of the noncommutative gauge theory, which may be
expressed as a sum over non-local electric dipole
configurations~\cite{pasz2}. In this sense the instanton expansion
is dual to an {\it open string} representation of the gauge
theory, in which the endpoints of strings are coupled to a
background magnetic field $B=(A\,\theta/2\pi)^{-1}$ and the
strings become polarized as dipoles. However, in contrast to the
instanton series, the dipole expansion is not very explicit and is
difficult to deal with analytically. In what follows we will
attempt to extract the string behaviour directly from the
instanton series, and among other things this will yield a direct
connection with the conventional string picture of ordinary
QCD$_2$, in which the exact partition function can be interpreted
as a sum over two-dimensional ramified covers of the underlying
torus $\torus^2$. In this section we will analyse the spectrum of
the noncommutative gauge theory and extract its high-energy
asymptotic behaviour.

We start by discussing the structure of the asymptotic density of
states in two-dimensional noncommutative gauge theory. The key to
extracting string characteristics is to note the extreme
differences in the nature of the partitions which contribute to
the instanton sum in the cases of rational and irrational values
of the noncommutativity parameter $\theta$~\cite{pasz1}. In the
rational case the contributing partitions have module dimensions
which are {\it a priori} bounded from below, with the fixed bound
depending on the value of the rational number $\theta$, while in
the irrational case partitions of arbitrarily small dimension
always contribute. We can see this difference qualitatively by
plotting the possible values of the instanton action
(\ref{SYMpart}) on a fixed Heisenberg module $\h_{p,q}$. In a
$2+1$ dimensional setting, we may also interpret these values as
Yang-Mills energies, and we will frequently adopt this point of
view in the following.

For example, Fig.~\ref{2compQ} illustrates the typical pattern of
energies in the case of rational $\theta$ from the contributions
of two component partitions. The vertical axis is the energy,
while the horizontal axis is the list of allowed partitions
$\underline{(p,q)}=\{(p_1,q_1)\,,\,(p_2,q_2)\}$ which contribute.
The plot traces out the positions of the energies on this list. It
is also possible to obtain similar plots for higher component
partitions by choosing an appropriate (lexicographic) ordering to
arrange partitions along the horizontal axis. As one can see, a
very regular pattern emerges, as there are only a small number of
well-defined trajectories in this case. Of course, by Morita
equivalence, this is the typical situation in ordinary Yang-Mills
theory. Since the gauge theory in this case is topological and
thereby possesses only finitely many degrees of freedom, we may
interpret the energy spectrum as that of some quantum mechanical
system consisting of different species of non-interacting
particles in a box. The full spectrum can be decomposed as the sum
of spectra from different classes of quantum mechanical systems,
the collection of which is in a one-to-one correspondence with the
partitions of $p-q\,\theta$ in this case.

\DOUBLEFIGURE{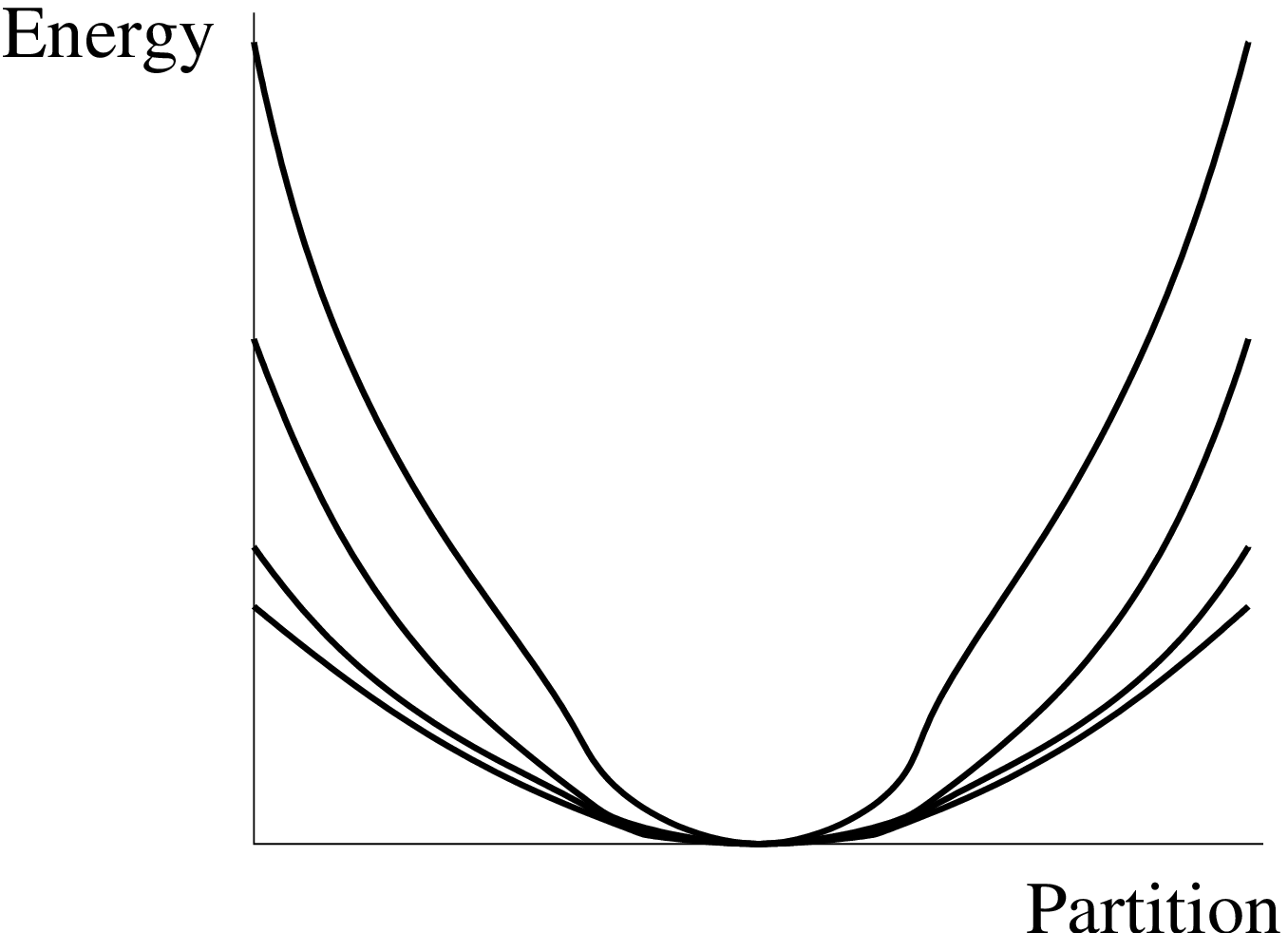,width=2in}{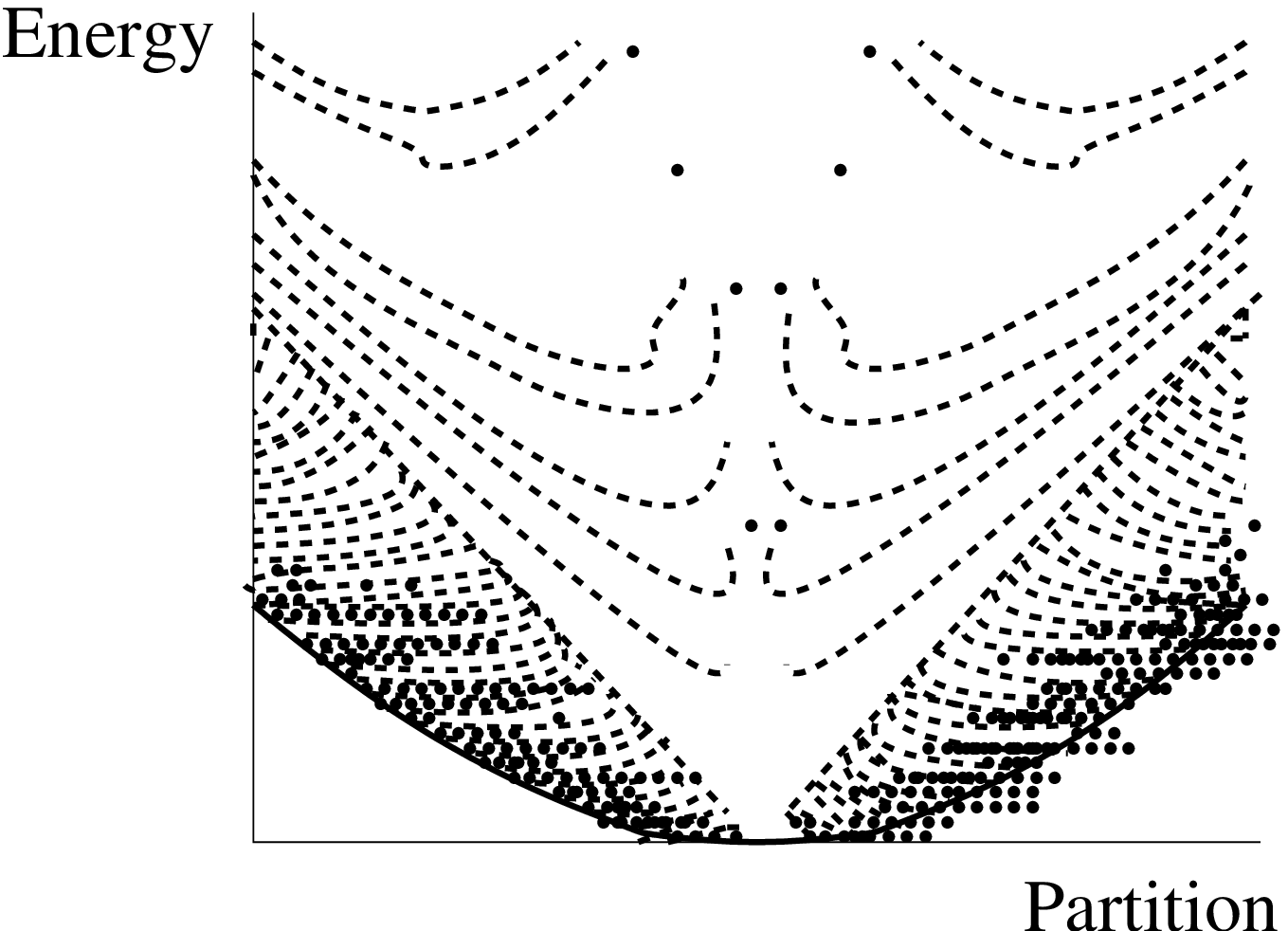,width=2in}{Allowed
energies for two component partitions of the Heisenberg module
${\cal E}_{2,2}$ for $\theta=\frac15$.\label{2compQ}}{Allowed
energies for two component partitions of the Heisenberg module
${\cal E}_{2,2}$ for $\theta=\frac1{\sqrt2}$.\label{2comp}}

In striking contrast, the corresponding situation for irrational
$\theta$ is depicted in Fig.~\ref{2comp}. Now the energy points
tend to fill the whole plane. This is especially evident in the
plots as one starts increasing the number of partition components.
The lists of energies from rational values of $\theta$ have far
more structure than those from irrational values. In other words,
while the lists for rational $\theta$ exhibit finitely many
trajectories, the irrational ones appear to display {\it fractal}
characteristics and tend to become dense in the plane as the
number of partition components increases. We will come back to
this point later on.

In view of these observations, we may expect that at high
energies the degeneracy of states can rise very rapidly,
leading to a Hagedorn behaviour of the spectrum. Because the
quantum field theory here is semi-classically exact, by a state of
the system we shall mean an instanton partition
$\underline{(p,q)}$. We will count the number of solutions of the
noncommutative Yang-Mills equations and find appropriate bounds at
high energies. These numbers coincide with the volumes of the symmetric
orbifold moduli spaces of classical solutions described
in~\cite{pasz1}. Let $N_{p,q}(E,\theta)$ be the number of submodules
of the Heisenberg module $\h_{p,q}$ which are available for making
up a partition $\underline{(p,q)}$ that has energy
$E=\frac{g^2A}{2\pi^2}\,S(\,\underline{(p,q)}\,;\,\theta)$, so
that \beq E=\sum_{k}\frac{q_k^2}{p_k-q_k\theta} \ .
\label{fixedenergy}\eeq Since the right-hand side of
(\ref{fixedenergy}) is a sum of positive terms, we may infer that
$q_k^2\leq E\,(p_k-q_k\theta)$ for each $k$. But from
the constraints (\ref{costri}) we also have $0<p_k-q_k\theta\leq
p-q\,\theta$, and hence the partition component integers are
generically bounded on the fixed energy slices as \bea
|q_k|&\leq&\sqrt{(p-q\,\theta)\,E} \ , \nonumber\\
q_k\theta\,<\,p_k&\leq&p-(q-q_k)\,\theta \label{qkpkbounds}\eea
for each $k$. They are further constrained by K-theory
charge conservation \beq \sum_{k}p_k=p \ , ~~
\sum_{k}q_k=q \label{Kchargecons}\eeq on the
module $\h_{p,q}$.

Let us consider first the case of rational noncommutativity
parameter $\theta=n/N$, where $n,N$ are relatively prime positive
integers. Submodule dimensions are then all bounded from below by
$\frac1N$~\cite{pasz1}, and by denoting $r_k:=N\,p_k-n\,q_k$,
$r:=N\,p-n\,q$, the second of the constraints (\ref{qkpkbounds})
reduces to $1\leq r_k\leq r$, while the first sum in
(\ref{Kchargecons}) may be replaced by $\sum_kr_k=r$. These are
just the defining conditions for classical solutions of the Morita
equivalent commutative gauge theory with structure group $U(r)$,
and they imply that the number of components $m$ of any given
partition in this case is bounded from above as $m\leq r$. Because
of the first bound in (\ref{qkpkbounds}), there are only finitely
many partitions which contribute to a fixed finite energy $E$.
The K-theory charges in this instance are completely decoupled,
and the counting of states is determined by taking the product of
the numbers of $q_k$'s and $r_k$'s obeying their respective
constraints. The total number of $r_k$'s is the number $\Pi(r)$ of
proper unordered partitions of the rank $r$ into natural numbers,
which can be computed from the generating function~\cite{milne1}
\beq
\frac{\e^{\pi\ii\tau/12}}{\eta(\tau)}=1+\sum_{r=1}^\infty\,\Pi(r)~
\e^{2\pi\ii r\,\tau} \label{partngenfn}\eeq where \beq
\eta(\tau):=\e^{\pi\ii\tau/12}\,\prod_{m=1}^\infty\left(1-
\e^{2\pi\ii m\,\tau}\right)
\label{Eulerseries}\eeq is the Dedekind function. For the $q_k$'s,
we will strictly bound their number from above by ignoring the
topological charge constraint and taking their totality in the
range $|q_k|\leq\sqrt{r\,E/N}$ for each $k=1,\dots,r$.

In this way we find that the total number of instanton solutions of
energy $E$ satisfies the upper bound
\beq
N_{p,q}(E,\theta=n/N)<\left(2\,\sqrt{r/N}\,\right)^r~\Pi(r)~
E^{r/2} \ , ~~ r=N\,p-n\,q \ .
\label{Npqrat}\eeq
The important feature of (\ref{Npqrat}) is that the corresponding
density of states
\beq
\rho_{p,q}(E,\theta)=\frac{\partial}{\partial E}N_{p,q}(E,\theta)
\label{DOS}\eeq
has at most a power-like growth $\rho_{p,q}(E,\theta=n/N)\sim
E^\alpha$ as the energy of the system increases. This is exactly the
result that one would anticipate from a gas of non-interacting
particles in $r$ dimensions, reflecting again the quantum mechanical
nature of the model in this case. The ``particles'' here are
pointlike instantons, corresponding to symmetric orbifold
singularities, of commutative $U(r)$ Yang-Mills theory on the torus
$\torus^2$~\cite{pasz1}.

Now let us analyse the degeneracy of states in the irrational
noncommutative gauge theory. The arguments just presented fail in
this case, because while in the rational case any submodule
dimension is a multiple of the dimension $\frac1N$ of the smallest
submodule, so that the spacing between submodule dimensions is
regular, in the irrational case there is no such lower bound and
the spacing between submodule dimensions is irregular. In
particular, the topological numbers $(p_k,q_k)$ of submodules no
longer decouple from each other. We can obtain a crude upper bound
on the number of states $N_{p,q}(E,\theta)$ in this instance,
starting by restricting the counting to $m$-component partitions.
We denote by ${\mbf\Pi}_m(p,q;\theta)$ the subset of partitions
$\underline{(p,q)}$, whose components $(p_k,q_k)$ obey the
constraints (\ref{costri}), with a fixed number
$m=|\,\underline{(p,q)}\,|$ of components on the Heisenberg module
$\h_{p,q}$ over the algebra $\alg={\cal S}(\torus_\theta^2)$. From
(\ref{qkpkbounds}) it follows that the topological partition
numbers of an element $\underline{(p,q)}\in{\mbf
\Pi}_m(p,q;\theta)$ lie in the ranges
$|q_k|\leq\sqrt{(p-q\,\theta)\,E}$ and
$|p_k|\leq|\theta|\,\sqrt{(p-q\,\theta)\,E}$ for large energies
$E$ and for each $k=1,\dots,m$. Ignoring the constraints
(\ref{Kchargecons}), the degeneracy of states
$N_{p,q}^{(m)}(E,\theta)$ in ${\mbf\Pi}_m(p,q;\theta)$ can thereby
be strictly bounded from above by the hypervolume containing these
integers as \beq
N_{p,q}^{(m)}(E,\theta)<\frac1{m!}\,\bigl(4\,|\theta|\,(p-q\,\theta)
\,E\bigr)^m \ , \label{Npqmupbd}\eeq where we have divided by the
permutation symmetry factor $m!$ to account for the fact that the
order of components in the partitions contributing to the vacuum
amplitude (\ref{Zpqpart}) is immaterial. By summing over all
possible partition lengths appropriate to the irrational theory,
we thereby arrive at the upper bound \beq
N^{~}_{p,q}(E,\theta)<\sum_{m=0}^\infty N_{p,q}^{(m)}(E,\theta)<
\e^{4\,|\theta|\,(p-q\,\theta)\,E} \ . \label{Npqirratup}\eeq

On the other hand, we may argue the form for a lower bound on the
number of classical solutions at high energies starting from the
explicit expression \beq
N_{p,q}(E,\theta)=\sum_{\stackrel{\scriptstyle{\rm
partitions}}{\scriptstyle\underline{(p,q)}}}\delta\left(
E-\mbox{$\frac{g^2A}{2\pi^2}$}\,S\bigl(\,\underline{(p,q)}\,;
\,\theta\bigr)\right) \label{rhopqexpl}\eeq for the degeneracy of
states. To get an idea of the behaviour, let us first consider
(\ref{rhopqexpl}) for the ``purely noncommutative'' Heisenberg
modules of K-theory charges $(p,q)=(0,q)$ (These projective
modules will play a prominent role in subsequent sections). The
counting of solutions in a module of the type $\h_{0,q}$ actually
takes care of the general case. A module $\h_{p,q}$ can always be
mapped into $\h_{0,{\rm gcd}(p,q)}$ via the Morita equivalence
generated by the $SL(2,\zed)$ transformation
\beq
\begin{pmatrix}0\\{\rm rank}\,\h_{p,q}\end{pmatrix}=
\begin{pmatrix}\frac q{{\rm rank}\,\h_{p,q}}&-\frac p{{\rm rank}\,
\h_{p,q}}\\c&d\end{pmatrix}\begin{pmatrix}~p~\\~q~\end{pmatrix} \ ,
\label{Morita0q}\eeq
where $c$ and $d$ are integers obeying the Diophantine relation
\beq
c\,p+d\,q={\rm gcd}(p,q) \ .
\label{Dio0q}\eeq
In particular, every module $\h_{p,q}$ for which $p$ and $q$ are
relatively prime is Morita equivalent to $\h_{0,1}$.

By truncating the sum (\ref{rhopqexpl}) to partitions for which
$p_k=0~~\forall k$ we can derive the lower bound \beq
N_{0,q}(E,\theta)>\sum_{\vq\,:\,q_k\in\nat_0}\delta_{
|q|\,,\,\sum_kq_k}~\delta\left(
E-\mbox{$\frac1{|\theta|}\,\sum_k$}\,q_k\right)=\Pi\bigl(|\theta|
\,E\bigr) \ . \label{rhopqlowbd}\eeq
When $|q|$ is very large, the right-hand side of (\ref{rhopqlowbd}) can be
thought of as a lower bound at asymptotically high energies, since then
$|q|=|\theta|\,E\to\infty$. The number $\Pi(|q|)$ of unordered
partitions of the positive integer $|q|$ into natural numbers has an
exact convergent series expansion for large $q$ provided by the
Rademacher formula~\cite{milne1}. The first term in this series is the
leading term of the asymptotic expansion of $\Pi(|q|)$ at
$|q|\to\infty$ and it yields the Hardy-Ramanujan formula \beq
\Pi\bigl(|q|\bigr)=\frac1{4\,\sqrt3\,|q|}~\e^{\pi\,\sqrt{2|q|/3}}\,
\left[1+O\left(\frac{\ln|q|}{|q|^{1/4}}\right)\right] \ .
\label{HRformula}\eeq
{}From (\ref{rhopqlowbd}) and (\ref{HRformula}) one arrives at the
asymptotic energy bound \beq
N_{0,q}(E,\theta)>\frac{\e^{\sqrt{2|\theta|\,E/3}}}{4\,\sqrt3\,
|\theta|\,E} \ . \label{asymptbds}\eeq Although only valid in the
limit $|q|\to\infty$, this expression
suggests that the corresponding density of states (\ref{DOS}) for
$\theta\in\real\,\backslash\,\rat$ has a high energy growth lying
between a sub-exponential $\e^{\alpha\,\sqrt E}$ behaviour and an
exponential $\e^{\beta\,E}$ behaviour, in marked contrast to the
power growth of the rational case. Note that ordinary
two-dimensional Yang-Mills theory at large $N$
exhibits exactly an $\e^{\alpha\,\sqrt E}$ rise in its density of
states~\cite{dougts}. The realization of the noncommutative
gauge theory as a particular kind of large $N$ limit of commutative
gauge theory thereby supports the general validity of
(\ref{asymptbds}).

With the belief that the lower bound in (\ref{asymptbds}) is
correct, we can present stronger evidence in favour
of an exponential-like behaviour, but with a different power behaviour
of the energy in the asymptotic growth. The idea is to embed in a
minimal way the previous result in a generic projective module
$\h_{p,q}$. The irrationality of $\theta$ allows for the appearence of
submodules of $\h_{p,q}$ with arbitrarily small dimension. Let \beq
\varepsilon=p'-q'\,\theta\label{smallm}\eeq be the dimension of
a ``small'' submodule $\h_{p',q'}$. For $\epsilon\to 0$ there are
always $n$ such submodules appearing in the instanton partition with
$n\,\varepsilon<p-q\,\theta$. The energy of these
configurations is easily estimated to be \beq E\simeq
\frac{n\,q^{\prime\,2}}{\varepsilon} \ . \eeq In order to respect the
constraint on the total dimension of the module we have to scale \beq
\varepsilon\simeq\frac1n\label{epscaling}\eeq in the limit
$n\to\infty$. Likewise, the smallness of $\varepsilon$ requires that
both $p',q'\to\infty$. The defining relation (\ref{smallm}) can in
fact be understood as an approximation procedure for the irrational
number $\theta$ by rational numbers as \beq
\frac{p'}{q'}-\theta=\frac{\varepsilon}{q'} \ . \eeq This equation
implies that in constructing the ``small'' submodules the integer $q'$
scales as $\varepsilon^{-\alpha}$ with $\alpha>0$. Taking into account
the scaling of $\varepsilon$ in (\ref{epscaling}) we
may then express the growth of $E$ in terms of the number of
components $n$ as \beq E\simeq n^{2+2\alpha} \ . \eeq

The degeneracy of these configuration is given by the number of
partitions $\Pi(n)$, and by applying the Hardy-Ramanujan formula
(\ref{HRformula}) we have \beq
\displaystyle{N_{p,q}(E,\theta)>\frac{a}{E^{\frac{1}{2+2\alpha}}}
\e^{b\,E^{\frac{1}{4+4\alpha}}}}\eeq
for some positive real constants $a,b$. In deriving this bound we have
taken into account the contribution of a very particular set of
configurations as $E\to\infty$, characterized by a peculiar smallness
in the submodule decompositions. This is a very crude approximation but, as we
have seen, it is sufficient to yield an exponential behaviour for the
asymptotic density of states. We have not succeeded in deriving an analog of
the lower bound (\ref{asymptbds}) for generic topological numbers
$(p,q)$, but we believe that at least the characteristic
$\e^{\alpha\,\sqrt E}$ behaviour could be obtained by a refined
estimate. This former growth is the standard behaviour of the
degeneracy of states in ordinary two-dimensional quantum field
theory (with finitely-many fields and propagating degrees of
freedom), while the behaviour $\e^{\beta\,E}$ in (\ref{Npqirratup}) is
characteristic of the Hagedorn spectrum in string theory. In this sense, the
point-like instantons of the commutative (rational) theory become
elongated in the irrational noncommutative gauge theory, and
behave more like the quanta in a system with infinitely-many
and/or extended degrees of freedom.

Let us now discuss the possible physical consequences of these
behaviours. The sum over partitions in (\ref{Zpqpart}) can be
rearranged and written in terms of the instanton density of states as
\bea Z_{p,q}\left(\tilde g^2\,,\,\theta\right)&=&\sum_{E>0}\e^{-E/2\tilde
g^2}\,\sum_{\stackrel{\scriptstyle{\rm
partitions}~\underline{(p,q)}} {\scriptstyle
S(\,\underline{(p,q)}\,;\,\theta)=E/2\tilde g^2}}
{W}_{p,q}\left(\,\underline{(p,q)}\,;\,\tilde g^2\,,\,\theta\right)\nonumber\\
&=&\int\limits_0^\infty\dd E~\rho_{p,q}(E,\theta)~
\e^{-E/2\tilde g^2}~{W}_{p,q}\left(E\,;\,\tilde g^2\,,\,
\theta\right) \ ,
\label{ZpqDOS}\eea
where $\tilde g=g\,\sqrt A/2\pi$ is the dimensionless Yang-Mills coupling
constant, and the function ${W}_{p,q}$ is the fluctuation
determinant in (\ref{Zpqpart}). In the rational case whereby the
high-energy density of states obeys a power-law increase with $E$, the
partition function (\ref{ZpqDOS}) is well-behaved for all $\tilde
g^2$. The lack of Hagedorn behaviour is
generic in local quantum field theories, so that if the lower bound of
(\ref{asymptbds}) represents the true behaviour of the asymptotic
density of states in the irrational noncommutative gauge theory, then
again there is no instability present in the system.

On the other hand, the upper bound in (\ref{Npqirratup}) can lead to an
instability if at high energies the exponential growth of the density
of states is sufficient to overcome the growth from the quantum
fluctuations ${W}_{p,q}$ in (\ref{ZpqDOS}). If the exponential
increase completely dominates over this entropy factor, then the partition
function (\ref{ZpqDOS}) will diverge at a critical coupling $\tilde g_{\rm
c}^2$
which from (\ref{Npqirratup}) may be estimated to be $\tilde g_{\rm
  c}^2\simeq1/8\,|\theta|\,(p-q\,\theta)$.
In this case, the exponential rise in the density of states would lead
to a phase transition beyond which the theory develops an
instability due to a condensation of instantons in the vacuum. This is
somewhat analogous to what happens in ordinary large $N$ QCD$_2$ on
the sphere~\cite{DK1}--\cite{grossmat}, although in the noncommutative
gauge theory we would not expect the result to rely on a large rank
limit, as the effective gauge group is then essentially the infinite
unitary group $U(\infty)$ (or more precisely an appropriate completion
thereof)~\cite{lsz1}. Note that $\tilde g_{\rm
c}^2\to\infty$ in the commutative limit $\theta\to0$. This
would then suggest that the physics beyond the transition point
is accurately described by the strong coupling expansion which can be
expressed in terms of contributions from the production of virtual
electric dipoles in the vacuum~\cite{pasz2}. In the commutative case
this would be exactly the regime in which a string representation of
the gauge theory is available.

We conclude this section by returning to the possibility of fractal
behaviour in the spectrum of noncommutative instantons. We expect
that the spacing between energy levels in the spectrum as the
energy increases becomes chaotic because there is no longer a
finite periodicity in the system, which was represented as the
decomposition into classes of spectra in the rational case, as
displayed in Fig.~\ref{2comp}. While the noncommutative gauge
theory for irrational $\theta$ is still topological~\cite{lsz1},
its fundamental degrees of freedom are dipoles with momentum
dependent lengths, and the collection of noncommutative instantons
in this case simulates a phase space of extended objects. Note
that, due to numerical limitations, the plots of the form in
Fig.~\ref{2comp} are indistinguishable from certain rational
plots, reflecting the fact that any irrational number can be
approximated by means of an infinite sequence of rational numbers.
Only in the irrational case do the trajectories become dense in
the plane as the number of partition components increases.

A possible strategy for studying this phenomenon at a more
quantitative level is as follows. Let ${\cal G}_m(p,q;\theta)$
be the graph of the Yang-Mills action
$S:{\mbf\Pi}_m(p,q;\theta)\to\real_+:=(0,\infty)$ defined in
(\ref{SYMpart}), and restricted to $m$-component partitions. Then
\beq {\cal
G}_m(p,q;\theta)\subset\left(\K^0(\torus_\theta^2)\right)^m\times
\real_+ \ . \eeq Denoting by
\beq\mathcal{G}(p,q;\theta):=\bigcup_{m=1}^\infty
\mathcal{G}_m(p,q;\theta)\label{graphYM}\eeq
the total graph, we can determine the irregularity of the
Yang-Mills energy curves by computing the fractal dimension of the
set (\ref{graphYM}). If the fractal dimension of the Yang-Mills
graph exceeds the topological dimension of the curves making
up the graph, we would observe evidence of fractal
behaviour in the instanton spectrum. We believe that this phenomenon
is a consequence of the denseness of ${\cal G}(p,q;\theta)$ in
$\real^2$, as suggested by the simplest case depicted in Fig.~\ref{2comp},
which ought to be a reflection of the denseness of the countable set
$\K^0(\torus_\theta^2)=\zed+\zed\,\theta$ on the real line
$\real$~\cite{HW1}, and it is reflected in the irregular structure of
the curve defined by the Yang-Mills action (\ref{SYMpart}) for
irrational values of the noncommutativity parameter $\theta$. We will
not attempt a rigorous proof here along the lines we have just
explained. Multifractal characteristics, of the finite temperature
phase diagrams, have also been noted in the supergravity dual of
noncommutative Yang-Mills theory in four dimensions~\cite{EPR1} and in
two-dimensional noncommutative open string theory~\cite{CHV1}.

The remainder of this paper is devoted to piecing the bits of
evidence from this section together. We shall seek a string theory
that has the same spectral characteristics as irrational
noncommutative Yang-Mills theory. Since the irrational theory may
be naturally regarded as a large $N$ commutative gauge theory,
there should be a natural string interpretation. While we shall
see that this is certainly the case, we find {\it no} evidence in
favour of a phase transition. In fact, noncommutativity appears to
even smoothen out the singularity that is present in the zero area
limit of commutative gauge theory on the torus. We will thereby
arrive at an open string representation of the noncommutative
gauge theory which is valid for all values of the gauge coupling,
and is thus in much better shape than its commutative cousin.

\section{Fluxon Contributions\label{Fluxon}}

To analyse the possibility of a phase transition as described in the
previous section, one would need to acquire a detailed understanding
of the entropy function ${W}_{p,q}(E;\tilde g^2,\theta)$
appearing in (\ref{ZpqDOS}). Unfortunately, this function is rather
complicated and in general difficult to deal with analytically. Instead, we
shall study a special limit of the gauge theory of Section~\ref{NCInsts}
which maps it onto a gauge theory on the noncommutative plane. Unlike
the commutative case, noncommutative gauge theory on $\real^2$
generically possesses natural topologically non-trivial gauge field
configurations and so even its partition function is non-trivial. The
topological configurations which dominate the partition function and
survive the limit in this instance themselves independently possess a
very natural and direct open string interpretation. This treatment
will also naturally connect with the large $N$ limit of commutative
gauge theory on $\torus^2$, in which the standard Gross-Taylor
string expansion will be exploited in the subsequent sections.

\subsection{Decompactification onto Fluxons\label{Decomp}}

The noncommutative instantons constructed in Section~\ref{NCInsts} live
on a torus $\torus^2$ of area $A$. In this
section we will analyse what becomes of these configurations in the
limit $A\to\infty$ whereby the noncommutative torus $\torus_\theta^2$
decompactifies onto the noncommutative plane $\real_\Theta^2$. As we
will see, the torus instantons map in a very precise and definite
way onto {\it fluxons}~\cite{gn}--\cite{gn2}, i.e. finite action
solutions of the noncommutative Yang-Mills equations on $\real^2$ which carry
non-zero magnetic flux. In fact, the usual properties of fluxons
arise very naturally from their origin as instantons on the
noncommutative torus, so that the noncommutative instantons may be
regarded as the ancestors of fluxons, or the corresponding topological
configurations appropriate to the compactified space.

In performing the decompactification limit,
we have to single out those partitions which survive the limit and yield a
finite classical action. Once the relevant configurations are
identified, we can also evaluate the exact partition function
(\ref{Zpqpart}) in the same limit and give a geometrical intepretation
to the instanton expansion presented in~\cite{gsv1}. We will first
describe in detail the case of a noncommutative torus with rational deformation
parameter $\theta$, as it is somewhat more transparent and will be
exploited in subsequent sections. The extension
to irrational values of $\theta$ is worked out afterwards.

Let us consider the projective module $\h_{p,q}$ with
noncommutativitity parameter $\theta=n/N$ where $n,N\in\nat_0$ are
relatively prime. Its dimension (\ref{dimhpq}) is given by
\begin{equation}
\displaystyle{\dim\h_{p,q}=\frac{r}{N}} \ , \,\,\,\,\,\, r=N\,p-n\,q
\ .
\end{equation}
In order to construct the classical solution space we have to find the
general decompositions (\ref{deco}) obeying eq.~(\ref{costri}).
The basic observation~\cite{pasz1} is that the minimal dimension of an
allowed submodule in the present case is $1/N$. This implies that all
dimensions of admissible submodules are quantized in units of
$1/N$, characterized by collections of non-negative integers
$\{r_k\}_{k=1}^{r}$ satisfying
\begin{equation}
\sum_{k=1}^{r}r_k=r
\label{summa}
\end{equation}
with $\dim\h_{p_k,q_k}=r_k/N$. As discussed in the previous section,
the number of such collections is the number $\Pi(r)$ of proper
unordered partitions of the natural number $r$.

The next step is to associate a pair of topological numbers
$(p_k,q_k)$ to each $r_k$, in order to completely specify the
submodules. The relevant condition is given by the Diophantine
equation
\begin{equation}
N\,p_k-n\,q_k=r_k \ .
\label{summa1}
\end{equation}
The general solution of eq.~(\ref{summa1}) is given by
\begin{eqnarray}
p_k&=&m_kr_k+n\,\wt{q}_k \ , \nonumber\\
q_k&=&s_kr_k+N\,\wt{q}_k \ ,
\label{pkqksolgen}\eea
where the integers $m_k,s_k$ satisfy the $SL(2,\zed)$ constraint
\beq
m_k N- s_k n = 1
\label{SL2Zrat}\eeq
and $\wt{q}_k\in\zed$. It is not difficult to prove that the only
degrees of freedom remaining are contained in the arbitrary integer
$\wt{q}_k$, as different choices of integer pairs $(m_k,s_k)$ obeying
the $SL(2,\zed)$ equation (\ref{SL2Zrat}) simply correspond to shifts
in $\wt{q}_k$. In particular, we can fix $(m_k,s_k)=(m,s)~~\forall k$
within a particular submodule decomposition (\ref{deco}). The relevant
decompositions can therefore be presented as
\begin{equation}
\h_{p,q}=\bigoplus_{k=1}^{r}\h_{n\,\wt{q}_k+m\,r_k\,,\,N\,\wt{q}_k+s\,r_k}
\ ,
\label{hpqratdec}\end{equation}
and the sum rule on $\wt{q}_k$ is easily derived to be
\begin{equation}
\sum_{k=1}^{r}\wt{q}_k=m\,q-s\,p \ .
\label{summwtqk}\end{equation}
Thus all classical solutions on the module $\h_{p,q}$, with rational
noncommutativity parameter $\theta=n/N$, are specified by the sets of
non-negative integers $\{r_k\}_{k=1}^{r}$ and the sets of
(generically negative) integers $\{\wt{q}_k\}_{k=1}^{r}$ obeying the
constraints (\ref{summa}) and (\ref{summwtqk}).

We are now ready to describe the decompactification limit which maps
the quantum gauge theory on the noncommutative torus to a quantum
gauge theory on the noncommutative plane. The important point is that, on the
noncommutative plane $\real_\Theta^2$, the relevant parameter $\Theta\in\real$
encoding noncommutativity  is dimensionful. On the torus it is given
by
\begin{equation}
\Theta=\frac{A\,\theta}{2\pi} \ ,
\label{Thetadef}\end{equation}
and consequently we have to perform the limit $A\to\infty$ while
keeping fixed the combination (\ref{Thetadef}). A natural way to
do this in the present case is to send $N\to\infty$ with
\begin{equation}
A=\displaystyle{\frac{2\pi\,N\,\Theta}{n}} \ ,
\label{relation}
\end{equation}
while keeping $\Theta$ and $n$ finite. As in the cases of matrix model
regularizations~\cite{amns1,LSZam1} and the Morita equivalence formulation of
noncommutative field theories~\cite{gsv1}, the large $N$ limit and the
infrared limit are correlated, resulting in a double scaling relation
on the classical solutions. The precise geometrical meaning of this
double scaling limit will be described at length in
Section~\ref{DSL}.

Let us now identify the configurations which have finite classical action
(\ref{SYMpart}) in this limit. It is straightforward to see that the
relevant sets of non-negative integers $\{r_k\}_{k=1}^{r}$ are given by
\begin{equation}
\{r_k\}_{k=1}^{r}=\{n\,\ell_k\}_{k=1}^\ell\cup\{r'\}~~~~
{\rm with}~~\sum_{k=1}^\ell\ell_k= \ell \ , ~~ r'=r-n\,\ell \ ,
\label{pesi}
\end{equation}
where $\ell$ is a positive integer which is of order~$1$ in the limit
$N\to\infty$. The associated values of $\wt{q}_k$ are given by the sets
\begin{equation}
\{\wt{q}_k\}_{k=1}^{r}=\{-m\,\ell_k\}_{k=1}^\ell\cup\{q'\}~~~~
{\rm with}~~q'=m\,(q+\ell)-s\,p \ .
\label{pesi1}
\end{equation}
These decompositions correspond to finite action solutions in the set
of all classical solutions of the Yang-Mills equations on the
projective module $\h_{p,q}$ of fixed topological numbers $(p,q)$. As
argued in~\cite{gsv1}, these are the dominant configurations in the
large $N$ limit, with all other contributions being exponentially
suppressed in the partition sum (\ref{Zpqpart}).

With the dominant configurations of topological numbers
(\ref{pesi},\ref{pesi1}), the submodule decompositions
(\ref{hpqratdec}) read
\begin{equation}
\h_{p,q}=\bigoplus_{k=1}^\ell\h_{0,-\ell_k}~\oplus~\h_{p,\ell+q} \ .
\label{hpqdomdec}\end{equation}
By using (\ref{SYMpart}) and (\ref{relation})--(\ref{pesi1}), one
finds that the classical solution corresponding to the decomposition
(\ref{hpqdomdec}) has action
\bea
S\left(\,\underline{(p,q)}\,;\,\theta=n/N
\right)&=&\displaystyle{\frac{2\pi^2N}{g^2A\,n}\,\sum_{k=1}^\ell
\ell_k+\frac{2\pi^2N}{g^2A}\,\frac{(\ell+q)^2}{Np-n(\ell+q)}}\nonumber\\
&=&\displaystyle{\frac{\pi\,\ell}{g^2\Theta}+\frac{\pi\,n}{g^2\Theta}\,
\frac{(\ell+q)^2}{Np-n(\ell+q)}} \ .
\label{SYMpartlargeN}\eea
In the large $N$ limit, the second term in (\ref{SYMpartlargeN})
vanishes. Thus only the instantons associated with the
first summand of (\ref{hpqdomdec}) contribute to the partition
function (\ref{Zpqpart}). In particular, the full partition function
${\cal Z}_\infty(g^2,\Theta)$ of ``physical'' gauge theory on
$\real_\Theta^2$, i.e. including a sum over all topological charges,
can be obtained from (\ref{Zpqpart}) in this limit, and it is dominated by the
same configurations as in eqs.~(\ref{pesi},\ref{pesi1}). Due to a
combination of large $N$ and large area suppression, only the instanton
partitions corresponding to the first summand of (\ref{hpqdomdec}) yield a
nonvanishing contribution, and the sum over partitions for fixed topological
numbers $(p,q)$ becomes a sum over $\ell\in\nat_0$ along with a sum
over the $\Pi(\ell)$ proper partitionings in (\ref{pesi}). One finds
\bea
{\cal Z}_\infty\left(g^2\,,\,\Theta\right)&:=&\lim_{A\to\infty}\,
Z_{p,q}\left(g^2A\,,\,\theta=2\pi\,\Theta/A\right)\nonumber\\&=&
\frac1{N\,\sqrt{g^2\Theta}}\,\sum_{\ell=0}^\infty\,(-1)^\ell~
\e^{-\pi\,\ell/g^2\Theta}\,\sum_{\stackrel{\scriptstyle\vnu\in
\nat_0^\ell}{\scriptstyle\sum_kk\,\nu_k=\ell}}~
\prod_{k=1}^\ell\frac{(-1)^{\nu_k}}{\nu_k!}\,\left(\frac{A^2}{2\pi\,
k^3g^2\Theta^3}\right)^{\nu_k/2} \ , \nonumber\\&&
\label{calZNCplane}\eea
where the (infinite) area $A$ of the noncommutative plane
plays the role of an infrared regularization of the quantum field theory.

This calculation suggests that, at the level of classical solutions
and the exact semi-classical expansion of the quantum gauge theory on
$\torus^2_\theta$, the decompactification limit of a {\it generic} projective
module $\h_{p,q}$ receives contributions only from submodules of the
type $\h_{0,-\ell_k}$, with the dimension positivity
constraints $p_k-q_k\theta>0$ in (\ref{costri}) requiring
$\ell_k\geq 0$ for $\theta>0$ (For $\theta<0$ one would instead find
$\h_{0,\ell_k}$ with $\ell_k\leq 0$). The classical action is
proportional to $\ell=\sum_k\ell_k$ with $\ell$ an arbitrary
integer. No information is retained of the original geometric
data of the gauge theory on the noncommutative torus, i.e. its
topological numbers $(p,q)$. This is simply the action of the
$|\ell|$-fluxon solution of $U(1)$ Yang-Mills theory on the
noncommutative plane~\cite{polly,gn2}.

It is important to note here
that the zero action configurations in the decompactification limit
are suppressed from the partition function by area factors coming from
the quantum fluctuation determinants. It is also remarkable that,
starting from any $(p,q)$, the resulting quantum theory on the
noncommutative plane naturally includes contributions from all
topological sectors labelled by the fluxon charge $\ell$. The sum over
$\ell\in\nat_0$, which labels the partitions surviving the
decompactification limit, arises in the vacuum amplitude
(\ref{Zpqpart}) over the module $\h_{p,q}$ because of the original sum
over partitions. In this sense, the quantum gauge theory in the limit is
``universal'', in that it is independent of the particular Heisenberg
module on which it is defined. Moreover, the contributions from any
finite action fluxon configuration are identical, as the spurious
dependences (e.g. on $\ell+q$ in (\ref{hpqdomdec})) are washed out
in the limit $\theta\to0$ in every topological sector of the
noncommutative gauge theory.

The same expressions were obtained by different means
in~\cite{gsv1}. There a $U(1)$ noncommutative gauge theory on a torus
of rational noncommutativity parameter $1/N$ was considered and Morita
equivalence was used to map the system onto a commutative one. The
decompactification limit was taken as in eq.~(\ref{relation}),
resulting in a peculiar large $N$ limit on the partition function of
(commutative) $SU(N)/\zed_N$ Yang-Mills theory on $\torus^2$, which can
be computed exactly by standard techniques~\cite{mig,rusakov1}. In that case
the size of the Morita equivalent torus vanishes as $1/N$, the
't~Hooft limit of the Yang-Mills coupling constant must be taken, and the
topological charge remains finite in the limit. In the original
$SU(N)/\zed_N$ theory, there are exactly $N$ topological sectors
characterizing the inequivalent principal $SU(N)/\zed_N$ gauge bundles over
$\torus^2$~\cite{hooft2}. The fluxon sum then appears as the sum over
configurations surviving the large $N$ limit in the $SU(N)/\zed_N$
partition function for a fixed topological sector in $\zed_N$. This
technique will be exploited later on within another, related context.

Finally, it is straightforward to see that precisely the same fluxon
configurations are recovered in the case of irrational deformation
parameter $\theta$. In the limit $A\to\infty$ with the combination
(\ref{Thetadef}) held finite, the field strengths (\ref{Fpqdef})
vanish for $p\neq0$, as do the fluctuation factors in
(\ref{Zpqpart}). Thus in the decompactification limit, only the
partitions associated to the Heisenberg submodules $\h_{0,-\ell_k}$
have finite action and survive in the quantum field theory. As in the
rational case, a generic instanton has vanishing classical action in
the limit. The only constant curvature connections, with non-trivial
action in the decompactification limit, derive from modules of the
type $\h_{0,-\ell_k}$.

\subsection{Decompactification of Instanton Configurations\label{DecompFields}}

While the analysis of the previous subsection provides circumstantial
evidence that the
instantons on noncommutative $\torus^2$ decompactify onto fluxons on
noncommutative $\real^2$, we can actually go further and obtain a
one-to-one correspondence between the two sets of non-trivial gauge
field configurations
in the limit. Let us assume for definiteness that $\theta>0$. Recall
that any projective module over the algebra $\alg_\infty={\cal
S}(\real_\Theta^2)$ of Schwartz functions on the noncommutative plane
is of the form ${\cal F}^\ell\oplus(\alg_\infty)^{\,m}$~\cite{ksrev},
and is thereby characterized by two non-negative integers $\ell$ and
$m$ which are, respectively, the magnetic flux quantum number and
gauge group rank of the corresponding gauge bundle over
$\real^2_\Theta$. It follows that the K-theory group of the
noncommutative plane is $\K^0(\real^2_\Theta)=\zed^2$. However, the
positive cone $\nat^2$ is rather different than that of
$\K^0(\torus_\theta^2)$, and as a consequence
one cannot realize modules with negative magnetic flux. There is no
natural way to distinguish between the deformations corresponding to
$\Theta$ and $-\Theta$ on the plane, as all deformations are Morita
equivalent in this case because the Heisenberg commutation relations
have a unique irreducible representation, the Schr\"odinger
representation of quantum mechanics, irrespective of the value of
$\Theta\neq0$ (This is the Stone-von~Neumann theorem). In the context
of fluxon solutions, noncommutativity breaks charge conjugation
symmetry and one cannot simply produce anti-vortices from vortices via
a change of orientation on $\real^2$. Thus any theory involving
fluxons will be chiral~\cite{ksrev,dnrev,gn2}. In the following we
will consider the rank~$1$ case $m=1$.

Let us now consider the contributing projective module decompositions
(\ref{hpqdomdec}) over the noncommutative torus in the
decompactification limit. Generally, from
(\ref{Fpqdef})--(\ref{WtHalg}) it follows that there is a natural isomorphism
$\h_{kp,kq}\cong(\h_{p,q})^k$, and so it suffices to consider the
total submodule
\beq
\h_{0,-\ell}=\bigoplus_{k=1}^\ell\h_{0,-\ell_k}=
{\cal F}\otimes\complex^\ell\cong{\cal F}^\ell
\label{h0ell}\eeq
for $\ell>0$. This isomorphism is a consequence of the fact that, for
$p=0$, the Weyl-'t~Hooft algebra (\ref{WtHalg}) reduces simply to the
requirement that $\Gamma_1\Gamma_2=\Gamma_2\Gamma_1$ which may be
solved generically (up to unitary isomorphism) by arbitrary
$\ell\times\ell$ diagonal unitary matrices
\beq
\Gamma_i=\e^{-2\pi\ii{\mbf\gamma}_i/\sqrt A}:=
\begin{pmatrix}\e^{-2\pi\ii\gamma_i^{(0)}/\sqrt A}& & \\ &\ddots& \\
& &\e^{-2\pi\ii\gamma_i^{(\ell-1)}/\sqrt A}\end{pmatrix} \ ,
\label{Gammaidiag}\eeq
with $\gamma_i^{(k)}\in\real$, $i=1,2$, $k=0,1,\dots,\ell-1$. Using
(\ref{Fpqdef}) and (\ref{Thetadef}), one identifies the field strength
of the global instanton of the module (\ref{h0ell}) in the
decompactification limit
as
\beq
F_{0,-\ell}=\frac\ii\Theta\,{\sf B}_\theta^\ell \ ,
\label{F0ell}\eeq
where we have explicitly represented the K-theory class of
$\h_{0,-\ell}$ through the Boca projection ${\sf
B}_\theta^\ell=\proj_{0,-\ell}$ on the noncommutative torus with $\Tr\,{\sf
B}_\theta^\ell=\ell\,\theta$~\cite{MM}--\cite{lls1}. The trace of the projector
$\proj_\ell^\prime=A\,{\sf B}_\theta^\ell$ over $\h_{0,-\ell}$ is
\beq
\Tr\,\proj^\prime_\ell=2\pi\,\Theta\,\ell \ ,
\label{Trprojell}\eeq
and it is possible to show~\cite{ks1,lls1} that its decompactification
limit yields the canonical rank $\ell$ projector
\beq
\frac1{2\pi\,\Theta}\,\proj^\prime_\ell=
\proj^{~}_\ell:=\sum_{k=0}^{\ell-1}|k\rangle\langle k|
\label{projellplane}\eeq
on the Fock module
\beq
\Fock=\bigoplus_{k=0}^\infty\,\complex\cdot|k\rangle
\label{Fockmodule}\eeq
over the noncommutative plane $\real^2_\Theta$.

We now identify the classical gauge field configurations in the
limit. In (\ref{U12commrel}) we wish to make an identification
of the form $U_i\sim\e^{2\pi\ii x_i/\sqrt A}$, with $x_i$, $i=1,2$ the
coordinate
generators of $\real_\Theta^2$ obeying the Heisenberg algebra
\beq
[x_1,x_2]=\ii\Theta \ .
\label{NCR2commrels}\eeq
However, one needs to be careful in making a naive equality, because
the $x_i$ are operators on the trivial rank~$1$ module $\alg_\infty$,
not on the Heisenberg module (\ref{h0ell}) on which
\beq
U_i=\e^{-\frac{2\pi}{\sqrt A}\,(\Theta\,\nabla_i\otimes\id_\ell+\ii\id\otimes
{\mbf\gamma}_i)}
\label{Uidecomp}\eeq
are represented. In fact, we can identify the Heisenberg module
(\ref{h0ell}) as a proper submodule of $\alg_\infty$ by using the fact that the
rank~$1$ free module can be decomposed as
\beq
\alg_\infty=\bigoplus_{k=0}^\infty|k\rangle\langle k|\cdot\alg_\infty \ ,
\label{freedecomp}\eeq
where the operator $|k\rangle\langle k|$, $k\in\nat_0$ is the orthogonal
projection
onto the one-dimensional subspace spanned by the vector $|k\rangle\in\Fock$,
and here we regard the algebra $\alg_\infty$ in its irreducible
representation on the Fock space (\ref{Fockmodule}) given by
functionals of the standard creation and annihilation operators. For
each $k\in\nat_0$ there is an isomorphism $|k\rangle\langle
k|\cdot\alg_\infty\cong\Fock$ given by the mapping
$|k\rangle\langle k|\cdot f\mapsto f|k\rangle\in\Fock$ for
$f\in\alg_\infty$. In particular, using this isomorphism we have
\beq
\h_{0,-\ell}\cong\proj_\ell\cdot\alg_\infty \ ,
\label{h0ellwtalg}\eeq
with $\proj_\ell$ the orthogonal projection (\ref{projellplane}) on
$\Fock\to\complex^\ell$. Eq.~(\ref{h0ellwtalg}) is just the
definition (\ref{stablemodule}) in the decompactification limit of the
torus.

Making the desired identification in the unitary generators $U_i$
requires that all operators be defined on a common domain. We
circumvent this difficulty by trivially embedding the pertinent
operators into the module $\h_{0,-\ell}\oplus\alg_\infty$ as
\bea
\hat\nabla_i&:=&\left(\nabla_i\otimes\id_\ell\right)\oplus0 \ , \nonumber\\
\hat{\mbf\gamma}_i&:=&\left(\id\otimes{\mbf\gamma_i}
\right)\oplus0 \ , \nonumber\\\hat x_i&:=&{\mbf 0}_\ell\oplus x_i \ .
\label{trivemb}\eea
We then represent all of these operators on the
free module $\alg_\infty$ by finding their images under a unitary
isomorphism of separable Hilbert spaces
\beq
\Omega_\ell\,:\,\h_{0,-\ell}\oplus\alg_\infty~
\stackrel{\approx}{\longrightarrow}~\alg_\infty \ .
\label{Deltaelliso}\eeq
That such an isomorphism between $\alg_\infty$-modules exists follows
from (\ref{h0ellwtalg}) and is just the usual statement that any gauge
bundle over the plane is trivial. In this way, we can then relate the
torus generators (\ref{Uidecomp}), regarded now as operators on
$\alg_\infty$, to exponentials of the plane generators obeying
(\ref{NCR2commrels}) through
\beq
U_i=\Omega_\ell^{~}~\e^{-\frac{2\pi}{\sqrt A}
\,(\Theta\,\hat\nabla_i+\ii
\hat{\mbf\gamma}_i)}~\Omega_\ell^{-1}:=\Omega_\ell^{~}~\e^{2\pi\ii
\hat x_i/\sqrt A}~\Omega_\ell^{-1} \ .
\label{Uiwtalgid}\eeq

The mapping $\Omega_\ell$ can be constructed explicitly by introducing the
standard shift operator
\beq
\shift_\ell=(\shift_1)^\ell:=\sum_{k=0}^\infty|k+\ell\rangle\langle k|
\label{shiftelldef}\eeq
on the Fock module (\ref{Fockmodule}), which is a partial isometry on
$\Fock$ obeying
\beq
\shift^\dag_\ell\,\shift_\ell^{~}=\id \ , ~~
\shift^{~}_\ell\,\shift^\dag_\ell=\id-\proj^{~}_\ell \ .
\label{shiftprojrel}\eeq
In particular, the operator $\shift_\ell$ is a unitary isomorphism on
the orthogonal complement in $\Fock$ to a finite dimensional cokernel,
i.e. $\ker\shift_\ell=\{0\}$ and $\ker\shift_\ell^\dag={\rm
im}\,\proj^{~}_\ell\cong\complex^\ell$. This implies that the submodule
$\shift_\ell\cdot\alg_\infty$ is the orthogonal complement in
$\alg_\infty$ of the submodule (\ref{h0ellwtalg}), and $\shift_\ell$ can
thereby be used to construct the isomorphism we seek. We thus define
(\ref{Deltaelliso}) as
\beq
\Omega_\ell\left(\,\mbox{$\sum\limits_{k=0}^{\ell-1}$}\,
f|k\rangle~\oplus~f'\,\right):=\proj_\ell
\cdot f+\shift_\ell\cdot f' \ ,
\label{Deltaelldef}\eeq
with inverse given by
\beq
\Omega_\ell^{-1}(f)=\sum_{k=0}^{\ell-1}f|k\rangle~\oplus~
\shift_\ell^\dag\cdot f
\label{Deltaellinv}\eeq
where $f,f'\in\alg_\infty$.

It is straightforward to now work out the images on
$\alg_\infty\to\alg_\infty$ of the operators (\ref{trivemb}) under
this isomorphism, and one finds
\bea
\Omega_\ell^{~}\,\hat x_i\,\Omega_\ell^{-1}(f)&=&\shift^{~}_\ell\,
x_i\,\shift^\dag_\ell\cdot f \ , \nonumber\\\Omega_\ell^{~}\,
\hat{\mbf\gamma}_i\,\Omega_\ell^{-1}(f)&=&\sum_{k=0}^{\ell-1}
\gamma_i^{(k)}~|k\rangle\langle k|\cdot f \ .
\label{Deltaellims}\eea
{}From (\ref{Uiwtalgid}) and (\ref{Deltaellims}) we arrive finally at an
expression for the gauge connection in the decompactification limit as
an operator on the free module $\alg_\infty$ given by
\beq
D_i:=\ii\Theta\,\Omega_\ell^{~}\,\hat\nabla_i\,
\Omega_\ell^{-1}=\shift^{~}_\ell\,x_i\,\shift^\dag_\ell+
\sum_{k=0}^{\ell-1}\gamma_i^{(k)}~|k\rangle\langle k| \ .
\label{Difluxon}\eeq
The operator (\ref{Difluxon}) is precisely the gauge field
configuration of the $\ell$-fluxon solution of $U(1)$ noncommutative
Yang-Mills theory on the Fock module (\ref{h0ell}) over
$\real^2_\Theta$~\cite{ksrev,dnrev,polly,agms1,gn2}.

\subsection{Properties of Fluxons\label{FluxonProps}}

The realization of fluxons in this way by their pre-images as
instantons on the noncommutative torus immediately implies many of the
peculiar features that these vortex solutions exhibit. These
properties are not so unusual in the parent theory where they are completely
transparent, as they follow from some of the basic properties of
Heisenberg modules. Because of the richer topology, there are of
course many more noncommutative instantons on $\torus^2$ than there
are noncommutative vortices on $\real^2$, but when the torus is
decompactified only a subset of the instantons survive and become
localized vortex solutions on the noncommutative plane. Note that this
is a quantum effect, as it arises from the suppression of all other
instanton modes in the fluctuation determinants of the vacuum
amplitude (\ref{Zpqpart}).

Let us now summarize some of the novel properties that the fluxons
naturally inherit in this way:
\begin{enumerate}
\item Fluxons only exist with one sign of the magnetic charge ${\rm
sgn}\,\ell={\rm sgn}\,\theta$. This is a consequence of the positivity
of the Murray-von~Neumann dimension (\ref{dimhpq}) of the module
$\h_{0,-\ell}$.
\item The quadratic dependence of the action (\ref{SYMpart}) on the
instanton charge becomes linear when $p_k=0$, explaining the linear
growth of the soliton mass with the topological charge $\ell$.
\item The $2\ell$-dimensional moduli space of solutions
for magnetic charge $\ell>0$ corresponding to the separations of the
vortices comes from the trivial representation (\ref{Gammaidiag}) on
$\complex^\ell$ of the generators (\ref{Uirep}) for the module
$\h_{0,-\ell}$. Thus the intrinsic structure of Heisenberg modules
induces the fluxon moduli space.
\item By using (\ref{F0ell}) and (\ref{Deltaellims}), the field
strength of the gauge field (\ref{Difluxon}) is easily worked out to
be
\beq
\frac{\ii A}{2\pi\,\Theta^2}\,\bigl[D_1\,,\,D_2\bigr]=-\frac{\ii A}
{2\pi\,\Theta}~\Omega_\ell^{~}
\,\hat F_{0,-\ell}\,\Omega_\ell^{-1}=\proj^{~}_\ell
\label{fluxoncurv}\eeq
with $\hat F_{0,-\ell}=[\hat\nabla_1,\hat\nabla_2]$,
implying that the configuration has quantized magnetic charge
$\Tr\,\proj_\ell=\ell$. By construction on $\torus_\theta^2$, the
field strength (\ref{F0ell}) {\it a priori} has the usual
noncommutative vortex form in terms of a projection operator on a Fock
module~\cite{ksrev,dnrev}, coming from its origin in terms of a
constant curvature connection over the noncommutative torus.
\item From (\ref{fluxoncurv}) it follows that the energy of the
configuration of vortices is independent of their position moduli
$\gamma_i^{(k)}\in\real$. As a consequence, the semi-classical
expansion of the quantum gauge theory about the fluxons
diverges~\cite{gn2}, and must be regulated by an infrared cutoff $A$,
the finite size of the torus $\torus^2$, as in (\ref{calZNCplane}). The exact
fluctuation spectrum around each fluxon is given by
the pre-exponential factors of the {\it exact} path integral
(\ref{calZNCplane}) of gauge theory on the noncommutative plane. It
contains, in particular, the correct moduli dependence
$A^{\nu_1+\dots+\nu_\ell}$ of the collections of $\nu_k$ elementary
constituent vortices of charge $k$ inside the $\ell$-fluxon solution,
along with the permutation symmetry factors $1/\nu_k!$ appropriate to the
identical vortices of equal charge inside each soliton~\cite{gsv1}.
\item All fluxons of charges $\ell\in\nat_0$ originate from
the {\it same} Heisenberg module $\h_{p,q}$ over the noncommutative
torus, with fixed topological numbers
$(p,q)\in\K^0(\torus_\theta^2)$. This fact has two important
implications. First of all, although the gauge connection
(\ref{Difluxon}) is constructed from the globally minimizing constant
curvature connection on the Heisenberg module (\ref{h0ell}), it
originates from the submodule decompositions (\ref{hpqdomdec}) which
generically correspond to unstable extrema of the gauge theory on
$\torus_\theta^2$. This naturally explains the instability of the
fluxon solutions on the noncommutative
plane~\cite{agms1,gn2}. Secondly, the fluxon charge $\ell$ is not a
conserved quantum number~\cite{dnrev}. Because of their metastability,
it is possible for a vortex
solution to interpolate between different topological sectors
characterized by total flux. This physical statement is in no
contradiction with the conventional definitions, provided we consider
it from the point of view of the instanton ancestors.
\end{enumerate}

Thus the torus instanton origin of fluxons described above naturally
explains their known properties in very simple settings, and moreover
implies new features of them. On the other hand, the
$\ell$-fluxon solution (\ref{Difluxon}) can be
interpreted~\cite{agms1} as a non-BPS configuration of $\ell$
Euclidean D$(-1)$-branes at positions
$(\gamma_1^{(k)},\gamma_2^{(k)})$, $k=0,1,\dots,\ell-1$ on a Euclidean
D1-brane, in the presence of a background $B$-field in the Seiberg-Witten
decoupling limit~\cite{sw}. Its fluctuation spectrum contains a tachyon and at
the endpoint of tachyon condensation the D-instantons are completely
dissolved in the D-string. It is natural to now examine if this string
origin of fluxons can be understood through the stringy properties of the torus
instantons that we unveiled in the previous section. We will begin
attacking this problem in the next section. Note that the charge
$\ell$ of a fluxon is proportional to its energy $E$, and for each
$\ell$ there are $\Pi(\ell)$ states in (\ref{calZNCplane}) which
contribute at fixed energy $E$. It then follows from the
Hardy-Ramanujan formula (\ref{HRformula}) that
the asymptotic density of states has the characteristic
sub-exponential behaviour $\e^{\alpha\,\sqrt E}$ of quantum field
theory. This strongly suggests that, despite their instability, the
fluxon contributions do {\it not} drive the system to a phase
transition. In Section~\ref{ISRNCGT} we will show that this is indeed
the case.

\subsection{The Fluxon Partition Function\label{TFPF}}

For later use, let us now explicitly sum the
fluxon expansion (\ref{calZNCplane}). For this, we unravel the
constraint on the partitions $\vnu$ via a contour integral
representation for the Kronecker delta-function given by
\beq
\delta_{mn}=\oint\limits_{{\cal C}_0}\,\frac{\dd z}{2\pi\ii z}~
z^{m-n} \ ,
\label{deltamncontint}\eeq
where $m,n\in\zed$ and the closed contour ${\cal C}_0$ encircles the
origin $z=0$ of the complex $z$-plane with counterclockwise
orientation. At this stage ${\cal C}_0$ is arbitrary, but later on we
will have to choose it carefully for convergence reasons. We may
thereby write the amplitude (\ref{calZNCplane}) as a sum over {\it
unconstrained} integers as
\bea
{\cal Z}_\infty\left(g^2\,,\,\Theta\right)&=&
\sum_{\ell=0}^\infty\,(-1)^\ell~\e^{-\pi\,\ell/g^2\Theta}
\,\sum_{\vnu\in\nat_0^\ell}~\prod_{k=1}^\ell\frac{(-1)^{\nu_k}}
{\nu_k!}\,\left(\frac{A^2}{2\pi\,k^3g^2\Theta^3}\right)^{\nu_k/2}
\nonumber\\&&\times\,
\oint\limits_{{\cal C}_0}\,\frac{\dd z}{2\pi\ii z^{\ell+1}}~
z^{\sum_ll\,\nu_l}
\label{calZunconstr}\eea
where we have discarded the overall multiplicative constant. The sums
can now be done explicitly thanks to the contour integration, and one
finds
\beq
{\cal Z}_\infty\left(g^2\,,\,\Theta\right)=\oint\limits_{{\cal C}_0}\,
\frac{\dd z}{2\pi\ii}~\frac1{z+\e^{-\pi/g^2\Theta}}~
\exp\left[-\frac A{\sqrt{2\pi\,g^2\Theta^3}}~{\rm Li}_{3/2}(-z)
\right] \ ,
\label{calZellsum}\eeq
where
\beq
{\rm Li}_\alpha(z):=\sum_{k=1}^\infty\frac{z^k}{k^\alpha}
\label{polylogdef}\eeq
is the Jonqui\`ere polylogarithm function of index
$\alpha\in\complex$. We will therefore restrict the contour of
integration ${\cal C}_0$ in (\ref{calZellsum}) to lie in the open
annulus $\e^{-\pi/g^2\Theta}<|z|<1$ in order to ensure convergence of
the series (\ref{polylogdef}) and to catch the pole at
$z=-\e^{-\pi/g^2\Theta}$. This constraint can be relaxed somewhat via
an appropriate analytic continuation which is described in
Appendix~\ref{appA}.

The residue theorem now implies our final expression
\beq
{\cal Z}_\infty\left(g^2\,,\,\Theta\right)=
\exp\left[-\frac A{\sqrt{2\pi\,g^2\Theta^3}}~{\rm Li}_{3/2}
\left(\e^{-\pi/g^2\Theta}\right)\right]
\label{calZNCplanefinal}\eeq
for the partition function of quantum gauge theory on the
noncommutative plane~\cite{gsv1}. This {\it exact} extensive form of the vacuum
amplitude is typical of a dilute instanton gas approximation. It is
consistent with the fact that fluxons are non-interacting solitons.

\section{Instanton String Expansions\label{SLISE}}

In the previous section we found that fluxons on the noncommutative
plane are dynamically induced by the topological configurations on the
noncommutative torus. As the noncommutative gauge theory can be
obtained from a commutative $SU(N)/\zed_N$ gauge theory through a
large $N$ double scaling limit~\cite{gsv1}, it is natural to explore the
problem of obtaining the open string interpretation of fluxons through
the conventional closed string expansion of ordinary Yang-Mills
theory in two spacetime dimensions. However, this string
representation is based on the strong-coupling expansion of the
commutative gauge theory. As we mentioned earlier, while such an
expansion is available for the gauge theory on the noncommutative
torus~\cite{pasz2}, it is rather complicated and does not
straightforwardly admit nice algebraic or geometric
characterizations. We must therefore try to extract the string
representation of two-dimensional noncommutative Yang-Mills theory
directly from its weak-coupling expansion. To understand this point,
in this section we will revisit the Gross-Taylor string expansion of
{\it commutative} $U(N)$ Yang-Mills theory on the torus $\torus^2$ and
analyse how it manifests itself in the instanton series.

\subsection{Gross-Taylor Series on the Torus\label{TG-TET}}

The chiral partition function of ordinary $U(N)$ Yang-Mills theory
on the torus $\torus^2$ is a variant of the usual Migdal
strong-coupling expansion~\cite{mig,rusakov1} given by \beq
Z_{U(N)}^+(\lambda)=\sum_{R\in{\rm
Rep}^+(U(N))}\e^{-\frac{\lambda}{N}\,C_2(R)} \ ,
\label{chiralsumR}\eeq where \beq \lambda=g^2A\,N/2
\label{tHooftconst}\eeq is the dimensionless 't~Hooft coupling constant and
$C_2(R)$ is the quadratic Casimir eigenvalue in the irreducible
representation $R=R(Y)$ of the gauge group $U(N)$. The restriction
to chiral representations means considering Young tableaux
$Y$ with positive number of boxes and dropping the constraint that
the number of rows be less than the rank $N$. In so doing, one
essentially assumes that only representations with small numbers
of boxes (compared to $N$) are relevant in the large $N$ limit,
the others being exponentially suppressed in $N$~\cite{gross1}.
More generally, a class of representations with box numbers of
order $N$, whose contribution is not exponentially damped, can be
found and used to construct the anti-chiral sector~\cite{gt1}. A
non-chiral coupled expansion, incorporating both contributions,
has been presented in~\cite{gt1,gt2} for the case of an $SU(N)$
gauge group and is widely accepted as the complete large $N$
description of the commutative gauge theory.

For a Young diagram $Y\in{\cal Y}_n$ of $n$ boxes, the quadratic
Casimir can be written as \beq
C_2\bigl(R(Y)\bigr)=N\,n+\wt{C}_2\bigl(R(Y)\bigr) \ ,
\label{C2tildeC2UN}\eeq where $\wt{C}_2(R(Y))$ is of order~$1$ in
the large $N$ limit. The partition function (\ref{chiralsumR})
then admits the asymptotic $\frac1N$ expansion \beq
Z_{U(N)}^+(\lambda)=\sum_{n=0}^\infty~\sum_{Y\in{\cal Y}_n}
\e^{-n\,\lambda}\,\sum_{m=0}^\infty\frac{\left[-\lambda\,
\wt{C}_2\bigl(R(Y)\bigr)\right]^m}{m!}~\frac1{N^m} \ .
\label{chiral1N}\eeq The irreducible representations $R(Y)$ are
labelled by non-increasing partitions $\infty>n_1\geq
n_2\geq\dots\geq n_n\geq 0$, $\sum_kn_k=n$ corresponding to the
conjugacy classes of the symmetric group $S_n$ and specifying the
lengths of rows in the Young diagram $Y\in{\cal Y}_n$ with \beq
\wt{C}_2\bigl(R(Y)\bigr)=\sum_{k=1}^nn_k(n_k+1-2k) \ .
\label{tildeC2expl}\eeq This quantity is the central character of
the corresponding irreducible representation of $S_n$.

In this way the expansion (\ref{chiral1N}) acquires the
form~\cite{gt1}
\beq Z_{U(N)}^+(\lambda)=1+\sum_{h=1}^\infty\frac1{N^{2h-2}}\,
\sum_{n=1}^\infty\e^{-n\,\lambda}\,\sum_{m=1}^\infty
\omega_h^{m,n}~(2\lambda)^m \ , \label{chiralHurwitz}\eeq where
the coefficients $\omega_h^{m,n}\in\nat_0$ give the number of
(topological classes of) $n$-sheeted holomorphic covering maps
without folds to the torus $\torus^2$. They count the number of
(disconnected) maps from a closed oriented Riemann surface of
genus $h$ to the torus with winding number $n$ and $m$ simple
branch points. The expansion (\ref{chiralHurwitz}) contains a
Nambu-Goto factor $\e^{-n\,\lambda}$ as well as a volume factor
$\lambda^m$ from the moduli space integration over the positions
of the branch points. It is important to note that this is only
the {\it chiral} expansion of the gauge theory. The full $SU(N)$
partition function also couples a sector of antiholomorphic maps,
along with singular geometrical contributions such as collapsed
handles and infinitesimal tubes connecting the sheets of the
holomorphic covers with those of the antiholomorphic covers. Such
singularities come from the boundary of moduli space where
worldsheet handles degenerate to a point in the spacetime
$\torus^2$, and they complicate the form of $\omega_h^{m,n}$.
However, at the present level of the chiral $U(N)$ theory, these
natural numbers are of the form
$\omega_h^{m,n}=\omega_h^n~\delta_{m,2h-2}$ by the Riemann-Hurwitz
theorem, and the explicit summation over Young diagrams in
(\ref{chiral1N}) yields \bea
\omega_h^n&=&\sum_{k=1}^n~\sum_{\stackrel{\scriptstyle \mbf
n\in\nat^k\,,\,\sum_ln_l=n}{\scriptstyle n_1\geq n_2\geq\cdots
\geq
n_k}}\,\left[\,\sum_{l=1}^k\frac{(n_l+k-l)(n_l+k-l-1)}2\right.
\nonumber\\&&\times\Biggl.~\prod_{l'\neq l}\,
\left(\frac{n_l-n_{l'}+l'-l-2}{n_l-n_{l'}+l'-l}\right)\Biggr]^{2h-2}
\ , \label{Hurwitzred}\eea in agreement with earlier results from
combinatorial group theory~\cite{Med1,MSS1}.

The free energy $F_{U(N)}^+(\lambda)=\ln Z_{U(N)}^+(\lambda)$ can
be written as \beq
F_{U(N)}^+(\lambda)=\sum_{h=1}^\infty\frac1{N^{2h-2}}\,
F^+_h(\lambda) \label{chiralfreeen}\eeq where
\beq F^+_h(\lambda)=(2\lambda)^{2h-2}\,
\sum_{n=1}^\infty\sigma_h^{n}~\e^{-n\,\lambda} \ .
\label{FglambdaA}\eeq The non-negative integers \beq
\sigma_h^{n}=\sum_{k=1}^n\frac{(-1)^k}k~\sum_{\stackrel{\scriptstyle
\mbf
n\in\nat^k}{\scriptstyle\sum_ln_l=n}}~\sum_{\stackrel{\scriptstyle
\mbf h\in\nat^k}{\scriptstyle\sum_lh_l=h}}\omega_{h_1}^{n_1}\cdots
\omega_{h_k}^{n_k} \label{Hurwitzirred}\eeq are called simple
Hurwitz numbers and they count the ${\it connected}$ (irreducible)
branched covering maps to $\torus^2$. In other words, the strong
coupling expansion for the chiral free energy of $U(N)$ gauge
theory on $\torus^2$ in the 't~Hooft limit is the generating
function for the Hurwitz numbers, and the expansion
(\ref{chiralfreeen}) can be identified as the partition function
of a string theory with torus target space. The coupling and
tension are given by \beq g_s=\frac1N \ , ~~
T=\frac1{2\pi\,\alpha'}=\displaystyle{\frac{\lambda}{A}}\ .
\label{gsalpha}\eeq There are two important and non-trivial
features in this context displayed by the expansion
(\ref{chiralfreeen}) of the gauge theory free energy. First of
all, only {\it even} powers $N^{2-2h}$ of the rank $N$ appear
weighted by the Euler characteristic of the covering surface, and
this is essential for a {\it closed} string interpretation.
Secondly, there is no term of order $N^{2}$ in the expansion,
consistent with the fact that there are no unfolded covers of a
torus by a sphere.

One of the most interesting properties of the QCD$_2$ string
partition function on $\torus^2$ is that the contributions (\ref{FglambdaA})
are
quasi-modular forms on the elliptic curve whose K\"ahler class is
dual to the modulus \beq \tau=-\frac{\lambda}{2\pi\ii} \ ,
\label{Kahlermoddef}\eeq with $F^+_h(\tau)$, $h\geq2$ of weight
$6h-6$ under the full modular group
$PSL(2,\zed)$~\cite{Rudd1}--\cite{zagier}. Quasi-modular forms are
polynomials in the basic holomorphic Eisenstein series
$E_2(\tau)$, $E_4(\tau)$ and $E_6(\tau)$, where \beq
E_k(\tau):=1-\frac{2k}{B_k}\,
\sum_{n=1}^\infty~\sum_{d\,|\,n}d^{k-1}~ \e^{2\pi\ii n\,\tau} \ ,
{}~~ k\in2\nat \label{Eisensteindef}\eeq and $B_k\in\rat$ is the
$k^{\rm th}$ Bernoulli number. The Eisenstein series $E_4(\tau)$
and $E_6(\tau)$ are respectively modular forms of weights 4 and 6,
\beq E_k\left(\mbox{$\frac{a\,\tau+b}{c\,\tau
+d}$}\right)=\left(c\,\tau+d\right)^k\,E_k\bigl(\tau\bigr) \ , ~~
k=4,6,\dots \eeq where \beq
\begin{pmatrix}~a~&~b~\\~c~&~d~\end{pmatrix}~\in~PSL(2,\zed) \ ,
\label{abcdSL2Z}\eeq
and they generate the ring $\mathcal{M}^0$ of all modular forms. On
the other hand, the form $E_2(\tau)$ is not quite of weight~$2$ but
has a modular anomaly given by
\beq
E_2\left(\mbox{$\frac{a\,\tau+b}{c\,\tau
+d}$}\right)=\left(c\,\tau+d\right)^2\,E_2\bigl(\tau\bigr)+
\mbox{$\frac6{\pi\ii}$}\,c\,\left(c\,\tau+d\right) \ .
\label{modanomaly}\eeq

The modular property of the Gross-Taylor series was first observed
in~\cite{Rudd1} by direct inspection of the Feynman diagram
expansion for the free energy within the conformal field theory
approach to large $N$ Yang-Mills theory proposed
in~\cite{douglas}. A rigorous proof was subsequently presented
in~\cite{dijkgraaf,zagier} directly from the equivalent free
fermion representation of the partition function~\cite{MiPo}. Another
proof, based on similar considerations, is presented in
Appendix~\ref{FFR} (see also~\cite{ochi1}). The quasi-modular character of the
$F^+_h(\tau)$ and their computability through the Feynman diagram
expansion in a Kodaira-Spencer field theory confirm two general
predictions of the mirror symmetry program in the special case of
elliptic curves~\cite{bcov}. In particular, the modular behaviour
of the string expansion provides an interesting perspective on
T-duality in the torus target space~\cite{douglas}.

Explicit formulas for the free energy contributions up to genus $h=8$
are given in~\cite{Rudd1} (see also~\cite{MSS1}). In a suitable basis
for the modular forms, obtained from the identities
\bea
E^{~}_4(\tau)&=&E^{~}_2(\tau)^2+12E_2^\prime(\tau) \ , \nonumber\\
E_6^{~}(\tau)&=&E_2^{~}(\tau)^3+18E_2^{~}(\tau)\,E_2^\prime(\tau)
+36E_2^{\prime\prime}(\tau) \ , \label{E46ids}\eea they are given
by the general expressions \bea
F_1^+(\lambda)&=&-\epsilon^{~}_{\rm F}-\ln\eta(\tau) \ , \nonumber
\\F_h^+(\lambda)&=&\frac{\lambda^{2h-2}}{(2h-2)!\,\rho_h}\,
{}~\sum_{k=0}^{3h-3}~\sum_{\stackrel{\scriptstyle l,m\in\nat_0}
{\scriptstyle
2l+3m=3h-3-k}}s_h^{kl}~E_2^{~}(\tau)^k\,E_2^\prime(\tau)^l
\,E_2^{\prime\prime}(\tau)^m
\label{FgUNexpl}\eea
with $h\geq2$ and $\rho_h,s_h^{kl}\in\nat$, where \beq
\epsilon^{~}_{\rm F}=-\frac{\lambda}{24}\,\left(N^2-1\right)
\label{Fermien}\eeq is the Fermi energy which in the Migdal
expansion is the contribution from the vacuum Young diagram
containing $i-1$ boxes in its $i^{\,\rm
  th}$ row, and $\eta(\tau)$ is the Dedekind
function~(\ref{Eulerseries}). By using the modular transformation
properties
\bea
\eta(-1/\tau)&=&\sqrt{-\ii\tau}~\eta(\tau) \ , \nonumber\\
E_2(-1/\tau)&=&\tau^2\,\left(E_2(\tau)+\mbox{$\frac6{\pi\ii\tau}$}
\right) \ , \nonumber\\E_k(-1/\tau)&=&\tau^k\,E_k(\tau) \ , ~~
k=4,6,\dots \ , \label{Eisenmodular}\eea it is possible to cast
the string representation (\ref{chiralfreeen},\ref{FglambdaA}) in
the weak-coupling regime $\lambda\ll1$ where the instanton
expansion of the gauge theory is appropriate. It was argued
in~\cite{Rudd1} that the contributions are of the generic form
\beq F_h^+(\lambda)=\lambda^{2h-2}\,
\sum_{k=3h-3}^{4h-3}\frac{r_{k,h}~\pi^{2(k-3h+3)}}{\lambda^k}
+O\left(\e^{-1/\lambda}\right) \label{Fggeneric}\eeq where
$r_{k,h}\in\rat$ are related to the simple Hurwitz numbers. The
precise geometrical meaning of the rational numbers $r_{k,h}$ will
be elucidated in the next section.

Explicit expressions up to genus $h=6$, consistent with this
structure, are found in~\cite{Rudd1}. The formula
(\ref{Fggeneric}) represents the free energy contribution to the
string expansion in the instanton picture, given as a coherent sum
over magnetic charges (Chern classes) of $U(N)$ gauge bundles over
$\torus^2$. Rather surprisingly, the free energy displays a very
mild singularity at $\lambda=0$ where the Douglas-Kazakov type phase
transition~\cite{DK1} would take place, indicating that the gauge string
theory in the present case may possess some special hidden
symmetries. In this limit, the gauge theory is related to a
topological sigma-model with torus target space coupled to
topological gravity~\cite{bcov}, and it thereby becomes a topological string
theory \cite{CMR1}. For generic values of the (dimensionless)
Yang-Mills coupling constant $\lambda$, the string theory is
described by a topological sigma-model perturbed by the K\"ahler
class of the elliptic curve.

The rational numbers $r_{k,h}$ appearing in (\ref{Fggeneric}) will be
computed in Section~\ref{SPA} using a saddle-point approximation of
the zero-instanton sector of the weakly-coupled gauge theory. An
interesting problem is then the possibility of reconstructing the
complete form of the free energy (\ref{FgUNexpl}) directly in terms of
the weak-coupling data, i.e. if it is possible, once the quasi-modular
structure is assumed, to determine the natural numbers $\rho_h,s_h^{kl}$
from only the knowledge of the $r_{k,h}$. We will show in
Section~\ref{CFP} that this is {\it not} possible and a certain number
of parameters, determined in a definite way by a closed subspace of
the vector space of modular forms, are therefore intrinsically related
to non-trivial higher instanton contributions. This is somewhat
surprising given the topological string character and the absence of a
phase transition at finite area. The solution to this problem would
yield the QCD$_2$ analog of solutions to the recurrence
relations encoded in the holomorphic anomaly equation for topological
strings on Calabi-Yau manifolds~\cite{bcov}.

\subsection{Instanton Expansion of Physical $U(N)$ Gauge Theory\label{IEUNGT}}

We will now work out the instanton expansion of the partition function
for commutative $U(N)$ Yang-Mills theory on
$\torus^2$~\cite{pasz1,gsv1,grig1}, in a way that will be amenable for
comparison later on with
the string representation of the previous subsection. Starting from
the general formula (\ref{Zpqpart}), we write the {\it physical}
commutative partition function, summed over all topological Chern
classes of $U(N)$ gauge bundles over the torus, as
\bea
Z_{U(N)}\left(g^2A\right)&:=&\e^{-\epsilon^{~}_{\rm F}}\,
\sum_{q=-\infty}^\infty(-1)^{N+(N-1)q}~Z_{N,q}\left(g^2A\,,\,\theta=0
\right)\nonumber\\
&=&(-1)^N~\e^{-\epsilon^{~}_{\rm F}}\,\sum_{\stackrel{\scriptstyle
\vnu\in\nat_0^{N}}{\scriptstyle\sum_kk\,\nu_k=N}}~
\prod_{k=1}^{N}\frac{(-1)^{\nu_k}}{\nu_k!}\,\left(
\frac{2\pi^2}{k^3g^2A}\right)^{\nu_k/2}\nonumber\\&&\times\,
\sum_{\vq\in\zed^{|\vnu|}}(-1)^{(N-1)\sum_kq_k}\,\exp\left[-
\frac{2\pi^2}{g^2A}\,\sum_{l=1}^N\frac1l\,\sum_{j=1+\nu_1+\dots+
\nu_{l-1}}^{\nu_1+\dots+\nu_l}q_j^2\right] \ , \nonumber\\&&
\label{ZUNinstexp}\eea
where we have defined $\nu_0:=0$ and the total number of partition components
$|\vnu|:=\nu_1+\dots+\nu_{N}$. In (\ref{ZUNinstexp}) it is
understood that if some $\nu_k=0$ then
$q_{1+\nu_1+\dots+\nu_{k-1}}=\dots=q_{\nu_1+\dots+\nu_k}=0$, while
$\epsilon^{~}_{\rm F}$ is the Fermi energy (\ref{Fermien}).

By defining the sequence of functions
\beq
\Xi_k\left(g^2A\right):=\sum_{q=-\infty}^\infty(-1)^{(N-1)q}~
\e^{-\frac{2\pi^2}{g^2A}\,\frac{q^2}k}
\label{calFkdef}\eeq
for $k=1,\dots,N$, we may bring the expansion (\ref{ZUNinstexp}) into
the more compact form
\beq
Z_{U(N)}\left(g^2A\right)=(-1)^N~\e^{-\epsilon^{~}_{\rm F}}\,
\sum_{\stackrel{\scriptstyle
\vnu\in\nat_0^{N}}{\scriptstyle\sum_kk\,\nu_k=N}}~
\prod_{k=1}^{N}\frac{(-1)^{\nu_k}}{\nu_k!}\,\left(
\frac{2\pi^2}{k^3g^2A}\right)^{\nu_k/2}~\left[\Xi_k\left(g^2A
\right)\right]^{\nu_k} \ .
\label{ZUNinstcomp}\eeq
Following the technique employed in
Section~\ref{TFPF}, we may carry out the sum over partitions
$\vnu$ in (\ref{ZUNinstcomp}) explicitly by resolving the constraint
using a contour integral representation (\ref{deltamncontint}), and
thereby arrive at the formula
\beq
Z_{U(N)}\left(g^2A\right)=\e^{-\epsilon^{~}_{\rm F}}\,
\oint\limits_{{\cal C}_0}\,\frac{\dd z}{2\pi\ii z^{N+1}}~
\exp\left[-\sqrt{\frac{2\pi^2}{g^2A}}~\sum_{k=1}^{\infty}\,\Xi_k
\left(g^2A\right)~\frac{(-z)^k}{k^{3/2}}\right] \ .
\label{ZUNcontint}\eeq
This representation of the weak-coupling partition function, valid for
all values of $N$ thanks to the contour integration, will be the key
to extracting its form in the desired large $N$ scaling limits.

For later comparison with the noncommutative setting, let us derive the
analog of the contour integral formula (\ref{ZUNcontint}) for the
strong-coupling expansion of the commutative $U(N)$ gauge theory.
For this, we use the Poisson resummation formula
\beq
\sum_{q=-\infty}^\infty f(q)=\sum_{n=-\infty}^\infty~\int
\limits_{-\infty}^\infty\dd s~f(s)~\e^{2\pi\ii n\,s}
\label{Poissonresum}\eeq
to rewrite the sequence of functions (\ref{calFkdef}) as
\beq
\Xi_k\left(g^2A\right)=\sqrt{\frac{k\,g^2A}{2\pi}}~
\sum_{n=-\infty}^\infty\e^{-\frac{g^2A}2\,(n-n^{~}_{\rm F})^2\,k} \ ,
\label{Xikresum}\eeq
where we have defined the Fermi surface level
\beq
n_{\rm F}^{~}=\frac{N-1}2  \ .
\label{Fermisurface}\eeq
This simple transformation rule is of course a consequence of the fact
that the functions (\ref{calFkdef}) define modular Jacobi
theta-functions on the elliptic curve of the previous subsection. By
substituting (\ref{Xikresum}) into (\ref{ZUNcontint}), the sum over
$k$ can be performed explicitly and one finally arrives at the
strong-coupling form
\beq
Z_{U(N)}\left(g^2A\right)=\e^{-\epsilon^{~}_{\rm F}}\,
\oint\limits_{{\cal C}_0}\,\frac{\dd z}{2\pi\ii z^{N+1}}~
\prod_{n=-\infty}^{\infty}\,\left(1+z~\e^{-\frac{g^2A}2\,
(n-n^{~}_{\rm F})^2}\right) \ .
\label{ZUNstrongcoupl}\eeq

The expression (\ref{ZUNstrongcoupl}) represents a concise
resummation of the Migdal expansion of commutative $U(N)$ gauge
theory on $\torus^2$, with the contour integration implementing
the constraint that the number of rows in the Young diagrams be
bounded from above by the rank $N$ of the gauge group. This may be
checked explicitly by expanding the product in
(\ref{ZUNstrongcoupl}) and evaluating the integral. This formula
agrees with the representation derived in \cite{dadda} by direct
group theory arguments and it is also similar to the contour
integral formula obtained in~\cite{douglas} from the free fermion
representation of the partition function. The free fermion formula
can also be derived directly from the combinatorics of branched
covers~\cite{dijkgraaf} and from a complex matrix model which
serves as a generating function for Hurwitz
numbers~\cite{KostStaud1}.

\subsection{Saddle Point Solution\label{SPA}}

We will now compute the partition function (\ref{ZUNcontint}) in
the large $N$ limit with the 't~Hooft coupling constant
(\ref{tHooftconst}) held fixed, and compare it to the string
representation of Section~\ref{TG-TET}. In the weak coupling
limit, the higher instanton contributions to the function
(\ref{calFkdef}) are of order $\e^{-N/\lambda}$ and hence can be
neglected to a first-order approximation at $N\to\infty$. We may
thereby focus on the zero-instanton sector with vanishing magnetic
charge $q=0$, and write the vacuum amplitude in terms of the
polylogarithm function (\ref{polylogdef}) as \beq {\cal
Z}_{U(N)}(\lambda)=\e^{-\epsilon^{~}_{\rm F}}\,
\oint\limits_{{\cal C}_0}\, \frac{\dd z}{2\pi\ii
z}~\exp\left[-N\ln z-\sqrt{\frac{\pi\,N}
{\lambda}}~\Li_{3/2}(-z)\right] \ . \label{largeNstringZUN}\eeq
Note that, in contrast to the case of Yang-Mills theory on the
sphere~\cite{MP1,grossmat}, the zero-instanton contribution on $\torus^2$ is an
infinite series when $N$ is large (through the expansion of
$\Li_{3/2}(-z)$) owing to the topological degeneracy of the
instanton configurations in this case. We will compute the
integral (\ref{largeNstringZUN}) in the large $N$ limit by means
of saddle-point techniques. The vanishing of the first derivative
with respect to $z$ of the exponential function gives the
saddle-point equation \beq
\Li_{1/2}(-z)=-\sqrt{\frac{N\,\lambda}{\pi}} \ .
\label{saddlepteqGT}\eeq By using the integral representation of
Appendix~\ref{appA}, eq.~(\ref{Lizint}), one can show that the
solutions $z=z_*$ of this equation are necessarily {\it real} for
all values of the coupling $\lambda$. In fact, the function
$\Li_{1/2}(-z)$ is a slowly decreasing negative function for
$z\geq1$, behaving as~\cite{Lewin1} $\Li_{1/2}(-z)\simeq-2\,\sqrt{\ln(z)/\pi}$
for $z$ real with $z\to\infty$ (Fig.~\ref{Li12plot}). Thus a
solution to (\ref{saddlepteqGT}) always exists and is located at
large $z\in[-1,\infty)$.

\EPSFIGURE{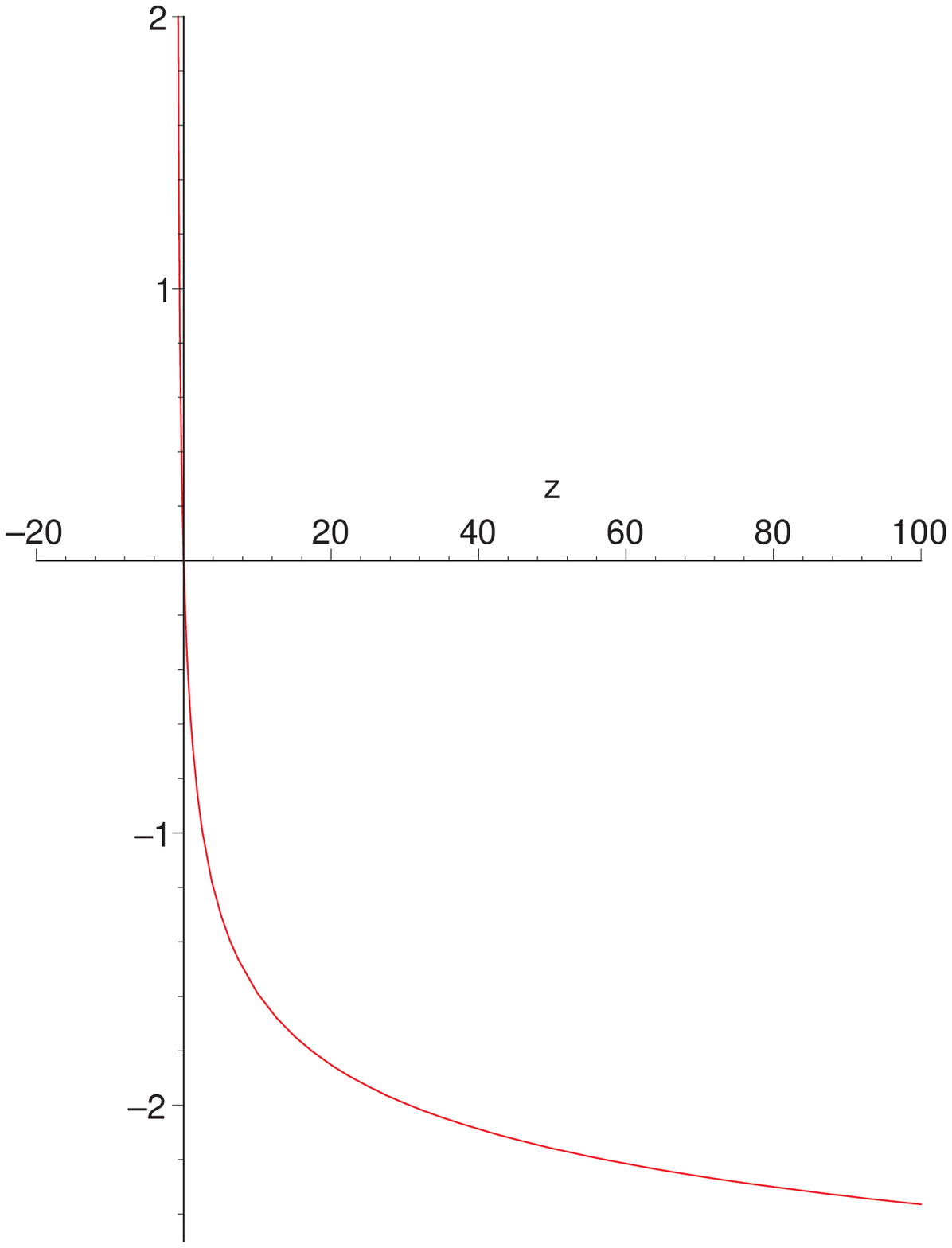,width=2.5in}{The Jonqui\`ere function $\Li_{1/2}(-z)$
  versus $z\in(-1,\infty)$.\label{Li12plot}}

The saddle-point equation (\ref{saddlepteqGT}) can be solved to
any order in $N$ by using the large $x$ asymptotic expansion of
the polylogarithm function \beq
\Li_\alpha\left(-\e^x\right)=2\,\sum_{k=0}^\infty
\frac{\left(1-2^{2k-1}\right)\,B_{2k}~\pi^{2k}}
{(2k)!~\Gamma(\alpha+1-2k)}~x^{\alpha-2k}
\label{Lialphaasympt}\eeq which we derive in Appendix~\ref{appA}.
Corrections to this formula are of order $\e^{-x}$. We seek a
solution of the form $z=z_*=\e^{x_*}$ with $x_*$ admitting an
asymptotic $\frac1N$ expansion \beq
x_*=x_{-1}\,N+\sum_{k=0}^\infty\frac{x_k}{N^k} \ .
\label{xstaransatz}\eeq The series (\ref{xstaransatz}) starts at
order $N$ due to the form of the leading term in the asymptotic
expansion (\ref{Lialphaasympt}). One can now proceed to obtain the
coefficients $x_k$ of the saddle-point solution recursively in
powers of $\frac1N$ by substituting (\ref{xstaransatz}) and
(\ref{Lialphaasympt}) into (\ref{saddlepteqGT}). The result of a
straightforward iterative evaluation up to order $1/N^{11}$ reads
\bea x_*&=&\frac{N\,\lambda}4+\frac{\pi^2}{3N\,\lambda
}+\frac{16\pi^4}{9\left(N\,\lambda\right)^3}
+\frac{448\pi^{6}}{9\left(N\,\lambda\right)^5}\nonumber\\&&
+\,\frac{1254656\pi^{8}}{405\left(N\,\lambda\right)^7}+
\frac{406598656\pi^{10}}{1215\left(N\,\lambda\right)^9}+
\frac{67556569088\pi^{12}}{1215\left(N\,\lambda\right)^{11}}+
O\left(\frac{1}{N^{13}}\right) \ . \label{xstarGTN11}\eea The
systematic vanishing of the $x_k$ for even powers of $\frac1N$ is
exactly what is expected from the form of the Gross-Taylor
expansion.

Armed with the solution (\ref{xstarGTN11}) of the saddle-point
equation, we can now proceed to compute the free energy ${\cal
  F}_{U(N)}(\lambda)=\ln{\cal Z}_{U(N)}(\lambda)$ by
parametrizing the integration variable in (\ref{largeNstringZUN})
as $z=\e^{x_*+x}$, where the variable $x$ contains contributions
from fluctuations about the saddle-point value. The saddle-point
itself is at $x=0$ and we can integrate the fluctuations over all
$x\in\real$, since the deviations from the results computed with
the correct integration domain appropriate to
(\ref{largeNstringZUN}) will be exponentially suppressed in the
small area limit. By using (\ref{Lialphaasympt}), the computation
of \beq {\cal F}_{U(N)}(\lambda)=\ln\left(\e^{-\epsilon^{~}_{\rm
F}}\, \int\limits_{-\infty}^\infty\dd x~\exp\left[-N\left(x_*+x
\right)-\sqrt{\frac{\pi\,N}
{\lambda}}~\Li_{3/2}\left(-\e^{x_*+x}\right)\right]\right)
\label{largeNfreenUN}\eeq as an expansion up to order $1/N^{10}$
is straightforward to do, but quite cumbersome. A numerical
evaluation using {\sl Mathematica} yields \bea {\cal
F}_{U(N)}(\lambda)&=&\frac12\,\ln(-2\pi\,\lambda)
+\frac{\pi^2}{6\lambda}+\left(\frac{2}{3\lambda
}-\frac{2\pi^2}{3\lambda^2}+\frac{8
\pi^4}{45\lambda^3}\right)\,\frac1{N^2}\nonumber\\&&
+\,\left(-\frac{8}{\lambda^2}+\frac{16\pi^2}
{\lambda^3}-\frac{100\pi^4}{9 \lambda^4}+\frac{224\pi^6}{81
\lambda^5}\right)\,\frac1{N^4}\nonumber\\&&
+\,\left(\frac{2272}{9\lambda^3}-\frac{2272\pi^2}
{3\lambda^4}+\frac{8096\pi^4}{9\lambda^5}-\frac{41504
\pi^6}{81\lambda^6}+\frac{48256\pi^8}{405
\lambda^7}\right)\,\frac1{N^6}\nonumber\\&&
+\,\left(-\frac{13504}{\lambda^4}+\frac{54016
\pi^2}{\lambda^5}-\frac{834304\pi^4}
{9\lambda^6}\right.\nonumber\\&&+\left.
\frac{7010816\pi^6}{81\lambda^7}-\frac{17887904
\pi^8}{405\lambda^8}+\frac{11958784\pi^{10}}
{1215\lambda^9}\right)\,\frac1{N^8}\nonumber\\&&
+\,\left(\frac{15465472}{15\lambda^5}-
\frac{15465472\pi^2}{3\lambda^6}+ \frac{105156608\pi^4}{9
\lambda^7}-\frac{418657280\pi^6}
{27\lambda^8}\right.\nonumber\\&&+\left.
\frac{572409344\pi^8}{45\lambda^9}- \frac{2467804672\pi^{10}}{405
\lambda^{10}}+\frac{33778284544\pi^{12}}{25515
\lambda^{11}}\right)\,\frac1{N^{10}}+ O\left(\frac1{N^{12}}\right)
\ . \nonumber\\&& \label{largeNfreenUNexp}\eea

The expression (\ref{largeNfreenUNexp}) matches {\it precisely}
eqs.~(3.43)--(3.47) in~\cite{Rudd1} which were obtained using the
conformal field theory representation of large $N$  two-dimensional
Yang-Mills theory. The saddle-point evaluation of the zero-instanton sector
thereby reproduces all non-exponentially suppressed terms in the weak-coupling
expansion of the chiral $U(N)$ free energy, giving the correct
rational numbers $r_{k,h}$ appearing in (\ref{Fggeneric}). An
important ingredient in this reproduction is the cancellation of
the ground state energy (\ref{Fermien}) that appears in
(\ref{largeNfreenUN}), since this term would otherwise dominate the
series in the large $N$ limit. In particular, the expansion starts
at order $N^0$ and there is no spherical contribution. The expansion
(\ref{largeNfreenUNexp}) also contains the correct leading modular
dependence of the Dedekind function coming from the genus~$1$ free
energy.

The derivation of the formula (\ref{largeNfreenUNexp}) that we
have presented here has three particularly noteworthy features.
First of all, it is a highly accurate check of the expansion
obtained in~\cite{Rudd1} by a completely independent method.
Secondly, it does not rely on any group theory or conformal field
theory techniques and is obtained directly in the instanton
representation. As far as we are aware, this is the first time
that stringy quantities are computed in two dimensional Yang-Mills
theory directly from the weak-coupling expansion, which is the
most natural one from a conventional quantum field theory point of
view. This is possible, of course, due to the absence of a phase
transition at finite area on the torus, as on the sphere the
Douglas-Kazakov transition prohibits the recovery of stringy
features from the weak-coupling
data~\cite{DK1}--\cite{grossmat}. Finally, we note that the dynamics
of the zero-instanton sector is surprisingly rich,
encoding properly the anticipated stringy features. This is most
likely related to the underlying structure of the topological
string theory governing the weak-coupling limit \cite{CMR1}.

However, despite this remarkable agreement, there is a serious flaw in
the computation above. Evidently, we have reproduced only the {\it
  chiral} part of the full $U(N)$ gauge theory. The saddle-point
technique has not picked up the coupling to the anti-chiral
sector. Furthermore, after some calculation one can find that even
in the exponentially suppressed contributions, there are terms
missing in the saddle-point analysis which prevent a resummation
to the full quasi-modular Dedekind function and Eisenstein series.
Part of the problem can be traced back to the fact that the
argument of the exponential integrand in (\ref{largeNstringZUN})
does not admit a nice large $N$ scaling of the form
$\e^{-N\,f(z;\lambda)}$, with the function $f(z;\lambda)$
independent of $N$. In evaluating the fluctuation integral
(\ref{largeNfreenUN}), there are Gaussian terms of the form
$\e^{-x^2/\lambda}$ which thus arise and are {\it independent} of
$N$. In order for the saddle-point technique to be reliable, the
widths of such Gaussians should vanish as $N\to\infty$ in order
for the solution to localize around the saddle-point value at
$x=0$. The difficulty stems from the linear growth in $N$ of the
position of the saddle-point (\ref{xstaransatz}).

Moreover, we eventually have to face up to the problem of evaluating the
non-zero higher instanton contributions. It is very likely that,
in spite of their exponential suppression order by order in
$\frac1N$, their collective behaviour will be crucial to recover
the complete string expansion. Recall that instantons are
essential to the recovery of the string picture of large $N$
Yang-Mills theory formulated on the sphere~\cite{grossmat}. We
have not accomplished an analytical evaluation of the full
instanton series at large $N$. However, we can address the problem
of how much information should be carried by the higher instantons
in order to recover the full string partition function. This is
the topic of the next subsection.

\subsection{Quasi-Modularity of Higher Instanton Configurations\label{CFP}}

On spacetimes of genus~$\geq2$, the partition function of
Yang-Mills theory in two dimensions is smooth in the limit of
vanishing area and it defines a topological string theory
\cite{CMR1}. This process can also be reversed, and the partition
function (and observables) at finite area can be expressed in
terms of correlators computed in this topological string theory
which are related to intersection numbers in the Hurwitz moduli
spaces of topological classes of branched covers. Instead, it is
not obvious that this procedure is reliable on the torus, since
the weak-coupling limit of the partition function is singular in
that case.  However, one may ask what amount of information is
encoded in the singular terms and if it is enough to reconstruct the
whole partition function. In what follows we shall answer this
question using only the knowledge that $F^+_{h}(\tau)$ is a
quasi-modular form of weight $6h-6$.

For this, let us first recall some basic facts about quasi-modular
forms~\cite{milne1}. Let $k\in2\nat_0$ and $s\in\nat_0$. A quasi-modular form
of
weight $k$ and level $s$ is a holomorphic function $f:\complex_+\to
\mathbb{C}$ on the upper complex half-plane such that
\begin{equation}
\label{qmod}
(c\,\tau+d)^{-k}\,f\left(\mbox{$\frac{a\,\tau+b}{c\,\tau
+d}$}\right)=\sum_{n=0}^s f_n\bigl(\tau\bigr)\,\left(\frac{c}{c\,
\tau+d}\right)^n \ ,
\end{equation}
where the functions $f_n(\tau)$, $1\leq n\leq s$ are independent of
the modular transformation matrices (\ref{abcdSL2Z}), and
$f_0(\tau):=f(\tau)$. A quasi-modular form of level $0$ is a modular
form. We denote the complex vector space of quasi-modular forms of
weight $k$ and level $\leq s$ by $\mathcal{M}_k^{s}$. The product of a
quasi-modular form in $\mathcal{M}_{k_1}^{s_1}$ with a quasi-modular
form in $\mathcal{M}_{k_2}^{s_2}$ is a quasi-modular form in
$\mathcal{M}_{k_1+k_2}^{s_1+s_2}$. Then the vector space
\begin{equation}
\mathcal{M}:=\bigoplus_{k\in2\nat_0}\,\bigoplus_{s=0}^\infty
\,\mathcal{M}_k^{s}
\label{gradedalg}\end{equation}
naturally becomes a graded algebra. The general structure of this
algebra is provided by the following central result.
\begin{theorem}
Every quasi-modular form $f(\tau)$ of weight $k$ and level $s$ can be
written as
$$
f(\tau)=\sum_{l=0}^s E_2(\tau)^l\,M_{k -2l}(\tau)
$$
where $M_{k-2l}(\tau)$ is a modular form of weight $k-2l$.
\label{quasimodthm}\end{theorem}
Theorem~\ref{quasimodthm} implies that the ring $\mathcal{M}$ of
quasi-modular forms is the graded algebra generated over the ring
$\mathcal{M}^0:=\bigoplus_{k\in2\nat_0}\mathcal{M}_k^0$ of modular
forms by the Eisenstein series of weight~$2$ as
\beq
\mathcal{M}=\mathcal{M}^0\otimes\complex[E_2] \ .
\label{gradedalgdecomp}\eeq

The modular anomaly of the Eisenstein series (\ref{modanomaly}) leads
to the requirement that the genus~$h$ free energy $F^+_h(\tau)$ be a
quasi-modular form of weight $k=6h-6$ and level
$s=3h-3$. Theorem~\ref{quasimodthm} then constrains the effects of the
gauge dynamics in a finite set of parameters, implying that the free
energy can be written as
\begin{equation}
\label{repres1}
F^+_h(\tau)=\sum_{l=0}^{3h-3} E_2(\tau)^{l}\,M_{6 h-6 -2l}(\tau)
\end{equation}
where $M_{6h-6-2l}(\tau)$ is a (true) modular form of weight
$6h-6-2l$ (or a weight $6h-6-2l$ combination of $E_2^\prime(\tau)$
and $E_2^{\prime\prime}(\tau)$ in the alternative basis of
Section~\ref{TG-TET}). The finite dimension of the $\complex$-linear
space $\mathcal{M}_{k}^0$ of modular forms of weight $k$ is given
by~\cite{milne1}
\begin{equation}
\dim\mathcal{M}_{k}^0= 1-\mbox{$\frac k2$}+\left\lfloor
\mbox{$\frac{k}{3}$}\right\rfloor+\left\lfloor
\mbox{$\frac{k}{4}$}\right\rfloor \ .
\end{equation}
The number of free parameters in the expansion (\ref{repres1})
can then be easily evaluated by computing the sum
\beq
K_h^\#:=\sum_{l=0}^{3h -3}\dim\mathcal{M}^0_{6 h-6 -2l}
=\left\{\begin{matrix} \mbox{$\frac{1}{4}$}\,\left(3 h^2+1\right)
&\mathrm{for}\ h \ \mathrm{odd}\\\mbox{$\frac{3}{4}$}\,h^2
&\mathrm{for}\ h \ \mathrm{even} \end{matrix}\right. \ .
\eeq

We are interested in this structure in the weak coupling limit
$\lambda\to0$. As in Section~\ref{TG-TET}, it is straightforward
to extract the small area behaviour of the free energy at genus
$h$ from the representation (\ref{repres1}) in terms of
quasi-modular forms, and for generic coefficients one finds the
general expansion
\begin{equation}
\label{zero2}
F^{+}_h(\lambda)=\sum_{l=3h-3}^{6h-6}\,\frac{m_{l,h}}
{\lambda^l}+O\left(\e^{-1/\lambda}\right)
\end{equation}
where $m_{l,h}\in\real$ are linear combinations of the $K_h^\#$
free parameters in eq.~(\ref{repres1}). This is to be compared
with the instanton representation (\ref{Fggeneric}), which singles
out the singular behaviour of the free energy at weak coupling as
a polynomial in $\frac{1}{\lambda}$ of degree $4 h-3$ with leading
term of order $3 h-3$. Thus matching the two expansions will
provide (at most) $3h-2$ conditions on the coefficients appearing
in the expansion (\ref{repres1}).

To determine the constraints arising from the small area behaviour, it
is convenient to choose an adapted basis for the space of modular
forms. For this, we decompose $\mathcal{M}_k^0$ as a vector
space in the form~\cite{milne1}
\begin{equation}
\mathcal{M}_{k}^0=\complex\cdot E^{~}_{k}~\oplus~\mathcal{M}^0_{k-12}\cdot
\triangle
\label{Mk0decomp}\end{equation}
where the Ramanujan modular form
\beq
\triangle(\tau):=E_4(\tau)^3-E_6(\tau)^2=1728\,\eta(\tau)^{24}
\label{curvediscr}\eeq
of weight~$12$ is the discriminant of the underlying elliptic
curve. Eq.~(\ref{Mk0decomp}) implies that
every modular form of weight~$k$ can be written as a complex linear
combination of the weight~$k$ Eisenstein series (\ref{Eisensteindef})
with a modular form which is the product of (\ref{curvediscr}) and a
modular form of weight~${k-12}$. The representation (\ref{repres1})
can be rewritten using this decomposition as
\begin{equation}
\label{bubba}
F^{+}_h(\tau)=\sum_{l=0}^{3 h-4} c_{l,h}~E_2(\tau)^{l}\,
E_{6 h-6 -2l}(\tau)+\triangle(\tau)\,
\sum_{l=0}^{3 h-3}  E_2(\tau)^{l}\,M_{6 h-18 -2l}(\tau) \ .
\end{equation}
The remarkable property of this expansion for the genus $h$ free
energy is that the small area behaviour is completely dictated by the
first contribution in (\ref{bubba}), as the discriminant $\triangle(\tau)$
is exponentially suppressed at both weak and strong coupling. Therefore,
matching eq.~(\ref{zero2}) with eq.~(\ref{Fggeneric}) fixes the $3h-3$
real coefficients $c_{l,h}$ in eq.~(\ref{bubba}). Note that this
procedure is not guaranteed to work, as there are $3h-3$ unknowns but only
$ 3h-2$ conditions. Some sort of miraculous reduction should
(and will) occur.

The weak coupling limit of eq.~(\ref{bubba}) can thereby be
straightforwardly worked out using the modular transformation
properties (\ref{qmod}), analogously to Section~\ref{TG-TET}, and
after some algebra one finds
\begin{eqnarray}
\label{bubba1} F^{+}_h(\lambda)&=&\sum_{l=0}^{3 h-5}
\frac{\left(-4 \pi^2 \right)^{l+2}\,(12)^{3h-5-l}}
{\lambda^{3h-1+l}}\,\left[{3 h-3\choose l+2}\,c_{3 h-4,h}+
\sum_{l'=0}^{l}{3 h-5-l'\, \choose l-l'}\,c_{3 h-5-l',h}
\right]\nonumber\\&&+\, \frac{(12)^{3h-3}\,c_{3h-4,h}}{\lambda^{3
    h-3}}-\frac{4\pi^2\,(3 h-3)\,(12)^{3h-4}\,c_{3 h-4,h}\,}
{\lambda^{3h-2}}+O\left(\e^{-1/\lambda}\right) \ .
\end{eqnarray}
{}From this expression it is clear that the expansion (\ref{bubba}) can
match eq.~(\ref{zero2}) if and only if the ratio of the two
coefficients of the lowest singularity in eqs. (\ref{bubba1}) and
(\ref{zero2}) is fixed and equal to $-\pi^2(h-1)$. One can check
explicitly using (\ref{repres1}) that this is indeed the case (This
is the miraculous reduction in the number of conditions that we
mentioned above). The fact that this criterion is satisfied is also a
check that the singularities we found through the saddle-point
computation of the previous subsection originate from a quasi-modular
form.

At this point, starting from the lowest singularity in (\ref{bubba1}),
we can compare with eq.~(\ref{zero2}) and recursively determine all
coefficients of the small area expansion. Then the number of free
parameters that remain undetermined is given by
\beq
\Delta K^\#_h= K^\#_h-(3h-3)=\left\{\begin{matrix}
\mbox{$\frac{1}{4}$}\,\bigl(3(h-2)^2+1\bigr)&\mathrm{for}\ h \
\mathrm{odd}\\\mbox{$\frac{3}{4}$}\,\bigl(h-2\bigr)^2&
\mathrm{for}\ h \ \mathrm{even} \end{matrix}\right. \ .
\eeq
For example, $\Delta K^\#_2=0$ and thus the genus~$2$ free energy is
completely determined by its behaviour at weak coupling. At genus~$3$ there
is $\Delta K^\#_3=1$ free parameter, at genus $4$ there are
$\Delta K^\#_4=3$ parameters, and so on. We are therefore led to
conclude that higher instantons should account for the remaining
undetermined dynamical parameters in the weak coupling expansion of
the gauge theory. Furthermore, the exponentially suppressed
contributions, which are not related to the small area behaviour and
are required to reinstate the full quasi-modular structure at genus $h$,
are parametrized by higher instanton contributions living in the
subspace
\beq
\mathcal{P}_h:=\bigoplus_{l=0}^{3h-3}\mathcal{M}^0_{6h-18-2l}\cdot
(E_2)^l\,\triangle
\label{expsubspace}\eeq
of the graded algebra (\ref{gradedalg}).

\section{The Double Scaling Limit\label{DSL}}

Motivated by the analysis of the previous section, we will now
analyse the $U(N)$ gauge theory in a large $N$ limit wherein the
saddle-point technique can capture the entire relevant story. It
is clear what to do. We should take the limit $N\to\infty$ while
keeping fixed the new scaled coupling constant \beq
\displaystyle{\mu:=\frac{N\,\lambda}{\pi}=\frac{N^2g^2A}{2\pi}} \ .
\label{mudoublescale} \eeq The partition function then assumes the
form \beq \hat{\cal Z}_{U(N)}(\mu)=\e^{\frac{\pi\,N\,\mu}{12}}\,
\oint\limits_{{\cal C}_0}\frac{\dd z}{2\pi\ii z}~
\exp\left[-N\left(\ln z+\frac1{\sqrt\mu}~\Li_{3/2}(-z)
\right)\right]=:\oint\limits_{{\cal C}_0}\frac{\dd z}{2\pi\ii z}~
\e^{N\,\hat F(z,\mu)} \label{partfndoubleUN}\eeq and hence has a
nice large $N$ limit. Corrections to this expression from higher
instanton configurations are of order $\exp(-N^2/\mu)$ and could
be completely suppressed in the $\frac1N$ expansion. In fact, at
$N=\infty$ the vacuum amplitude is given by the leading planar
term $\hat{\cal
  Z}_{U(N)}(\mu)=\e^{N\,\hat F(z_*,\mu)}$ in the $\frac1N$ expansion of
the integral (\ref{partfndoubleUN}), which can be rigorously computed
in the saddle-point approximation. We will refer to this new limit of
the gauge theory as the ``double scaling limit'', because it is the
limit appropriate for the mapping onto noncommutative Yang-Mills
theory which will be carried out in the next section.

This limit is very different from the conventional planar large
$N$ limit used to derive the Gross-Taylor expansion. In this
section we will analyse this double scaling in the simpler $U(N)$
theory in order to explore the fate of the string expansion in
this new limit of the gauge theory. Again we shall see that there
is a non-trivial saddle point governing the large $N$ behaviour of
the zero-instanton sector. We will indeed find that an open string
representation potentially emerges with expansion coefficients
completely determined, in the limit of infinite winding number, by
the geometrical quantities (Hurwitz numbers) parametrizing the
closed strings on the torus. The open string theory is deeply tied
to the geometry of a particular class of moduli spaces of Riemann
surfaces which we describe in detail. This will be the crux of the
string expansion of noncommutative gauge theory that we will
describe in the next section.

\subsection{Saddle Point Solution\label{DSLsaddle}}

Starting from (\ref{partfndoubleUN}), we derive the saddle-point
equation \beq \Li_{1/2}(-z)=-\sqrt\mu \ ,
\label{saddleptdoubleUN}\eeq and it can be solved exactly, at
least in principle. In contrast to the conventional 't~Hooft
limit, this equation does not depend on $N$ and therefore its
solution $z=z_*(\mu)$ does not rely on any approximation in
general. It can be simply written as
$z_*(\mu)=-\Li^{-1}_{1/2}(-\sqrt\mu\,)$, the inverse function
being uniquely defined in the region of interest thanks to the
monotonic behaviour of the polylogarithm function $\Li_{1/2}(-z)$.
The double scaling free energy can then be evaluated as $\hat{\cal
F}_{U(N)}(\mu)=N\,\hat F(z_*(\mu),\mu)$ and is depicted in
Fig.~\ref{DoubleU(N)}. It is a smooth function of $\mu>0$, with a
logarithmic singularity in the weak-coupling limit $\mu\to 0$ where
the Douglas-Kazakov type phase transition takes place. In
fact, using properties of the Jonqui\`ere function (see
Appendix~\ref{appA}), we can write down an {\it exact} relation
that gives the free energy directly as a primitive of the position
of the saddle point. By parametrizing the saddle point solution as
before in the form $z_*(\mu)=\e^{x_*(\mu)}$ and defining
$y:=\sqrt{\mu}$, one can easily derive the equation \beq \hat
F\bigl(z_*(\mu)\,,\,\mu\bigr)=\frac{1}{\sqrt{\mu}}~
\int\limits_{\sqrt{\mu}}^\infty\dd y~\left[x_*\left(y^2\right)-
\frac{\pi\,y^2}{4}\right]\ . \label{bella}\eeq This explicit
representation is useful because it exhibits the structure of the
free energy straightforwardly in terms of the properties of the
saddle point solution.

\EPSFIGURE{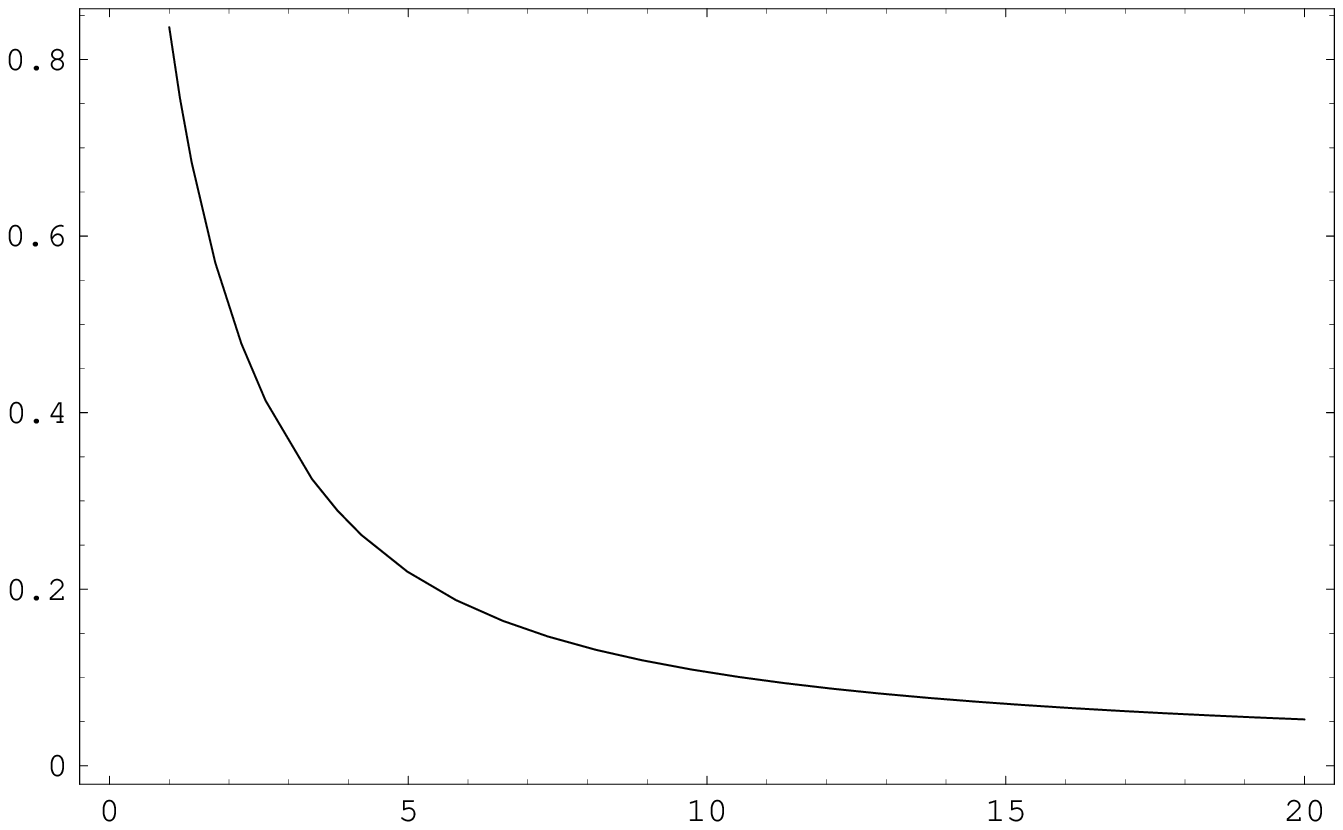,width=3.5in}{The double scaling free
energy function $\hat F(z_*(\mu),\mu)$
versus~$\mu$.\label{DoubleU(N)}}

The physical meaning of the double scaling limit can be understood
by expanding the free energy for large $\mu$. It is in this regime
that we expect to see a relation with the string picture derived
previously in the 't~Hooft scaling limit. The strong-coupling
saddle point solution is given by the expansion (\ref{xstarGTN11})
which is naturally written as a series in the double scaling
parameter given by \beq
x_*(\mu)=\pi\,\sum_{k=0}^{\infty}\,\frac{\xi_{2k-1}}{\mu^{2k-1}}
\label{expa}\eeq
with $\xi_{2k-1}\in\rat$. The double scaling free energy $\hat{\cal
F}_{U(N)}$ may then be computed directly from (\ref{bella}) to get
\beq \hat{\cal
F}_{U(N)}(\mu)=\pi\,N\,\left[\,\sum_{k=1}^{\infty}\frac{\xi_{2k-1}}
{4k-3}~\frac1{\mu^{2k-1}}+O\left(\e^{-\mu}\right)\right] \ ,
\label{freenDSdirect}\eeq where we note the cancellation of the
vacuum energy contribution. From the explicit expression in
(\ref{xstarGTN11}) the first few terms are found to be given by
\bea \hat{\cal F}_{U(N)}(\mu)&=&2\pi\,N\,
\left[\frac{1}{6\mu}+\frac{8}{45\mu^3}+\frac{224}{81\mu^5}
+\frac{48256}{405 \mu^7}\right.\nonumber\\&&+\left.\frac{11958784}
{1215\mu^9}+\frac{33778284544}{25515
\mu^{11}}+O\left(\frac1{\mu^{13}}\right)+O\left(\e^{-\mu}\right)\right]
\ . \label{largeNfreenUNdouble}\eea

By momentarily disregarding the exponentionally suppressed
contributions to (\ref{largeNfreenUNdouble}) and comparing with
(\ref{largeNfreenUNexp}), we see that at strong-coupling the
double scaling limit has extracted the most singular terms, as
$\lambda\to0$, at each order of the original $\frac1N$ expansion.
In other words, the double scaled gauge theory at strong-coupling
presents a resummation of the most singular terms in the
weak-coupling limit of the chiral Gross-Taylor string expansion.
By substituting (\ref{mudoublescale}) into (\ref{Fggeneric}), the
leading contribution there at $N=\infty$ is given at $k=4h-3$, and
thus the general form of this expansion can be written as \beq
\hat{\cal F}_{U(N)}(\mu)=\pi\,N\,\left[\,\sum_{h=1}^\infty
\frac{r_{4h-3,h}}{\mu^{2h-1}}+O\left(\e^{-\mu}\right)\right] \ .
\label{Fggenericdouble}\eeq {}From (\ref{freenDSdirect}) it then
follows that the rational numbers $r_{4h-3,h}$ are completely
determined by the strong-coupling solution (\ref{expa}) of the
saddle point equation to be \beq
r_{4h-3,h}=\frac{\xi_{2h-1}}{4h-3} \ . \label{relationsaddle}\eeq

We will now compute the explicit forms of the coefficients of the saddle-point
expansion (\ref{expa},\ref{freenDSdirect}) as polynomials in
Bernoulli numbers, and thereby write down the {\it exact} solution
of the double scaling gauge theory in the strong-coupling limit.
For this, we set $x_*=1/w^2$ and use (\ref{Lialphaasympt}) to
write the saddle-point equation (\ref{saddleptdoubleUN}) as \beq
w=-\frac2{\sqrt\mu}\,\sum_{k=0}^\infty\frac{\left(1-2^{2k-1}\right)\,
B_{2k}~\pi^{2k}}{(2k)!~\Gamma\left(\frac32-2k\right)}~w^{4k}=:
\frac1{\sqrt\mu}~L(w) \ . \label{saddleptTaylor}\eeq
The solution of the saddle-point equation to all orders in the
strong-coupling regime is thereby reduced to the formal inversion of a
Taylor series. This problem may be solved by standard Lagrange
inversion~\cite{Whitt1}, which gives the formal solution of
(\ref{saddleptTaylor}) as the infinite series
\beq
w(\mu)=\sum_{k=0}^\infty\frac1{(k+1)!}~\left.\frac{\dd^k}{\dd z^k}
L(z)^{k+1}\right|_{z=0}~\left(\frac1{\sqrt\mu}\right)^{k+1} \ .
\label{Lagrangesoln}\eeq
The solution (\ref{Lagrangesoln}) can be rewritten in a more compact
form by means of the contour integral
\beq
w(\mu)=\oint\limits_{{\cal C}_0}\frac{\dd z}{2\pi\ii}~z~
\frac{1-\frac1{\sqrt\mu}\,L'(z)}{z-\frac1{\sqrt\mu}\,L(z)} \ ,
\label{Lagrangecontint}\eeq
where now the contour ${\cal C}_0$ encircles both the origin $z=0$ and
the zero of the denominator in (\ref{Lagrangecontint}). The
equivalence of the two expressions (\ref{Lagrangesoln}) and
(\ref{Lagrangecontint}) is straightforwardly established by writing a
formal Taylor series expansion of the integrand in
(\ref{Lagrangecontint}) in powers of $\frac1{\sqrt\mu}$ and computing
the integral.

For our purposes it will be more useful to employ the B\"urmann
generalization of the Lagrange inversion formula~\cite{Whitt1}. The
contour integral representation (\ref{Lagrangecontint}) can be easily
generalized to evaluate any analytic function $G$ of the solution
$w(\mu)$ by writing
\beq
G\bigl(w(\mu)\bigr)=\oint\limits_{{\cal C}_0}\frac{\dd z}{2\pi\ii}~G(z)~
\frac{1-\frac1{\sqrt\mu}\,L'(z)}{z-\frac1{\sqrt\mu}\,L(z)} \ .
\label{Burmanngen}\eeq
To compute $x_*(\mu)=1/w(\mu)^2$ we should take
$G(z)=1/z^2$, but this cannot be directly inserted into the
formula (\ref{Burmanngen}) as it would introduce a spurious
contribution from the double pole at $z=0$. The simplest way to
deal with this problem is to subtract the undesired contribution
by hand, and thereby write the solution $x_*(\mu)$ of the
saddle-point equation as \beq x_*(\mu)=\oint\limits_{{\cal C}_0}
\frac{\dd z}{2\pi\ii
z^2}~ \frac{1-\frac1{\sqrt\mu}\,L'(z)}{z-\frac1{\sqrt\mu}\,L(z)}-
\frac{L(0)\,L''(0)-\bigl(\sqrt\mu-L'(0)\bigr)^2}{L(0)^2} \ .
\label{xmucontint}\eeq
By formally expanding (\ref{xmucontint}) in powers of
$\frac1{\sqrt\mu}$ and using the definition of the function $L(z)$ in
(\ref{saddleptTaylor}), we arrive at the strong-coupling solution
\bea
x_*(\mu)&=&-\frac{L(0)\,L''(0)-\bigl(\sqrt\mu-L'(0)\bigr)^2}{L(0)^2}
-2\,\sum_{k=1}^\infty\frac1{(k+2)!\,k}\,
\left.\frac{\dd^{k+2}}{\dd z^{k+2}}L(z)^{k}\right|_{z=0}~
\frac1{\mu^{k/2}}\nonumber\\&=&\frac{\pi\,\mu}4-2\,\sum_{k=1}^\infty
\frac{L_{k+2}^{(k)}}k~\frac1{\mu^{k/2}}
\label{xmuexpansion}\eea
where the coefficients
\beq
L_{k+2}^{(k)}=\sum_{\stackrel{\scriptstyle\mbf k\in\nat_0^k}
{\scriptstyle4\sum_ik_i=k+2}}~\prod_{i=1}^k\frac{\left(2-2^{2k_i}
\right)\,B_{2k_i}~\pi^{2k_i}}{(2k_i)!~\Gamma\left(\frac32-2k_i
\right)}
\label{Lcoeffs}\eeq
are non-zero only when $k=4h-2$ for some $h\in\nat$.

Written in the form (\ref{expa}), one has $\xi_{-1}=\frac14$ as
anticipated, while
\beq
\xi_{2h-1}=-\frac{L_{4h}^{(4h-2)}}{\pi\,(2h-1)}=
-\frac{\pi^{2h-1}}{2h-1}~\sum_{\stackrel{\scriptstyle
\mbf h\in\nat_0^{4h-2}}{\scriptstyle
\sum_kh_k=h}}~\prod_{k=1}^{4h-2}\,\frac{\left(2-2^{2h_k}\right)\,
B_{2h_k}}{(2h_k)!~\Gamma\left(\frac32-2h_k\right)}
\label{xiexact}\eeq for $h\geq1$, and we have thereby found the
complete strong-coupling expansion of the double scaled gauge
theory. We can simplify the sum over ordered partitions $\mbf
h\in\nat_0^{4h-2}$ of the integer $h$ by reducing it to a sum over
partitions of $h$ into $m$ positive integers. By inserting~$0$
into all possible positions we obtain ${4h-2}\choose m$ partitions
of the original type in (\ref{xiexact}), and we find \bea
\xi_{2h-1}&=&-\frac{\pi^{2h-1}}{2h-1}\,\sum_{m=1}^h{4h-2\choose m}
\left(\frac{B_0}{\Gamma\left(\frac32\right)}\right)^{4h-2-m}\,
\sum_{\stackrel{\scriptstyle\mbf h\in\nat^m}
{\scriptstyle\sum_kh_k=h}}~\prod_{k=1}^m\,\frac{\left(2-2^{2h_k}\right)\,
B_{2h_k}}{(2h_k)!~\Gamma\left(\frac32-2h_k\right)}
\nonumber\\&=&-\sum_{m=1}^h\frac{(-1)^m\,2^{2h+m-1}\,(4h-3)!}
{(4h-2-m)!\,m!}~\sum_{\stackrel{\scriptstyle\mbf h\in\nat^m}
{\scriptstyle\sum_kh_k=h}}~\prod_{k=1}^m\,\frac{\left(2^{2h_k-1}-1
\right)\,(4h_k-3)!!~B_{2h_k}}{(2h_k)!} \ . \nonumber\\&&
\label{xireduce}\eea
Finally, by exploiting the symmetry of the second summand in
(\ref{xireduce}) we can reduce the sum over ordered partitions of
$h$ with $m$ components to a sum over conjugacy classes and cycles
of the symmetric group $S_h$, i.e. over unordered partitions of
$h$. An unordered partition of $h$ is specified by $h$
non-negative integers $\nu_k$ with $\sum_kk\,\nu_k=h$, while the
condition that the partition contain only $m$ parts is implemented
by requiring that $\sum_k\nu_k=m$. By inserting the combinatorial
factor $\frac{m!}{\nu_1!\cdots\nu_h!}$ which counts the number of
different ordered partitions that originate from the same
unordered partition, we may bring (\ref{xireduce}) into our final
equivalent form.
\begin{theorem}
The coefficients of the asymptotic expansion as $\mu\to\infty$ for
the free energy (\ref{freenDSdirect}) of $U(N)$ gauge theory on
$\torus^2$ in the double-scaling limit are given by \bea
\xi_{2h-1}&=&(4h-3)!\,\sum_{m=1}^h\frac{(-1)^{m-1}\,2^{m+2h-1}}
{(4h-2-m)!}\nonumber\\&&\times\,
\sum_{\stackrel{\scriptstyle\mbf\nu\in\nat_0^h}
{\scriptstyle\sum_kk\,\nu_k=h\,,\,\sum_k\nu_k=m}}~
\prod_{k=1}^h\,\frac1{\nu_k!}\,\left(\frac{\left(2^{2k-1}-1\right)\,
(4k-3)!!~B_{2k}}{(2k)!}\right)^{\nu_k} \ . \nonumber\eea
\label{xifinal}\end{theorem}
The rationality of these numbers will be important for the geometric
interpretation of Section~\ref{MSHD}.

\subsection{Asymptotics of Simple Hurwitz Numbers\label{ASHN}}

We will now elucidate the geometrical meaning of the rational
numbers (\ref{relationsaddle}) in order to pave the way towards
the string interpretation of the double scaling limit. For this,
let us return to the chiral expansion
(\ref{chiralfreeen},\ref{FglambdaA}) of the gauge theory in the
ordinary 't~Hooft limit. The $\lambda\to0$ behaviour of the series
(\ref{FglambdaA}) is controlled by the large $n$ asymptotics of
the simple Hurwitz numbers $\sigma_h^n$ defined by
(\ref{Hurwitzred},\ref{Hurwitzirred}), which geometrically count
the number of holomorphic maps from a closed oriented Riemann
surface of genus $h$ to the torus $\torus^2$ with winding number
$n$ and $2h-2$ simple branch points. The first thing to realize is
that, at fixed genus $h$, the $\frac1\lambda$-type singularities of
the free energy (\ref{FglambdaA}) as $\lambda\to 0$ are related to a power-like
growth \beq \sigma_h^n\simeq\beta_h~n^{\alpha_h} \label{grow}
\eeq of the number of covering maps with large winding number $n$,
where $\alpha_h,\beta_h>0$. The leading
singularity of the series (\ref{chiralfreeen},\ref{FglambdaA}) as
$\lambda\to 0$ is extracted by substituting (\ref{grow}) to get
\bea \lim_{\lambda\to0}\,F_{U(N)}^+(\lambda)&=&
\sum_{h=1}^\infty\left(\frac{2\lambda}N\right)^{2h-2}\,\beta_h~
\sum_{n=1}^\infty n^{\alpha_h}~\e^{-n\,\lambda}\nonumber\\&=&
\sum_{h=1}^\infty\left(\frac{2\lambda}N\right)^{2h-2}\,\beta_h~
\Li_{-\alpha_h}\left(\e^{-\lambda}\right) \ .
\label{FUNsingsubs}\eea In the limit $\lambda\to0$, we can
substitute the singular behaviour of the Jonqui\`ere function
$\Li_{-\alpha_h}(z)$ for $z\to1^-$ (see Appendix~\ref{appA},
eq.~(\ref{Liseriesexpz1})) and to leading order we find
\beq \lim_{\lambda\to 0}\,F_{U(N)}^+(\lambda)=
\sum_{h=1}^\infty\,\frac{2^{2h-2}~\beta_h~\Gamma(\alpha_h+1)}{N^{2h-2}}~
\lambda^{2h-3-\alpha_h} \ . \label{FUNsinglead}\eeq

In the double scaling limit, we insert (\ref{mudoublescale}) into
(\ref{FUNsinglead}). Comparing this with the weak coupling
expansion (\ref{Fggenericdouble}) of the free energy, we find the
power of the growth (\ref{grow}) as the natural number \beq
\alpha_h=4h-4 \ , \label{alphahexpl}\eeq while the positive
numbers $\beta_h$ are given by
\beq
\beta_h=\frac{r_{4h-3,h}~\pi^{2h}}{2^{2h-2}\,(4h-4)!}=\frac{\xi_{2h-1}~
\pi^{2h}}{2^{2h-2}\,(4h-3)!} \ . \label{slope}\eeq From the
elementary power series identity \beq
\frac1{1-\e^{-\lambda}}\,\sum_{n=1}^\infty\sigma_h^n~
\e^{-n\,\lambda}=\sum_{n=1}^\infty\e^{-n\,\lambda}\,
\sum_{k=1}^n\sigma_h^k \label{powerid}\eeq we also obtain the
simple relation \beq
\beta_h=(4h-3)\,\lim_{M\to\infty}\,\frac1{M^{4h-3}}\,
\sum_{n=1}^M\sigma_h^n \label{betahasympts}\eeq which will be
crucial in the next subsection. That the explicit knowledge of the
small area behaviour of the gauge theory allows one to reconstruct
the asymptotic forms of the simple Hurwitz numbers is our first
main geometric characterization.
\begin{proposition}
The asymptotic expansion as $\mu\to\infty$ for the free energy
(\ref{freenDSdirect}) of $U(N)$ gauge theory on $\torus^2$ in the
double-scaling limit is the generating function for the asymptotic
Hurwitz numbers with
$$
\lim_{n\to\infty}\,\sigma_h^n=\frac{\pi^{2h}}{2^{2h-2}\,(4h-3)!}~\xi_{2h-1}~n^{4h-4}
\ .
$$
\label{Hurwitzprop1}\end{proposition}

We have thereby found that, in the strong coupling expansion of the
$U(N)$ gauge theory, the double scaling limit of the zero-instanton
sector picks up the contributions, genus by genus, of branched
covering maps over $\torus^2$ with infinite degree and it sums the
asymptotic behaviour of the simple Hurwitz numbers $\sigma_h^n$. Thus
the saddle-point equation for the zero-instanton double scaling free
energy solves the combinatorial problem of determining the asymptotics
of Hurwitz numbers. Proposition~\ref{Hurwitzprop1} combined with
Theorem~\ref{xifinal} produces a formula which coincides {\it
precisely} with Theorem~7.1 of~\cite{eskin1}, whereby the
asymptotics of simple Hurwitz numbers are evaluated directly by
involved combinatorial techniques. This computation, which follows from
the standard free fermion representation of
two-dimensional Yang-Mills theory, is sketched in Appendix~\ref{FFR}. We
stress that here the saddle-point equation provides a very efficient
and much simpler method for extracting these numbers.

\subsection{Principal Moduli Spaces of Holomorphic
  Differentials\label{MSHD}}

We will now demonstrate that the double scaling limit of the
$U(N)$ gauge theory on $\torus^2$ is in fact, rather remarkably,
related to the geometry of some very special moduli
spaces~\cite{eskin1}--\cite{MZE},
whose ``integer lattice" points (and thus
their volumes) are ``counted'' by the strong coupling expansion
coefficients of Section~\ref{DSLsaddle}. Let $\mathcal{H}_h$ be
the moduli space of (topological classes of) pairs
$(\Sigma,\dd\mbf u)$, where $\Sigma$ is a compact Riemann surface
of genus $h$ and $\dd\mbf u$ is a holomorphic one-form on $\Sigma$
with exactly $m=2h-2$ simple zeroes. The zero locus of $\dd\mbf u$
is the divisor
\beq
 \Delta_{\dd\mbf
u}=\sum_{i=1}^{2h-2}[u_i]\in{\rm Pic}(\Sigma) \ , ~~ u_i\in\Sigma
\ .
\label{zerolocus}\eeq
We call $\mathcal{H}_h$ a principal moduli space of holomorphic
differentials, or a principal stratum. Usually the points $u_i$ are
taken to be ordered (distinguishable), so that the generalizations to
the cases of zeroes of higher orders is straightforward. However, in
our simpler situation, this choice will produce a redundant
combinatorial factor to be introduced in the analysis below. The space
$\mathcal{H}_h$ admits a natural
action of the group $GL_+(2,\real)$ of $2\times2$ real matrices of
positive determinant acting as \beq
\begin{pmatrix}{\rm Re}(\dd\mbf u)\\{\rm Im}(\dd\mbf u)\end{pmatrix}
{}~\longmapsto~\begin{pmatrix}~a~&~b~\\~c~&~d~\end{pmatrix}
\begin{pmatrix}{\rm Re}(\dd\mbf u)\\{\rm Im}(\dd\mbf u)\end{pmatrix}
\ , ~~\begin{pmatrix}~a~&~b~\\~c~&~d~\end{pmatrix}\in GL_+(2,\real) \ .
\label{SL2Raction}\eeq
It may also be regarded as a fibration over the
moduli space $\mathcal{M}_h$ of Riemann surfaces of genus $h$, with
the fiber over $[\Sigma]\in\mathcal{M}_h$ equal to the vector space
$\Gamma(\Sigma,\Omega^{1,0}(\Sigma))$ modulo the action of the
automorphism group ${\rm Aut}(\Sigma)$ of the curve.

We can coordinatize the moduli space $\mathcal{H}_h$ as
follows. Consider the relative homology group $H_1(\Sigma,\Delta_{\dd\mbf
  u};\zed)\cong\zed^{4h-3}$, and choose a basis of relative one-cycles
$\{\gamma_i\}_{i=1}^{4h-3}\subset H_1(\Sigma,\Delta_{\dd\mbf
  u};\zed)$ such that $\gamma_i$ for $i=1,\dots,2h$ form
a canonical symplectic basis of one-cycles for the ordinary
homology group $H_1(\Sigma,\zed)$, while the open contours
$\gamma_{2h+i}$ for $i=1,\dots,2h-3$ connect the zeroes $u_{i+1}$
to $u_1$ on $\Sigma$, i.e.
$\partial\gamma_{2h+i}=[u_{i+1}]-[u_1]$. This basis may be
conveniently chosen in such a way that each element is represented
by a geodesic with respect to the flat metric ${\rm Re}(\dd\mbf
u~\overline{\dd\mbf u}~)$ induced by the
holomorphic differential $\dd\mbf u$. This choice naturally cuts
the surface into a union of flat polygons. We define the
corresponding period map \beq
\phi\,:\,\mathcal{H}_h~\longrightarrow~\complex^{4h-3}
\label{periodmap}\eeq by the formula \beq \phi(\Sigma,\dd\mbf
u):=\left(\,\mbox{$\oint_{\gamma_1}\dd\mbf u$}
\,,\,\dots\,,\,\mbox{$\oint_{\gamma_{2h}}\dd\mbf u$}\,,\,
\mbox{$\int_{\gamma_{2h+1}}\dd\mbf u$}\,,\,\dots\,,\,
\mbox{$\int_{\gamma_{4h-3}}\dd\mbf u$}\right) \ .
\label{periodmapdef}\eeq This map is holomorphic and locally
injective, and so it defines a local system of complex coordinates
on $\mathcal{H}_h$. This makes $\mathcal{H}_h$ a complex orbifold
of dimension \beq \dim\mathcal{H}_h=4h-3 \ .
\label{orbifolddim}\eeq The area $A_{\dd\mbf u}(\Sigma)$ of the
surface $\Sigma$ with respect to the metric defined by the
holomorphic one-form $\dd\mbf u$ can also be expressed in terms of
the periods via the Riemann bilinear relation \beq A_{\dd\mbf
u}(\Sigma):= \frac{\ii}{2}\,\int\limits_{\Sigma}\dd\mbf u
\,\wedge\, \overline{\dd\mbf
u}=\frac{\ii}{2}\,\sum_{i=1}^{h}\left(\,\oint
\limits_{\gamma_i}\dd\mbf u~\oint\limits_{\gamma_{h+i}}
\overline{\dd\mbf u}-\oint\limits_{\gamma_i}\overline{\dd\mbf u}~
\oint\limits_{\gamma_{h+i}}\dd\mbf u\right) \ . \label{volu}\eeq

By using the period map we can define a smooth measure  on the
moduli space $\mathcal{H}_h$ using the pull-back of the Lebesgue
measure from $\complex^{4h-3}$ to $\mathcal{H}_h$ under $\phi$. We
normalize this measure so that the volume of the unit cube in
$\complex^{4h-3}$ is $1$. This measure is independent of the
choice of basis $\{\gamma_i\}_{i=1}^{4h-3}$. However, the total
volume of $\mathcal{H}_h$ with respect to this measure is
infinite. To cure this, we restrict to the subspace
$\mathcal{H}_h'\subset\mathcal{H}^{~}_h$ consisting of pairs
$(\Sigma,\dd\mbf u)$ of unit area $A_{\dd\mbf u}(\Sigma)=1$.
This subset, which according to (\ref{volu}) defines a hyperboloid in
$\mathcal{H}_h$ with respect to the coordinates chosen above, is
invariant under the $SL(2,\real)$ subgroup of the $GL_+(2,\real)$
action (\ref{SL2Raction}). The volume element $\dd{\rm
  vol}(\mathcal{H}_h)$ on $\mathcal{H}_h$ naturally induces a volume
element $\dd{\rm vol}(\mathcal{H}_h')$ on the hypersurface
$\mathcal{H}_h'$ in the following way. Any element in $\mathcal{H}_h$
can be represented as $(\Sigma,r~\dd\mbf u)$ where $r$ is a positive
real number and $(\Sigma,\dd\mbf u)$ represents an element of
$\mathcal{H}_h'$. We define the volume element $\dd{\rm
  vol}(\mathcal{H}_h')$ on the hyperboloid  $\mathcal{H}_h'$ by
disintegration of the volume element $\dd{\rm vol}(\mathcal{H}_h)$ on
$\mathcal{H}_h$ as
\beq
\label{dis}
\dd{\rm vol}\bigl(\mathcal{H}_h\bigr)=r^{4h-4}~\dd r~\dd{\rm vol}
\left(\mathcal{H}_h'\right) \ .
\eeq
It is an important result of ergodic theory that the volume of
$\mathcal{H}_h'$ is finite with respect to this
measure~\cite{ergodic1,ergodic2}. In particular, the action of the
diagonal subgroup of $SL(2,\real)\subset GL_+(2,\real)$ on
$\mathcal{H}_h'$ is ergodic.

We will now explicitly compute the volumes of these moduli spaces.
This can be done through an indirect procedure. One starts by
introducing a sequence of cones
 \beq {\rm
  C}_M\phi\left(\mathcal{H}_h'\right):=\left
\{t\,\phi\left(\mathcal{H}_h'\right)~\left|~0\leq t\leq\sqrt M\,
\right.\right\} \label{coneN}\eeq
over $\phi(\mathcal{H}_h')$ with vertex at the origin of
$\complex^{4h-3}$. We express the volume of the moduli
space $\mathcal{H}_h'$ in terms of the volume of the unit cone by the
formula
\beq {\rm
vol}\left(\mathcal{H}_h'\right):=(4h-3)~{\rm vol}^{~}_{\complex^{4h-3}}
\Bigl({\rm
  C}_1\phi\left(\mathcal{H}_h'\right)\Bigr)=(4h-3)\,\int\limits_{{\rm
  C}_1\phi(\mathcal{H}_h')}\dd\nu~1
\label{volumedef}\eeq
where $\dd\nu$ is the Lebesgue measure on $\complex^{4h-3}$. The
normalization factor takes into account the integration over radial
part in eq.~(\ref{dis}). Computing the volume of the unit cone is
actually simpler, as one can use an old trick. Consider an integer
lattice $\Lambda\cong\zed^{2(4h-3)}$ inside
$\real^{2(4h-3)}\cong\complex^{4h-3}$ and count the intersections of
this lattice with ${\rm C}_M\phi(\mathcal{H}_h')$ for $M$ large. Their
cardinality scales as
\beq
\label{lat1}
\Bigl|{\rm C}_M\phi\left(\mathcal{H}_h'\right)~\cap~\Lambda\Bigr|
\simeq M^{4h-3}~{\rm vol}^{~}_{\complex^{4h-3}}\Bigl({\rm
  C}_1\phi\left(\mathcal{H}_h'\right)\Bigr) \ ,
\eeq
and thus the asymptotic behavior of the left-hand side of (\ref{lat1})
yields the volume we are looking for.

The relationship between these principal moduli spaces and the double
scaling limit of two-dimensional Yang-Mills theory on the torus
$\torus^2$ emerges when we have to select and geometrically interpret
the integer lattice $\Lambda$. Any point $(\Sigma,\dd\mbf
u)\in\mathcal{H}_h$, whose first $2h$ coordinates $\phi_i(\Sigma,\dd\mbf u)$,
$i=1,\dots,2h$ are integral (i.e. $\phi_i(\Sigma,\dd\mbf
u)\in\mathbb{Z}^2$), is in one-to-one correspondence with a covering
of the torus with only simple branch points and degree given by the
area of $\Sigma$~\cite{eskin1}. Consider a simple branched covering
\beq
\varpi\,:\,\Sigma~\longrightarrow~\torus^2
\label{branchedpi}\eeq
of the torus by a Riemann surface of genus~$h$, with simple
ramification over distinct points
$z_1,\dots,z_{2h-2}\in\torus^2$. With $\dd z$ denoting the canonical
holomorphic differential on $\torus^2$, one can use the pull-back
under (\ref{branchedpi}) to associate the point
\beq
\bigl(\Sigma\,,\,\varpi^*(\dd z)\bigr)~\in~\mathcal{H}_h
\label{piptHh}\eeq
with simple zeroes $u_i=\varpi^{-1}(z_i)$ corresponding to the
ramification points of the cover. The integration of $\varpi^*(\dd z)$
over the homology cycles of $\Sigma$ produces integer values. Conversely, given
$(\Sigma,\dd\mbf u)\in\mathcal{H}_h$ we can define a covering map
(\ref{branchedpi}) by
\beq
z=\varpi(u):=\int^u\dd\mbf u~~~~{\rm mod}~\zed^2
\label{convbranchedpi}\eeq
which is well-defined since $\oint_\gamma\dd\mbf u\in\zed^2$ for any
one-cycle $\gamma\subset\Sigma$. The critical points of $\varpi$ are
precisely the simple zeroes $u_i$ of the holomorphic differential
$\dd\mbf u=\varpi^*(\dd z)$, and $\varpi$ thereby has simple ramification at
$u_i\in\Sigma$. The degree of the covering map (\ref{convbranchedpi}) is
the area (\ref{volu}). We have thereby arrived at a one-to-one
correspondence between simple branched covers of the torus, and hence
terms in the chiral Gross-Taylor string expansion, and integral points
in the principal moduli spaces of holomorphic differentials.

We will now show that the numbers (\ref{volumedef}) are computed
by the strong-coupling saddle-point expansion of the gauge theory
that we obtained in Section~\ref{DSLsaddle}. As alluded to above, the basic
idea is to use a ``refined" version of (\ref{lat1}). By definition
we have
\beq
{\rm vol}\left(\mathcal{H}_h'\right)=(4 h-3)\,\lim_{M\to\infty}\,
\frac1{M^{4h-3}}\,\left|\,{\rm C}_M\phi\left(\mathcal{H}_h'\right)
{}~\cap~\left(\zed^{2(4h-3)}+\mbf b\right)\,\right|
\label{volumelimit}
\eeq
where the vector $\mbf
b=(b_i)\in\complex^{4h-3}$ has components $b_i\in\zed^2$ for
$i=1,\dots,2h$ while $b_i\neq b_j~~{\rm
  mod}~\zed^2$ for $i,j>2h$ with $i\neq j$ (This vector is inserted in
conjunction with the quantization properties of the period map
(\ref{periodmapdef})). On the other hand, from
the above correspondence it follows that each point of the
intersection ${\rm
C}_M\phi(\mathcal{H}_h')\cap(\zed^{2(4h-3)}+\mbf b)$ corresponds
to a simple branched cover $\varpi$ of $\torus^2$ with winding
number $\leq M$, and thus the volume (\ref{volumelimit}) may be
computed via the asymptotics of simple Hurwitz numbers as
\beq {\rm
vol}\left(\mathcal{H}_h'\right)=(2h-2)!\,(4 h-3)\,\lim_{M\to\infty}\,
\frac1{M^{4h-3}}\,\sum_{n=1}^M\sigma_h^n \ .
\label{volumeHurwitz}
\eeq
The combinatorial factor $(2 h-2)!$ accounts for the fact that the
ramification points are considered indistinguishable in the counting
of the coverings. Comparing with (\ref{betahasympts}) and
(\ref{slope}) we thereby find that the volumes are completely
determined in terms of the coefficients $\xi_{2h-1}$ of the
saddle-point solution to be
\beq {\rm
vol}\left(\mathcal{H}_h'\right)=\frac{2^{2-2h}\,(2h-2)!
}{(4h-3)!}~\xi_{2h-1}~\pi^{2h} \ . \label{volumesaddle}
\eeq
Using the formula (\ref{volumesaddle}) and the explicit expansion
(\ref{largeNfreenUNdouble}), the first few volumes can be readily computed and
are summarized in Table~\ref{Hurwitztable}. (These agree with the
values for low genus computed in~\cite{MZE}). Their general expressions
are provided by Theorem~\ref{xifinal}. In particular, the
saddle-point computation explicitly demonstrates the rationality
property $\pi^{-2h}~{\rm
  vol}(\mathcal{H}_h')\in\rat$~\cite{kont1,kont2} and gives a precise
geometrical meaning to the rational numbers that we first encountered
in the weak-coupling expansions of the previous section. We may
summarize the geometric interpretation of the double scaling limit as
follows.
\begin{proposition}
The asymptotic expansion as $\mu\to\infty$ for the free energy of
$U(N)$ gauge theory on $\torus^2$ in the double-scaling limit is the
generating function for the volumes of the principal moduli spaces of
holomorphic differentials given by
$$
\hat{\mathcal{F}}_{U(N)}(\mu)=N\,\sum_{h=1}^\infty\,
\frac{2^{2h-2}\,(4h-4)!}{(2h-2)!}~\left(\frac1{\pi\,\mu}\right)^{2h-1}~{\rm
vol}\left(\mathcal{H}_h' \right)+O\left(\e^{-\mu}\right) \ .
$$
\label{Hurwitzprop2}\end{proposition}

\TABULAR{|c|c|}{\hline \ $h$ \ & \
 {${\rm vol}(\mathcal{H}_h')/\pi^{2h}$} \ \\ \hline\hline
$1$ & {\mbox{$\frac13$}}
\\ \hline $2$ & {\mbox{$\frac1{135}$}} \\ \hline $3$ & {
\mbox{$\frac1{4860}$}} \\ \hline
$4$ & {
\mbox{$\frac{377}{67359600}$}} \\ \hline
$5$ & {
\mbox{$\frac{23357}{157621464000}$}} \\ \hline
$6$ & {
\mbox{$\frac{16493303}{42765855611248000}$}}
\\ \hline}{The normalized volumes of the principal moduli spaces of
 holomorphic differentials up to genus $h=6$.\label{Hurwitztable}}

\subsection{Open String Interpretation\label{OSI}}

At this stage we can speculate on the string interpretation of the
double scaling limit of the gauge theory that we have constructed
above, with the foresight of the connection
to a fluxon theory on the noncommutative plane that we will
describe in the next section. First of all, only the chiral part
of the Gross-Taylor series appears to contribute in the double
scaling limit of commutative $U(N)$ gauge theory on $\torus^2$.
This is not surprising upon examination of the explicit form for the
non-chiral part of the free energy presented in~\cite{Rudd1}. It
is not difficult to check that non-chiral contributions are always
subleading in the small area limit (An alternative perspective on
these non-chiral couplings has been presented recently
in~\cite{vafa1}). It is in complete harmony with the fact, discussed
at the beginning of Section~\ref{DecompFields}, that any theory of fluxons is
necessarily chiral. Alternatively, we may regard this fact as
being the statement that the holomorphic and anti-holomorphic
sectors are identified with one another, so that the expansion is
in terms of {\it open} strings. Being a chiral series, the
original strong-coupling expansion is given as a sum over {\it
all} Young diagrams, labelling representations of the infinite
unitary group $U(\infty)$, which is the gauge group of a {\it
noncommutative} gauge theory. In the strong coupling regime, the
double scaling limit involves the asymptotic Hurwitz numbers, of
branched covering maps to $\torus^2$ in the limit of infinite
winding number, again reflecting the open nature of the string
degrees of freedom. This latter feature also implies that the
toroidal spacetime is effectively decompactified onto the plane in
the double scaling limit, corresponding to the equi-anharmonic
limit $\tau\to\ii\infty$ of the underlying elliptic curve. This is
certainly true of the limit analysed in Section~\ref{Fluxon} and
is responsible for the loss of modular structure in the free
energy~(\ref{Fggenericdouble}).

Let us now note that the Gross-Taylor series is an expansion in
$\frac1N$, i.e. it is {\it perturbative} in the string coupling
$g_s$, and as such it ignores non-perturbative corrections of the
form $\e^{-N\,\lambda}$. From (\ref{gsalpha}) we see that these
contributions have a natural interpretation~\cite{LMR1} (see
also~\cite{MM1}) as coming from (Euclidean) D1-branes of tension \beq
T_1=\frac1{\pi\,\alpha'g_s} \label{T1tension}\eeq which wrap
around the target space torus $\torus^2$ without foldings. There
are no stable D0-branes in this picture~\cite{LMR1}, and so the
non-perturbative structure is typical of a Type~0B string theory.
Clearly, any consistent treatment of the new double
scaling limit should be able to capture the D-string contributions
which are of order $\e^{-\mu}$, with the double scaled coupling
constant
\beq\mu=\displaystyle{\frac{T_1A}{2\pi}\,}\label{doubletension}\eeq
related to the brane tension. The free energy (\ref{bella})
computed from the exact solution of the saddle-point equation
encodes contributions of this type. They are the constituents of
the exponentially suppressed part at strong-coupling in
(\ref{freenDSdirect}) and they appear consistently resummed by the
saddle-point equation. Alternatively, the necessary presence of
these contributions can be understood from the point of view of
the strong-coupling expansion. The series (\ref{freenDSdirect})
truly diverges from the string theory perspective, i.e. in the
double scaling limit of the Gross-Taylor expansion, because the
coefficients of the series grow as $(4k-4)!~\Pi(k)$. It is
unlikely that this series is even Borel summable, since
$\Pi(k)\simeq\e^{\alpha\,\sqrt k}$ for large $k$
(c.f.~eq.~(\ref{HRformula})). Instead, in our
saddle point approach this is just an artifact of the
strong-coupling approximation, the exact solution being completely
regular except at $\mu=0$ where the Douglas-Kazakov type phase
transition could take place (c.f.~Fig.~\ref{DoubleU(N)}). Thus the
exponentially suppressed contributions of order $\e^{-\mu}$,
interpreted in the string theory picture as contributions of
D-strings, have to come into the game in order to ensure
regularity of the complete result. It would be interesting to have
a direct physical understanding in the field theory for the
necessity of these non-perturbative D1-brane corrections.

These corrections will be particularly important in the
noncommutative gauge theory because they will be {\it enhanced} in
the required double scaling limit of the $SU(N)/\zed_N$ theory.
The problem of computing the noncommutative free energy will
thereby boil down to the problem of resumming these open string
degrees of freedom. We expect that the D-strings which emerge in
this way are the remnants of the electric dipole configurations in
the instanton expansion.

With this line of reasoning, it is now tempting to conjecture a
more direct open string interpretation of the expansion
(\ref{Fggenericdouble}). In order to select the relevant
geometrical structures we start by rewriting the strong-coupling
limit of the free energy (\ref{Fggenericdouble}) in terms of open
string parameters as
\beq \mathcal{Z}_{\rm str}(g_s,T_1A):=\lim_{\mu\to\infty}\,\hat
{\mathcal F}_{U(N)}(\mu)=\frac1{g_s}\,\sum_{k=1}^\infty
\frac{2^{4k-3}\,(4k-3)!}{(2k-2)!}~\frac{{\rm
vol}\left(\mathcal{H}_k'\right)}{(T_1A)^{2k-1}} \ .
\label{openstrdouble}\eeq We interpret this expansion as the
remnant of a resummation of the D1-branes. The effective action
(\ref{openstrdouble}) is of order $1/g_s$, and it thereby
represents an open string {\it disk} amplitude. Thus the double
scaling limit of commutative $U(N)$ gauge theory on $\torus^2$, in
the strong-coupling regime, can be interpreted as a theory of
D1-branes at tree-level in open string perturbation theory. The
truncation of the dynamics to tree level is presumably related to
the fact that the strong coupling regime of the double scaling
gauge theory is described by some sort of topological open string
theory. In the next section, similar considerations in the
$SU(N)/\zed_N$ gauge theory, appropriate to the description of
noncommutative Yang-Mills theory, will be used to argue that the
fluxon expansion on $\real_\Theta^2$ is a certain noncommutative
deformation of the open string theory described here.

Denoting by $\mathcal{H}':=\coprod_{h\in\nat}\mathcal{H}_h'$ the total
moduli space of holomorphic differentials, we may write the string
partition function (\ref{openstrdouble}) as the moduli space integral
\beq
\mathcal{Z}_{\rm str}(g_s,T_1A)=\int\limits_{\mathcal{H}'}
\dd{\rm vol}_{\mathcal{H}'}^{~}(\phi)~W(\phi;g_s,T_1A)
\label{Zstrmodint}\eeq
where the density $W(\phi;g_s,T_1A)$ is given by appropriate symmetry
factors. From the general principles of topological string
theory~\cite{CMR1}, we expect that this large area result computes the
orbifold Euler characteristic $\chi(\mathcal{H}'\,)$ of the moduli space
$\mathcal{H}'$. We can present some evidence in favour of this as
follows.

The moduli space $\mathcal{H}^{~}_h$ is connected~\cite{kont2} and can be
related to $\mathcal{H}_h'$ through the fibration
\bea
\real_+~\longrightarrow~&\mathcal{H}^{~}_h&
\nonumber\\&\downarrow&\nonumber\\&\mathcal{H}_h'& \ ,
\label{Hhfibration}\eea
where the multiplicative group $\real_+=(0,\infty)$ is identified with
the one-parameter subgroup $\{\e^t\,\id_2~|~t\in\real\}\subset
GL_+(2,\real)$ acting as in (\ref{SL2Raction}). Since the Euler
character of a fiber bundle with connected base is the product of the
Euler characters of the base and fiber, by taking $\chi(\real_+)=1$ we
find the equality
\beq
\chi\left(\mathcal{H}_h^{~}\right)=\chi\left(\mathcal{H}_h'\right) \ .
\label{chiHhequal}\eeq
Next we apply the fibration noted in the previous subsection,
\beq
\begin{matrix}\Gamma\bigl(\Sigma\,,\,\Omega^{1,0}(\Sigma)\bigr)
&\longrightarrow&\mathcal{H}^{~}_h
\\ & &\downarrow&\\ & &\mathcal{M}^{~}_h\end{matrix} \ ,
\label{Mhfibration}\eeq
along with the Harer-Zagier-Penner formula~\cite{Penner1} for the
Euler characters of Riemann moduli spaces to compute
\beq
\chi\left(\mathcal{H}_h'\right)=\chi\left(\mathcal{M}_h^{~}
\right)~\chi\left[
\Gamma\bigl(\Sigma\,,\,\Omega^{1,0}(\Sigma)\bigr)\right]=
\frac{B_{2h}}{2h\,(2h-2)}~\chi\left[
\Gamma\bigl(\Sigma\,,\,\Omega^{1,0}(\Sigma)\bigr)\right] \ .
\label{chiHhGamma}\eeq
Given the explicit formulas for the volumes ${\rm
  vol}(\mathcal{H}_h')$ obtained in this section, the relation
  (\ref{chiHhGamma}) strongly suggests that an explicit realization of
  the string partition function (\ref{Zstrmodint}) as an Euler
  sigma-model of the form
$\mathcal{Z}_{\rm str}\simeq\chi(\mathcal{H}'\,)=
\int_{\mathcal{H}'}e(T\mathcal{H}'\to\mathcal{H}'\,)$ should hold,
with $e(T\mathcal{H}'\to\mathcal{H}'\,)$ the Euler characteristic
class of the tangent bundle over the moduli space
$\mathcal{H}_h'$. This would directly connect the double scaling
string theory with a certain topological string theory~\cite{CMR1}.
We may also expect the form (\ref{Zstrmodint})
to be generically valid at finite area, so that the
string path integral localizes in the usual way onto finite dimensional
moduli spaces. The double scaling limit replaces the Hurwitz moduli
space of simple branched covers of $\torus^2$, arising in the 't~Hooft
limit of QCD$_2$ on the torus~\cite{CMR1}, by the principal moduli space of
holomorphic differentials.

This realization is the starting point for an investigation of the
rewriting of the double scaling expansion as some sort of topological
field theory. The action principle for this open string theory is
given by the topologically twisted $\mathcal{N}=2$ superconformal
field theory coupled to gravity that describes the closed string
expansion of two-dimensional Yang-Mills theory~\cite{bcov}, by taking
the torus target space to be equi-anharmonic after a modular
transformation. The specification of vacua in this
context requires a choice of holomorphic vector bundle over the
underlying elliptic curve, and the one-loop open string
partition function is computed as the corresponding holomorphic
Ray-Singer analytic torsion, represented concisely in terms of an
index integral as in the closed string case. This would thus identify
the open string action in terms of a deformed topological sigma-model
coupled to topological gravity. The double scaling limit replaces the
counting of holomorphic maps $\Sigma\to\torus^2$, arising in the
't~Hooft limit of QCD$_2$ on the torus~\cite{bcov}, by the counting of
holomorphic differentials on complex curves $\Sigma$. This may be loosely
interpreted as replacing a topological A-model string theory with
elliptic curve target space by a topological B-model string
theory. An alternative action principle for the
ordinary QCD$_2$ string is proposed in~\cite{Horava1}.

The relation between two-dimensional Yang-Mills theory and string
theory has been exploited recently in connection with ${\cal N}=2$
supersymmetric gauge theories in four dimensions~\cite{MMO}. The large
$N$ limit of the partition function of $U(N)$ Yang-Mills theory on the
sphere $\mathbb{S}^2$ reproduces the instanton counting of
four-dimensional ${\cal N}=2$ supersymmetric gauge theories introduced
in~\cite{nek1,nek2}. The relevant instanton moduli spaces, appropriate
for the counting, have been also shown to map onto the moduli spaces
of punctured Riemann spheres~\cite{mato}, suggesting a relation
between instantons in four dimensions and Hurwitz numbers. Because the
rationale behind this circle of ideas is still topological
string theory, it could be interesting to explore the connection of
our construction with these recent results.

We can now argue that the indications of a string theory that we
have found are consistent, at least at a qualitative level, with
the spectrum of instanton strings that we found in
Section~\ref{InstStrings}. First of all, the moduli spaces
$\mathcal{H}_h'$ admit an ergodic action of the group
$SL(2,\real)$. This action is the Teichm\"uller geodesic flow which
gives the Euler-Lagrange equations for geodesics with respect to the
Teichm\"uller metric on the moduli space $\mathcal{M}_h$ of complex curves of
genus~$h$~\cite{kont1}. The chaotic nature of these flows on the
moduli space $\mathcal{H}'$ suggests that the topological string
theory amplitudes (\ref{Zstrmodint}) may admit an interpretation
as computing certain fractal dimensions, tied to the structure of
the orbifold Euler characteristics described above. This property
may be related to the fractal characteristics of the
noncommutative instanton spectrum that we conjectured in
Section~\ref{InstStrings}. Secondly, the free fermion formalism
described in Appendix~\ref{FFR} shows that the string states which
contribute to the string expansion of the gauge theory are
represented by the partition basis (\ref{partbasis}) of the
fermionic Fock space. In the double scaling limit strings of very
large winding number dominate and are represented by very large
partitions $\mbf n$, i.e. $\sum_ln_l=n$ with $n\to\infty$. From
the Hardy-Ramanujan formula (\ref{HRformula}) it follows that the
degeneracy of such string states grows like
$\Pi(n)\simeq\e^{\alpha\,\sqrt n}$, producing the familiar
exponential form of worldsheet densities of states. This is
completely analogous to the asymptotic behaviour of the instanton
density of states that we described in Section~\ref{InstStrings}.

We should also explore the possibility that a Douglas-Kazakov type
phase transition could take place for some finite value of the
coupling $\mu>0$ in the double-scaling limit, in contrast to what
occurs in the ordinary 't~Hooft limit. Note that this limit
necessarily restricts us to the weak-coupling regime of the gauge
theory, in which only the instanton expansion is available. To
answer the question of a non-trivial phase transition requires an
accurate analysis of {\it all} higher instanton contributions to
the partition function and one must really face up to the entire
instanton series.

\section{Noncommutative Instanton String Representation\label{ISRNCGT}}

In this section we will examine the instanton expansion of
noncommutative Yang-Mills theory on $\torus^2$ with rational
valued noncommutativity parameter $\theta=1/N$ ($n=1$ in the
notations of Section~\ref{Decomp}). By Morita equivalence, this
theory is completely identical to commutative $U(N)$ gauge theory
on $\torus^2$ in a certain Chern class $q\in\zed$. Because of the
global group isomorphism $U(N)\cong U(1)\times SU(N)/\zed_N$, the
partition function of this theory receives a trivial contribution
$\exp\bigl(-\frac{2\pi^2q^2}{g^2A\,N}\bigr)$ from the $U(1)$
sector, which is exactly cancelled by the background gauge field
generated in the Morita transformation~\cite{szrev,gsv1}. The remainder is the
partition function of $SU(N)/\zed_N$ gauge theory on $\torus^2$
with Chern number $q$. The Chern classes are labelled by elements
of the fundamental group of the gauge group, which in the present
case is \beq
\pi_1\bigl(SU(N)/\zed_N\bigr)=\pi_0\bigl(\zed_N\bigr)=\zed_N \ .
\label{pi1SUN}\eeq Thus there are $N$ topological 't~Hooft flux
sectors labeled by $q=0,1,\dots,N-1$~\cite{hooft2}. The Morita transformation
also shrinks the area of the torus by a factor of $N^2$, $A\to
A/N^2$, and rescales the Yang-Mills coupling constant as $g^2\to
g^2/N$~\cite{szrev,gsv1}.

The large $N$ limit required to produce gauge theory on the noncommutative
plane involves a double scaling limit defined by taking $A\to\infty$
in (\ref{relation}) with $\Theta$ fixed and $n=1$. Because of the
rescalings generated by Morita equivalence, it follows that it is
precisely the coupling constant (\ref{mudoublescale}) which is kept
finite in this limit, and it is given in terms of the parameters of
the noncommutative gauge theory as
\beq
\mu=g^2\Theta \ .
\label{mugTheta}\eeq
It is possible to carry out a double scaling limit in which the size
$\sqrt A$ of the torus is also kept finite in the large $N$
limit~\cite{amns1}, by fixing instead the 't~Hooft coupling constant
(\ref{tHooftconst}) and sending $n\to\infty$ in (\ref{relation}) with
$n/N$ fixed, but here we shall deal only with the somewhat simpler
case where the limiting gauge theory lives on $\real_\Theta^2$ rather
than on $\torus_\theta^2$.

In this section we will analyse the noncommutative gauge theory by
generalizing the technique of the previous section and use it to
describe a possible open string representation for gauge theory on
the noncommutative plane $\real_\Theta^2$. We will attempt to interpret
this string theory, in the strong coupling regime of the gauge
theory, as a tree-level theory of D1-branes similar to the one
described in Section~\ref{OSI}. This analysis may shed light on
the nature of the full string theory that reproduces the spectrum
of noncommutative instantons that we analysed at length in
Section~\ref{InstStrings}. Heuristically, because $SU(N)$ Yang-Mills
theory on $\torus^2$ is a string theory at large $N$, it induces a
corresponding string representation of the noncommutative gauge theory
obtained through the double scaling limit described above. We will
find that this gives a certain {\it closed} gauge string theory
description of the origin of fluxons on the noncommutative plane.

\subsection{Instanton Expansion of $SU(N)/\zed_N$ Gauge
  Theory\label{IESUNZNGT}}

We begin by rewriting the the instanton representation of commutative
$SU(N)/\zed_N$ gauge theory on $\torus^2$, in the $q^{\rm th}$
't~Hooft sector, in a manner analogous to what we did for the $U(N)$
theory in Section~\ref{IEUNGT}. The partition function can be read off
from the general instanton series (\ref{Zpqpart}) at $\theta=0$,
$p=N-1$ by inserting the appropriate background flux shifts in the
Chern numbers and the appropriate terms which arise from the Migdal
expansion. It is given explicitly by~\cite{pasz1,gsv1}
\bea
Z_{SU(N)}^{(q)}\left(g^2A\right)&=&(-1)^{N+(N-1)q}~
\e^{-\epsilon^{~}_{\rm F}}\,\Biggl\{\,\sqrt{\frac{2\pi^2}{g^2A\,N}}+
{\sum_{\stackrel{\scriptstyle
\vnu\in\nat_0^{N-1}}{\scriptstyle\sum_kk\,\nu_k=N}}}'~
\prod_{k=1}^{N-1}\frac{(-1)^{\nu_k}}{\nu_k!}\,\left(\frac{2\pi^2}
{k^3g^2A}\right)^{\nu_k/2}\Biggr.\nonumber\\&&\times\Biggl.
\sum_{\stackrel{\scriptstyle\vq\in\zed^{|\vnu|}}
{\scriptstyle\sum_jq_j=q}}\exp\left[-\frac{2\pi^2}
{g^2A}\,\sum_{l=1}^N\frac1l\,\sum_{j=1+\nu_1+\dots+
\nu_{l-1}}^{\nu_1+\dots+\nu_l}\left(q_j-\frac{l\,q}N\right)^2\right]
\Biggr\}
\label{ZNMorita}\eea
where we use the same conventions as in (\ref{ZUNinstexp}). The prime
on the first sum in (\ref{ZNMorita}) means that it omits the trivial partition
of
$N$ which has $\nu_1=N$, $\nu_k=0~~\forall k>1$. This is the
contribution from vacuum fluctuations represented by the first
non-exponential term in (\ref{ZNMorita}), and it is thereby associated
with the contribution from the zero-instanton sector of the
theory. The infinite series then contains the contributions from all
multi-instanton configurations.

We proceed as in Section~\ref{IEUNGT}, except that we now introduce a
{\it double} contour integration to resolve both the partition and
topological charge constraints in (\ref{ZNMorita}). One finds
\bea
Z_{SU(N)}^{(q)}\left(g^2A\right)&=&(-1)^N~\e^{-\epsilon^{~}_{\rm
    F}+\frac{2\pi^2q^2}{g^2A\,N}}\,\oint
\limits_{{\cal C}_0(z)}\frac{\dd z}{2\pi\ii z^{N+1}}~
\oint\limits_{{\cal C}_0(w)}\frac{\dd w}{2\pi\ii w^{q+1}}~
\sum_{\vnu\in\nat_0^{N-1}}~\prod_{k=1}^{N-1}\frac{(-1)^{\nu_k}}{\nu_k!}
\nonumber\\&&\times\,\left(\frac{2\pi^2}{k^3g^2A}\right)^{\nu_k/2}\,
z^{k\,\nu_k}\,\left[\Xi_k\left(w\,;\,g^2A\right)\right]^{\nu_k} \
, \label{ZNunconstr}\eea where we have absorbed the identity
contribution to (\ref{ZNMorita}) into the sum over all
$\vnu\in\nat_0^{N-1}$ in (\ref{ZNunconstr}), and defined the
sequence of functions \beq
\Xi_k\left(w\,;\,g^2A\right):=\sum_{m=-\infty}^{\infty}
(-1)^{(N-1)m}~w^m~\e^{-\frac{2\pi^2}{g^2A}\,\frac{m^2}k} \ , ~~
k=1,\dots,N-1 \label{XikwgTheta}\eeq which coincide with
(\ref{calFkdef}) at $w=1$. Carrying out the sums in
(\ref{ZNunconstr}) thereby leads to the formula \bea
Z_{SU(N)}^{(q)}\left(g^2A\right)&=&\oint \limits_{{\cal
C}_0(z)}\frac{\dd z}{2\pi\ii z^{N+1}}~ \oint\limits_{{\cal
C}_0(w)}\frac{\dd w}{2\pi\ii w^{q+1}}~\exp\left[
-\sqrt{\frac{2\pi^2}{g^2A}}~\sum_{k=1}^\infty\,\Xi_k\left(w\,;\,g^2A\right)~
\frac{(-z)^k}{k^{3/2}}\right]\nonumber\\&&\times~\e^{-\epsilon^{~}_{\rm
    F}+\frac{2\pi^2q^2}{g^2A\,N}} \ .
\label{ZNnusummed}\eea
By changing integration variables $z\to z/w$ and then rescaling
$z\to\e^{2\pi^2/g^2A}\,z$, $w\to\e^{4\pi^2/g^2A}\,w$, one finds that
the partition function (\ref{ZNnusummed}) depends only on the value of
the 't~Hooft flux $q$ modulo $N$,
\beq
Z_{SU(N)}^{(q+N)}\left(g^2A\right)=Z_{SU(N)}^{(q)}\left(g^2A\right) \
,
\label{modNsym}\eeq
as it should.

Let us now compare the $SU(N)/\zed_N$ gauge theory with the
strong-coupling expansion of the $U(N)$ gauge theory that we
obtained in Section~\ref{IEUNGT}. Following the steps which led to
(\ref{ZUNstrongcoupl}), the only changes in the present case are
that the Fermi level (\ref{Fermisurface}) is shifted as
$n^{~}_{\rm F}\to n_{\rm F}^{~}+\ln(w)/2\pi\ii$ and the Fermi
energy (\ref{Fermien}) by the $U(1)$ charge contribution of the
$q^{\rm th}$ flux sector, along with the accompanying contour
integration over ${\cal C}_0(w)$. In this way we arrive at the
strong-coupling form \bea Z_{SU(N)}^{(q)}\left(g^2A\right)&=&\oint
\limits_{{\cal C}_0(z)}\frac{\dd z}{2\pi\ii z^{N+1}}~
\oint\limits_{{\cal C}_0(w)}\frac{\dd w}{2\pi\ii w^{q+1}}~
\prod_{n=-\infty}^\infty\,\left(1+z~\e^{-\frac{g^2A}2\,(n-n^{~}_{\rm
      F}-\frac{\ln w}{2\pi\ii})^2}\right)\nonumber\\&&\times~
\e^{-\epsilon^{~}_{\rm F}+\frac{2\pi^2q^2}{g^2A\,N}} \ .
\label{ZSUNstrongcoupl}\eea
Again, the Migdal expansion is recovered by expanding the product and
doing the integrals, with the contour integration over $w$
implementing the appropriate $U(1)$ charge subtraction in the $q^{\rm
  th}$ topological sector. In terms of the Young tableaux variables of
Section~\ref{TG-TET}, the 't~Hooft fluxes are given by $N\,q=\sum_kn_k-n$
while the quadratic Casimir (\ref{C2tildeC2UN}) is shifted as
$C_2(R(Y))\to C_2(R(Y))-\frac{(N\,q+n)^2}N$ for $Y\in{\cal Y}_n$. The
ordinary $SU(N)$ partition function is given by
(\ref{ZSUNstrongcoupl}) in the trivial sector $q=0$ with
$Z^{~}_{SU(N)}(g^2A)=N\,Z_{SU(N)}^{(0)}(g^2A)$.
On the other hand, the $U(N)$ partition function is recovered by
reinstating the $U(1)$ charge subtraction and summing
$\exp\bigl(-\frac{2\pi^2q^2}{g^2A\,N}\bigr)\,Z_{SU(N)}^{(q)}(g^2A)$ over all
$q\in\nat_0$. Using the identity $\sum_{q\in\nat_0}w^{-q}=w/(w-1)$
and evaluating the residue of the simple pole arising in
(\ref{ZSUNstrongcoupl}) at $w=1$, we arrive at
(\ref{ZUNstrongcoupl}).

\subsection{Double Scaling Limit\label{NCDSL}}

Starting from the contour integral representation
(\ref{ZNnusummed}), we shall now study the large $N$ double
scaling limit of the $SU(N)/\zed_N$ gauge theory in which the
coupling constant (\ref{mudoublescale},\ref{mugTheta}) is held
fixed. In this limit, we reproduce noncommutative gauge theory in
two dimensions. Let us assume for simplicity that the rank $N$ is
odd, so that the Fermi level (\ref{Fermisurface}) is an integer.
This assumption has no bearing on the final results obtained in
the $N\to\infty$ limit. We also rescale the integration variable
$w\to\e^{N\,\pi\,q/{g^2\Theta}}\,w$ and thereby write
(\ref{ZNnusummed}) in the form \bea \hat{\cal
Z}_{SU(N)}^{(q)}\left({g^2\Theta}\right)
&=&\e^{\frac{\pi\,N\,g^2\Theta}{12}}\, \oint\limits_{{\cal
C}_0(z)}\frac{\dd z}{2\pi\ii z^{N+1}}~
\exp\left[-N\,\sqrt{\frac{\pi}{g^2\Theta}}~\Li_{3/2}(-z)\right]
\nonumber\\&&\times \,\oint\limits_{{\cal C}_0(w)}\frac{\dd
w}{2\pi\ii w^{q+1}}~
\exp\left[-N\,\sqrt{\frac{\pi}{g^2\Theta}}\,\sum_{k=1}^\infty
\frac{(-z)^k}{k^{3/2}}\right.\nonumber\\&&\times\left.
\sum_{m=1}^\infty\e^{-\frac{N^2\pi}{g^2\Theta}\,\frac{m^2}k}\,\left(
\e^{N\,\pi\,q\,m/{g^2\Theta}}\,w^m+\e^{-N\,\pi\,q\,m/{g^2\Theta}}
\,w^{-m}\right)\right] \ . \label{ZSUNqnewform}\eea This form of
the partition function exhibits it as a factorization into the
zero-instanton vacuum amplitude of the $U(N)$ theory that we
encountered in the previous section, plus two infinite series
comprising all the positively and negatively charged higher
instanton contributions. When the contour ${\cal C}_0(w)$ is the unit
circle we have $w^{-1}=\overline{w}$, suggesting a sort of
chiral/anti-chiral factorization for the instanton contributions.

With respect to the zero-instanton sector, we can single out those
higher instanton configurations which yield contributions that are not
exponentially suppressed in $N$. The positive and negative instanton
contributions in (\ref{ZSUNqnewform}) are weighted respectively by
Boltzmann factors with the modified actions
\beq
S^{(\pm)}_{\rm inst}\left(m\,,\,k\,;\,{g^2\Theta}
\right):=\frac{N\,\pi\,q\,m}{g^2\Theta}\,\left(
\frac{N\,m}{q\,k}\mp1\right) \ .
\label{Sinstpm}\eeq
By using the $\zed_N$ symmetry (\ref{modNsym}) of the partition function, we
can
choose
the flux $q$ to lie in the range $-n^{~}_{\rm F}\leq q\leq n_{\rm
  F}^{~}$ without loss of generality. Suppose first that $q>0$. Then the
solutions of the Diophantine equation
\beq
N\,m-q\,k=n_+ \ ,
\label{DioorderN}\eeq
with $n_+\in\nat$ of order~$1$ in the large $N$ limit, select an
  infinite collection of pairs of integers $(m,k)$ for which the
  instanton action $S^{(+)}_{\rm
  inst}(m,k;{g^2\Theta})$ is finite at $N\to\infty$. If $n_+<0$ then no finite
solution is possible, because from (\ref{ZNunconstr}) it follows that
the large $N$ limit must be taken with $0<k<N$. In addition, the
action $S^{(-)}_{\rm inst}(m,k;{g^2\Theta})$ is infinite, because the
analogous condition $N\,m+q\,k=n_-\in\nat$ for it necessarily requires
$n_-$ to be of order $N$ due to the positivity of the integer $k$.
{}From (\ref{DioorderN}) it follows that the surviving action admits the
$\frac1N$ expansion
\beq
S_{\rm inst}^{(+)}\left(m\,,\,k\,;\,{g^2\Theta}
\right)=\frac{N\,\pi\,q\,m}{g^2\Theta}\,
\frac{n_+}{q\,k}=
\frac{\pi\,q\,n_+}{g^2\Theta}\,\frac1{1-\frac{n_+}{N\,m}}=
\frac{\pi\,\ell}{g^2\Theta}+O\left(\frac1N\right) \ .
\label{SinstpluslargeN}\eeq
The leading non-vanishing term in (\ref{SinstpluslargeN}) at
$N=\infty$ is simply the action for a fluxon of charge
$\ell=q\,n_+$. For $q<0$ it is instead the action $S^{(-)}_{\rm
  inst}(m,k;{g^2\Theta})$ which survives the large $N$ double scaling
limit. Therefore, in a manner very similar to that of
Section~\ref{Decomp}, the large $N$ double scaling limit of the
$SU(N)/\zed_N$ gauge theory becomes dominated by the fluxon solutions
of noncommutative gauge theory.

As discussed in Section~\ref{Fluxon}, the limiting gauge theory on the
noncommutative plane $\real_\Theta^2$ is independent of the choice of
non-trivial projective module for the parent gauge theory on the
torus $\torus^2$, and hence we may restrict our attention to the flux
sector $q=-1$ without loss of generality. We rewrite the partition
function $\hat{\cal Z}_\infty^{~}({g^2,\Theta}):=\hat{\cal
  Z}_{SU(N)}^{(-1)}({g^2\Theta})$ in (\ref{ZSUNqnewform}) by replacing the sum
over $k\in\nat$ with a sum over the fluxon charge $\ell=N\,m-k$.
Since $k\geq1$, the charge is bounded from above and in the limit
$N\to\infty$ one has \bea \hat{\cal
Z}_\infty\left({g^2\,,\,\Theta}\right)&=&
\e^{\frac{\pi\,N\,g^2\Theta}{12}}\,\oint \limits_{{\cal
C}_0(z)}\frac{\dd z}{2\pi\ii z^{N+1}}~
\exp\left[-N\,\sqrt{\frac{\pi}{g^2\Theta}}~\Li_{3/2}(-z)\right]\\&&
\times\,\oint\limits_{{\cal C}_0(w)}\frac{\dd w}{2\pi\ii}~
\exp\left[-N\,\sqrt{\frac{\pi}{g^2\Theta}}\,\sum_{m=1}^\infty
w^{-m}~ \sum_{\ell=-\infty}^{N\,m-1}\e^{-\pi\,\ell/{g^2\Theta}}~
\frac{(-z)^{N\,m-\ell}}{(N\,m-\ell\,)^{3/2}}\right] \ .
\nonumber\label{ZNCfluxon}\eea This expression demonstrates
that the approximation we have employed keeps only the chiral part
of the full instanton series, in the sense that only negative
magnetic charges (i.e. negative powers of $w$) appear. This is in
beautiful harmony with the chiral nature of the double scaling
limit that we discovered in the previous section, and also with
the fact that any theory involving fluxons is necessarily chiral.

We now expand the factor $(N\,m-\ell\,)^{-3/2}$ in (\ref{ZNCfluxon})
for $\ell$ of order~$1$ in the large $N$ limit and rescale the second contour
integration variable as $w\to z^N\,w$. We then expand the second
exponential function and perform the integral over $w$ to arrive at
\bea
\hat{\cal Z}_\infty\left({g^2\,,\,\Theta}\right)&=&
\sum_{\ell=-\infty}^{N-1}(-1)^\ell~\e^{-\pi\,\ell/{g^2\Theta}}\,
\oint\limits_{{\cal C}_0}\frac{\dd z}{2\pi\ii z^{\ell+1}}~
\exp\left[-N\,\sqrt{\frac{\pi}{g^2\Theta}}~\Li_{3/2}(-z)\right]
\nonumber\\&&\times\,\sqrt{\frac\pi{N\,{g^2\Theta}}}~
\e^{\frac{\pi\,N\,g^2\Theta}{12}} \ .
\label{ZNCsumell}\eea
To sum the series in (\ref{ZNCsumell}), we decompose the integration
contour into two connected, closed components as ${\cal C}_0^{~}={\cal
  C}_0^{+}\cup{\cal C}_0^{-}$. The contour ${\cal C}_0^{+}$ is
situated in the region $\e^{-\pi/{g^2\Theta}}<|z|<1$ of the
complex $z$-plane and along it the sum over $\ell=0,1,\dots,N-1$
can be performed, while the contour ${\cal C}_0^-$ is located at
$|z|<\e^{-\pi/{g^2\Theta}}$ and along it we may carry out the sum
over all $\ell\leq-1$. After dropping the irrelevant
multiplicative constant, we arrive in this way at the formula \bea
\hat{\cal Z}_\infty\left({g^2\,,\,\Theta}\right)&=&
\e^{\frac{\pi\,N\,g^2\Theta}{12}}\,\Biggl\{~\oint \limits_{{\cal
C}_0^-}\frac{\dd z}{2\pi\ii\,z}~
\frac{z~\e^{\pi/{g^2\Theta}}}{1+z~\e^{\pi/{g^2\Theta}}}~
\exp\left[-N\,\sqrt{\frac{\pi}
{g^2\Theta}}~\Li_{3/2}(-z)\right]\Biggr.\nonumber\\&&-\Biggl.
\oint\limits_{{\cal C}_0^+}\frac{\dd z}{2\pi\ii\,z}~
\frac{1+\left(\frac{\e^{-\pi/{g^2\Theta}}}z\right)^N}
{1+\frac{\e^{-\pi/{g^2\Theta}}}z}~\exp\left[-N\,\sqrt{\frac{\pi}{g^2\Theta}}~
\Li_{3/2}(-z)\right]\Biggr\} \ . \label{ZNCcontints}\eea Since the
contour ${\cal C}_0^-$ does not catch the pole at
$z=-\e^{-\pi/{g^2\Theta}}$, the first integral in
(\ref{ZNCcontints}) vanishes by Cauchy's theorem.
To evaluate the contribution of the second integral, we expand it into
a sum of two contour integrals, the first one having a trivial dependence on
$N$. Since by construction ${\cal C}_0^+$ {\it does} encircle the
pole, this first contribution can be safely evaluated by using the
residue theorem.

By using the relations (\ref{relation}) and (\ref{mugTheta}), we
finally arrive in this way at the expression \bea \hat{\cal
Z}_\infty\left({g^2\,,\,\Theta}\right)&=&
\e^{\frac{\pi\,N\,g^2\Theta}{12}}\,{\cal Z}_\infty
\left(g^2\,,\,\Theta\right)\nonumber\\&&+\,
\e^{\pi\,N\,(\frac{g^2\Theta}{12}-\frac1{g^2\Theta})}\,\oint
\limits_{{\cal C}_0^+}\frac{\dd z}{2\pi\ii z^{N+1}}~ \frac
z{z+\e^{-\pi/{g^2\Theta}}}~\exp\left[-N\,\sqrt{\frac{\pi}{g^2\Theta}}~
\Li_{3/2}(-z)\right]\nonumber\\&& \label{ZNCfinalexp}\eea for the
gauge theory partition function on the noncommutative plane
$\real_\Theta^2$. The first term in (\ref{ZNCfinalexp}) is exactly
the resummation of the weak-coupling semi-classical expansion of
the gauge theory in terms of fluxons, as obtained in
(\ref{calZNCplanefinal}). The second term is similar to the
contribution of the $U(N)$ zero-instanton sector that was studied
in the previous section, with the important difference that in
evaluating the saddle-point position there are corrections coming from
the function $\frac z{z+\e^{-\pi/{g^2\Theta}}}$
occuring in the integrand. In the $\frac1N$ expansion these
contributions can be considered subleading and we may discard them
in a first approximation. We can therefore identify, at leading
order, the second contribution with the previously computed one
in the $U(N)$ gauge theory, up to the relevant multiplicative factors.
Note that the vacuum energy is not cancelled out in the fluxon
contribution.

\subsection{Weak and Strong Coupling Limits\label{PPT}}

\EPSFIGURE{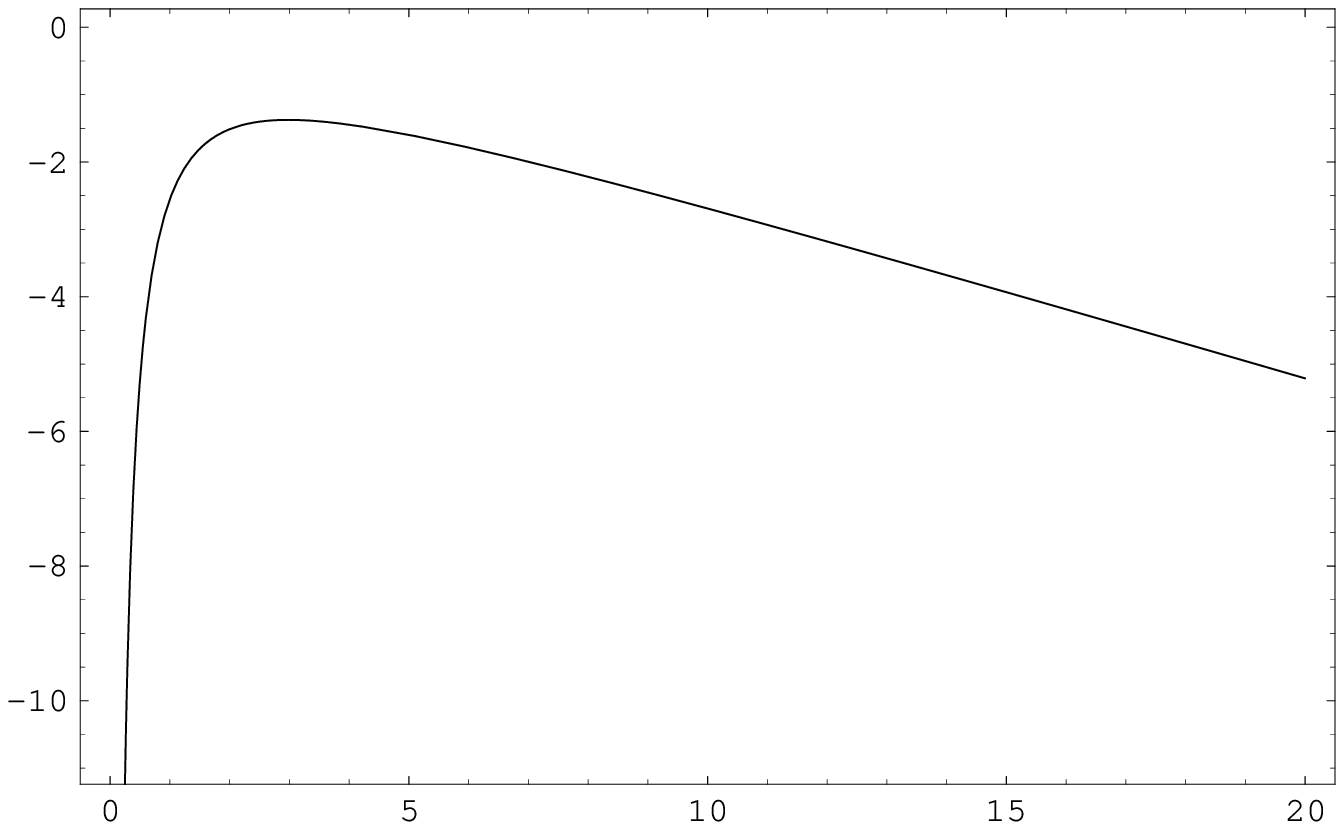,width=3.5in}{The zero-instanton free energy
  fluctuation $\Delta{\cal F}_\infty$ as a function of the double
  scaling parameter $\mu=g^2\Theta$. \label{Notransit}}

The two terms in the decomposition (\ref{ZNCfinalexp}) represent,
respectively, the contributions from the zero-instanton sector and
the collective resummation of higher instanton sectors of the
$SU(N)/\zed_N$ gauge theory on $\torus^2$ in the double scaling
limit. Since $\Li_{3/2}(0)=0$, in the weak coupling limit
${g^2\Theta}\to0$ the partition function (\ref{calZNCplanefinal})
is of order~$1$ and thus the first term in (\ref{ZNCfinalexp})
dominates the full vacuum amplitude. This is of course the regime
wherein we anticipate the noncommutative instanton expansion to be
valid. We can now analyse if there exists some finite value of the
double scaling coupling $\mu=g^2\Theta>0$ for which the second
term in (\ref{ZNCfinalexp}) dominates over the fluxon partition
function. For this, we rewrite (\ref{ZNCfinalexp}) in terms of the
fluxon free energy
$\mathcal{F}_\infty(g^2,\Theta)=\ln\mathcal{Z}_\infty(g^2,\Theta)$
as \beq \hat{\mathcal{Z}}_\infty\left(g^2\,,\,\Theta\right)=:
\e^{\frac{\pi\,N\,g^2\Theta}{12}+\mathcal{F}_\infty(g^2,\Theta)}\,
\left(1+\e^{\Delta\mathcal{F}_\infty(g^2,\Theta)}\right) \ ,
\label{freenfluct}\eeq and seek regions in parameter space where
the free energy fluctuation function
$\Delta\mathcal{F}_\infty(g^2,\Theta)$ yields a dominant
contribution to the partition function (\ref{freenfluct}). The
result of a numerical evaluation of this function is depicted in
Fig.~\ref{Notransit}. We see that
$\Delta\mathcal{F}_\infty(g^2,\Theta)<0$ over the full range of
parameters, and thus the corrections to the fluxon contribution in
(\ref{ZNCfinalexp}) are exponentially suppressed in the double
scaling limit. Thus the phase transition described earlier,
potentially driven by the enhancement of the collective higher
instanton configurations, does {\it not} occur, at least in our
approximations. The semi-classical expansion is completely stable against
vacuum-like fluctuations, and this stability is a highly robust
property of the noncommutative gauge theory.

The quantum field theory is thus always dominated by the fluxons
of noncommutative gauge theory which in the string representation
correspond to unstable D-instantons in the two dimensional target
spacetime. In fact, the fluxon contributions are completely
well-defined for {\it all} values of the coupling $\mu=g^2\Theta$,
as emphasized by Fig.~\ref{InstantonsPart} which illustrates that
the fluxon free energy is generically a smooth function. Using
properties of the Jonqui\`ere function, one can show that all
derivatives of $\mathcal{F}_\infty(g^2,\Theta)$ with respect to
$\mu$ vanish at $\mu=0$. The weak-coupling limit of the
noncommutative gauge theory seems therefore completely regular and
the expected singularity, representing a sort of Douglas-Kazakov
phase transition at vanishing coupling, is resolved by
noncommutativity. Again, this non-singular behaviour is a highly
non-trivial feature of the noncommutative gauge theory.

\EPSFIGURE{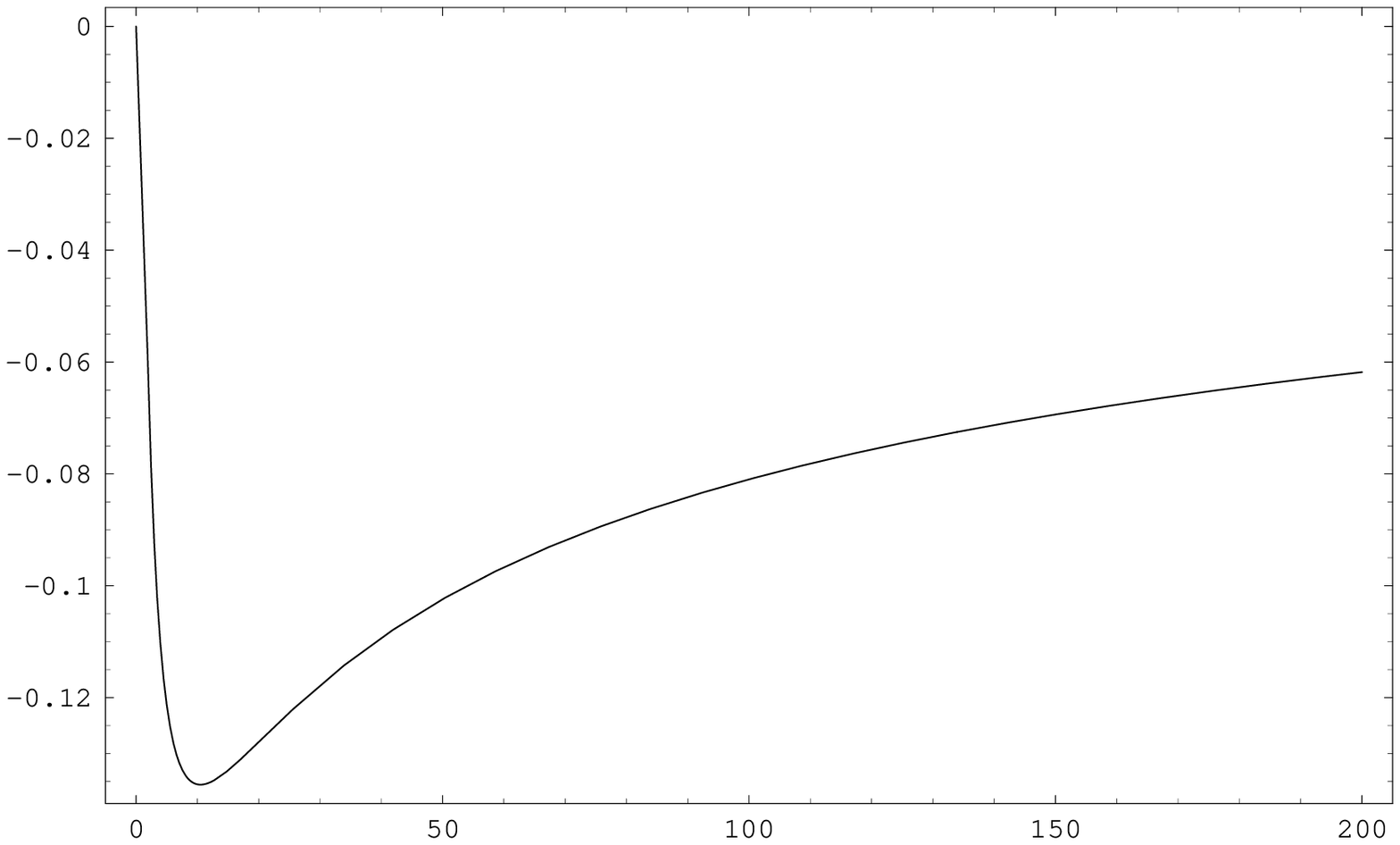,width=4.5in}{The fluxon free energy ${\cal
  F}_\infty$ as a function of the double scaling
  parameter $\mu=g^2\Theta$. \label{InstantonsPart}}

At the other end of the parameter region, we can expand the fluxon
contribution (\ref{calZNCplanefinal}) in the strong-coupling limit
$g^2\Theta\to\infty$, which is the regime in which we can compare
to the commutative string expansion of the previous section. In
fact, since the large $\Theta$ expansion of any noncommutative
field theory is known to organize itself much like the large $N$
expansion of an ordinary multicolour field theory, this is
precisely the limit where we would hope to be able to explicitly
extract stringy behaviour of the noncommutative gauge theory. The
free energy
$\hat{\mathcal{F}}_\infty(g^2,\Theta)=\ln\hat{\mathcal{Z}}_\infty(g^2,\Theta)$
can be computed as an asymptotic series in $\frac1\Theta$ by using the singular
expansion of the Jonqui\`ere function $\Li_{3/2}(z)$ around its
finite branch point at $z\to1^-$ (see Appendix~\ref{appA},
eq.~(\ref{Liseriesexpz1})), and one finds \beq
\lim_{\Theta\to\infty}\,\hat{\mathcal{F}}_\infty\left(g^2\,,\,\Theta\right)=
\frac A{2\Theta}\,\left[\frac{g^2\Theta}{12}+\frac{2\,\sqrt\pi}
{g^2\Theta}-\frac1{\sqrt\pi}\,\sum_{k=0}^\infty\frac{(-1)^k\,
\zeta\left(\mbox{$\frac32$}-k\right)}{k!}~\frac1
{\left(g^2\Theta\right)^{k+\frac12}}\right] \ .
\label{strongcouplfluxon}\eeq Non-perturbative corrections to this
expansion (for finite ${g^2\Theta}<\infty$) are of order
$\e^{-{g^2\Theta}}$. It is amusing to note that, starting from the
gauge theory on $\torus_\theta^2$ and assuming continuity in
$\Theta$, the noncommutative gauge theory naively approaches the
$SU(N)/\zed_N$ gauge theory at large $N$ in the limit
$\Theta,A\to\infty$ with $g\,\Theta/\sqrt A$ held fixed. This
suggests another large $\Theta$ limit in which string behaviour
could potentially be extracted from the gauge dynamics.

\subsection{Open String Interpretation\label{NCOSI}}

In the strong-coupling phase analysed in the previous subsection,
the physical degrees of freedom are dominated by a tree-level
resummation of the D1-branes described in Section~\ref{OSI}, of
tension (\ref{T1tension}), which can be thought of as dual to the
usual electric dipole quanta of noncommutative gauge theory. We
will now reconcile this interpretation, and in particular the open
string worldsheet perturbation expansion of Section~\ref{OSI},
with the standard string interpretation of fluxons in
noncommutative Yang-Mills theory. Consider a D-string extended
along $\real^2$ and subjected to a constant, perpendicularly
applied background NS-NS field $B$. The low-energy effective field
theory of this system is described by noncommutative Yang-Mills
theory with 16 supercharges and the open string parameters \beq
\Theta=\frac1B \ , ~~ g^2=\frac{2\pi\,\alpha'\,g_sB}A \ .
\label{openstrpars}\eeq This identification holds in the
Seiberg-Witten decoupling limit $(\alpha'\,)^2B\gg A$ wherein the
string scale is small compared to the scale of noncommutativity~\cite{sw}.
Using it we may identify the strong coupling expansion
(\ref{strongcouplfluxon}) with the string partition function \bea
\mathcal{Z}^\infty_{\rm str}(B,T_1A)&:=&\lim_{\Theta\to\infty}\,
\hat{\mathcal{F}}_\infty\left(g^2\,,\,\Theta\right)\nonumber\\&=&
\frac{A\,B}2\,\left[\frac1{6T_1A}+\sqrt\pi\,T_1A-
\sqrt{\frac{T_1A}{2\pi}}\,\sum_{k=0}^\infty
\frac{(-1)^k\,\zeta\left(\frac32-k\right)}{2^k\,k!}~
(T_1A)^k\right] \ . \nonumber\\&& \label{NCstrpartfn}\eea

The expansion (\ref{NCstrpartfn}) is to be compared with
(\ref{openstrdouble}). We see that in the presence of the background
$B$-field the open string vacuum amplitude becomes a non-analytic
function of the worldsheet expansion parameter $T_1A$. The most
striking distinction though is that, since
$\zeta(\frac32-k)\in\real\,\backslash\,\rat~~\forall k\in\nat_0$, the
coefficients in the expansion of the partition function are now {\it
  irrational} numbers. In this way the $B$-field deforms the moduli
spaces that arise in the double scaling expansion of the $U(N)$ gauge
theory of the previous section to ones with generically irrational
normalized volumes. It is tempting to conjecture that, at least in the
strong-coupling regime, these deformed moduli spaces, and the
noncommutative open string theory, could be described by some sort of
noncommutative topological sigma-model. Noncommutative versions of the
standard topological field theories may be constructed by deforming
the BRST algebras of observables~\cite{IKS1}--\cite{G-CP1}.

Thus the D-string interpretation of ordinary two-dimensional
Yang-Mills theory is consistent with that of the noncommutative case,
and in the double scaling limit it provides a toy laboratory to
explore the stringy aspects of noncommutative gauge theory. In this
language, the mass $E_\ell$ of a fluxon of charge $\ell\in\zed$ can be
written as
\beq
E_\ell=\frac{\pi\,|\ell|}{g^2\Theta}=\frac\pi2\,|\ell|\,T_1A \ .
\label{fluxonmassT1}\eeq
This coincides with the Nambu-Goto action for $|\ell|$ D-strings, with
$\ell$ the bound D-instanton charge. It
is intriguing to now examine the lift of our two-dimensional
constructions to $3+1$-dimensions, whereby the unstable fluxon
solutions are known to correspond to full stable BPS solutions of
$\mathcal{N}=4$ noncommutative gauge theory describing $|\ell|$
infinite D-strings piercing a D3-brane~\cite{gn,gn1,gn2}. In the
context of the present paper, they are obtained by embedding our
noncommutative instantons on $\torus_\theta^2$ in two of the three
spatial dimensions of the low-energy effective field theory, and
interpreting the resulting object as a string infinitely stretched in
the remaining spatial dimension. After decompactification via the
double-scaling limit, we obtain a configuration of static Euclidean strings on
$\real^2\times\real_\Theta^2$. From (\ref{fluxonmassT1}) it follows
that these instanton strings are equivalent to the D-strings that
appear in the non-perturbative sector of the large $N$ expansion of
ordinary Yang-Mills theory in the double-scaling limit. Given the
S-duality between these string solitons and the string of
$3+1$-dimensional noncommutative open string theory~\cite{NCOS1}, it
is natural to expect that the electric dipole degrees of freedom of
the gauge theory on $\torus_\theta^2$ map in an analogous way onto
noncommutative open strings. This prospect is consistent with the S-duality, in
$1+1$-dimensions, between noncommutative open string theory and
ordinary Yang-Mills theory with non-zero electric flux~\cite{NCOS2}.

\acknowledgments

We thank L.~Paniak for early collaboration on the issues
described in Section~\ref{InstStrings}. We are grateful to
A.~Bassetto, G.~Cicuta, F.~Colomo, Ph.~Flajolet, D.~Johnston,
M.~Matone, A.~Polychronakos, S.~Ramgoolam and J.~Wheater for helpful
discussions and correspondence. R.J.S. would like to thank the
Department of Physics at the University of Parma and the Department of
Physics at the University of Florence for their hospitality during
various stages of this work. The work of R.J.S. was supported in part
by an Advanced Fellowship from the Particle Physics and Astronomy
Research Council~(U.K.).

\appendix

\section{Properties of the Jonqui\`ere Function\label{appA}}

In this appendix we will describe some elementary properties of the
polylogarithm function
\beq
\Li_\alpha(z)=\sum_{k=1}^\infty\frac{z^k}{k^\alpha}
\label{Liapp}\eeq
for $z,\alpha\in\complex$. It can be regarded as a generalization of
the Riemann zeta-function which is obtained in the limit $z=1$,
$\Li_s(1)=\zeta(s)$. In particular, we will derive a new asymptotic
expansion formula for it, which is the crux of the saddle-point
analyses of the main text. Some related results are found
in~\cite{flajolet1}.

\subsection{Analytic Continuation\label{ACPolyLog}}

The series (\ref{Liapp}) is absolutely convergent everywhere on the
open unit disk $|z|<1$. To extend it analytically into the entire
complex plane $z\in\complex$, we use a Mellin transform to write the
integral representation
\beq
\Li_\alpha(z)=\sum_{k=1}^\infty\,\frac1{\Gamma(\alpha)}\,
\int\limits_0^\infty\dd t~t^{\alpha-1}~\e^{-t\,k}~z^k=
\frac z{\Gamma(\alpha)}\,\int\limits_0^\infty\dd t~
\frac{t^{\alpha-1}}{\e^t-z}
\label{Lizint}\eeq
which is valid when ${\rm Re}(\alpha)>0$. By using the derivative
identity
\beq
\Li_\alpha(z)=z\,\frac{\dd}{\dd z}\Li_{\alpha+1}(z)
\label{Liderivid}\eeq
along with the integral representation (\ref{Lizint}) for the function
$\Li_{\alpha+1}(z)$, the change of variables $s=\e^t$ leads to the
alternative form
\beq
\Li_\alpha(z)=\frac z{\Gamma(\alpha+1)}\,\int\limits_1^\infty
\dd s~\frac{\ln^\alpha s}{(s-z)^2}
\label{Lisintrep}\eeq
valid when ${\rm Re}(\alpha)>-1$. This expression exhibits a
branch cut singularity of $\Li_\alpha(z)$ along $z\in[1,\infty)$.
Its behaviour near the branch point at $z=1$ is given by the series
expansion~\cite{flajolet1}
\beq
\Li_\alpha(z)=\Gamma(1-\alpha)\,(-\ln z)^{\alpha-1}
+\sum_{k=0}^\infty\frac{\zeta(\alpha-k)}{k!}~\ln^kz \ , ~~
z\to1^- \ .
\label{Liseriesexpz1}\eeq

\subsection{Asymptotic Expansion\label{AEPolyLog}}

We are now ready to derive our new asymptotic expansion of the
  Jonqui\`ere function, which we have also checked numerically using
  {\sl Mathematica}.
\begin{theorem}
The asymptotic expansion as $x\to\infty$ of the polylogarithm function
$\Li_\alpha(-\e^x)$ is given by
$$
\Li_\alpha\left(-\e^x\right)=-2\,\sum_{k=0}^\infty
\frac{\left(1-2^{1-2k}\right)\,
\zeta(2k)}{\Gamma(\alpha+1-2k)}~x^{\alpha-2k}+O\left(\e^{-x}\right) \
{}.
$$
\label{asymptthm}\end{theorem}

\noindent
{\sc Proof:} For $x\in[0,\infty)$, we start from the integral
representation (\ref{Lisintrep}) and change integration variables as
$y=\e^{-x}\,s$ to write
\beq
\Li_\alpha\left(-\e^x\right)=-\frac1{\Gamma(\alpha+1)}~
\int\limits_{\e^{-x}}^\infty
\dd y~\frac{(\ln y+x)^\alpha}{(y+1)^2} \ .
\label{Liyintrep}\eeq
Let us decompose the domain of integration in (\ref{Liyintrep}) as the
disjoint union $[\e^{-x},\e^x)\cup[\e^x,\infty)$. It is
straightforward to show that the contribution to the integral
(\ref{Liyintrep}) from the second component is exponentially
suppressed. For this, we use the fact that $y\in[\e^x,\infty)$ is
large for $x\to\infty$ to expand
\bea
\int\limits_{\e^x}^\infty\dd y~\frac{(\ln y+x)^\alpha}{(y+1)^2}
&=&-\sum_{n=1}^\infty(-1)^n\,n\,\int\limits_{\e^x}^\infty\,
\frac{\dd y}{y^{n+1}}~(\ln y+x)^\alpha \nonumber\\&=&
-\sum_{n=1}^\infty(-1)^n\,n~\e^{n\,x}\,\int\limits_{2x}^\infty
\dd u~u^\alpha~\e^{-n\,u}\nonumber\\&=&-\sum_{n=1}^\infty\frac{(-1)^n}
{n^\alpha}~\e^{n\,x}~\Gamma(\alpha+1,2n\,x) \ .
\label{2ndintexp}\eea
Let us now recall the large $x$ behaviour of the incomplete Euler
gamma-function given by~\cite{GR1}
\beq
\Gamma(\nu,z)=z^{\nu-1}~\e^{-z}\,\left[\,\sum_{m=0}^{l-1}\frac{(-1)^m}
{z^m}\,\frac{\Gamma(1-\nu+m)}{\Gamma(1-\nu)}+O\bigl(|z|^{-l}\bigr)\right]
\label{incomplGammasympt}\eeq
with $|z|\to\infty$, $-\frac{3\pi}2<{\rm arg}\,z<\frac{3\pi}2$, and
$l\in\nat$. By keeping only the lowest order $l=1$ term in
(\ref{incomplGammasympt}), one finds that the integral
(\ref{2ndintexp}) is given approximately for large $x$ as
\beq
\int\limits_{\e^x}^\infty\dd y~\frac{(\ln y+x)^\alpha}{(y+1)^2}
\simeq2^\alpha~x^\alpha~\e^{-x} \ ,
\label{2ndintapprox}\eeq
exhibiting the claimed exponential suppression at $x\to\infty$.

It remains to compute the integral (\ref{Liyintrep}) over the first
component. For $y\in[\e^{-x},\e^x)$ one has $|\ln(y)/x|\leq1$, and so
  we may expand
\beq
\int\limits_{\e^{-x}}^{\e^x}\dd y~\frac{(\ln y+x)^\alpha}{(y+1)^2}
=\sum_{n=0}^\infty\frac{\Gamma(\alpha+1)}{\Gamma(n+1)\,\Gamma(\alpha+1-n)}
{}~x^{\alpha-n}\,\int\limits_{\e^{-x}}^{\e^x}\dd y~\frac{\ln^ny}
{(y+1)^2} \ .
\label{1stintexp}\eeq
By performing the change of variable $y\to1/y$, one can deduce that
the integrals in (\ref{1stintexp}) vanish for $n$ odd. For $n=2k$
even, the integrals can be computed by means of the change of
integration variable $y=\e^v$, and neglecting exponentially suppressed
terms in the limit $x\to\infty$ one finds after integrating by parts
the result
\bea
\int\limits_{\e^{-x}}^{\e^x}\dd y~\frac{\ln^{2k}y}
{(y+1)^2}&=&4k\,\int\limits_0^\infty\dd v~\frac{v^{2k+1}}{\e^v+1}+O
\left(\e^{-x}\right)\nonumber\\&=&2\left(1-2^{1-2k}\right)\,
\Gamma(2k+1)\,\zeta(2k)+O\left(\e^{-x}\right) \ .
\label{evencontr1st}\eea
By substituting this into (\ref{1stintexp}), we arrive finally at the
claimed asymptotic expansion of the polylogarithm function
(\ref{Liyintrep}). \hfill{$\Box$}

\noindent
This formula can be expressed in terms of Bernoulli numbers by using
the Euler identity~\cite{GR1}
\beq
\zeta(2k)=\frac{2^{2k-1}\,B_{2k}~\pi^{2k}}{(2k)!} \ , ~~ k\in\nat_0 \ .
\label{zetaBer}\eeq

\section{Free Fermion Representation\label{FFR}}

Proposition~\ref{Hurwitzprop1} combined with Theorem~\ref{xifinal}
determines the asymptotic behaviour as $n\to\infty$ of the
combinatorial formula (\ref{Hurwitzred},\ref{Hurwitzirred}) via the
strong coupling expansion of the gauge theory in the double
scaling limit. The explicit calculation was performed by
solving a saddle-point equation that avoids the very difficult
task of obtaining these asymptotic combinatorics directly. In this
appendix we sketch an alternative derivation based on the standard free
fermion representation of two-dimensional Yang-Mills
theory~\cite{douglas,MiPo}, which gives a useful fermionic picture of
the double scaling limit whose potential extension to the case of
non-trivial fluxes could provide a free fermion formulation of
noncommutative gauge theory. Moreover, it immediately yields another
proof of the quasi-modularity of the free energy as described in
Section~\ref{CFP}. The idea is to consider a more general amplitude
which contains the pertinent information about the chiral partition
function~\cite{eskin1}, but which has a much simpler and tractable
behaviour in the double scaling limit.

Let us introduce free chiral fermion fields \beq
\psi(z)=\sum_{r\in\zed+\frac12}\psi^{~}_r~z^{-r}~\sqrt{\frac{\dd
z}{ 2\pi\ii z}} \ , ~~
\psi^*(z)=\sum_{r\in\zed+\frac12}\psi^*_r~z^{-r}~ \sqrt{\frac{\dd
z}{2\pi\ii z}} \label{freeFermifields}\eeq for $z\in\complex$
obeying the non-vanishing canonical anticommutation relations \beq
\left\{\psi^{~}_r\,,\,\psi^*_s\right\}=\delta_{r+s,0} \ .
\label{ccrnot0}\eeq They act on the standard fermionic Fock space
defined by \beq \mathcal{F}_{\rm F}=\bigoplus_{k,l=1}^\infty~
\bigoplus_{\stackrel{\scriptstyle\mbf r\in\nat^k}
{r_1<r_2<\cdots<r_k}}~\bigoplus_{\stackrel {\scriptstyle\mbf
s\in\nat^l}{s_1<s_2<\cdots<s_l}}~\complex\cdot
\prod_{i=1}^k\psi^{~}_{-r_i}\,\prod_{j=1}^l\psi^*_{-s_j}|0\rangle
\label{FermiFockspace}\eeq where $|0\rangle$ is the Fermi vacuum
defined by the annihilation conditions \beq
\psi^{~}_r|0\rangle=\psi^*_r|0\rangle=0 ~~~~ \forall r>0 \ .
\label{Fermivac}\eeq Using the fermion fields
(\ref{freeFermifields}) we define the $U(1)$ charge operator \beq
J_0:=\oint\limits_{\mathcal{C}_0}
\NO\psi(z)\,\psi^*(z)\NO=\sum_{r\in\zed+\frac12}\NO\psi^{~}_r\,
\psi^*_{-r}\NO \label{affinecurrent}\eeq where $\NO\cdot\NO$
denotes normal ordering with respect to the vacuum $|0\rangle$.
The Fock space (\ref{FermiFockspace}) then admits a natural
$\zed$-grading by total $U(1)$ charge as $\mathcal{F}^{~}_{\rm
  F}=\bigoplus_{q\in\zed}\mathcal{F}^q_{\rm F}$, where
$\mathcal{F}_{\rm F}^q$ is the eigenspace of the zero mode
operator (\ref{affinecurrent}) with eigenvalue $q\in\zed$. A set
of spanning vectors for this subspace is provided by the
``partition basis'' with \beq \mathcal{F}_{\rm
F}^q=\bigoplus_{n=0}^\infty~\bigoplus_{\stackrel{ \scriptstyle\mbf
n\in\nat_0^n\,,\,\sum_ln_l=n}{\scriptstyle n_1\geq
n_2\geq\cdots\geq n_n}}\complex\cdot\prod_{l=1}^n
\psi_{-q-\frac12+l-n_l}|q-n\rangle \ , \label{partbasis}\eeq where
$|q\rangle$ is the Fermi vacuum of overall $U(1)$ charge $q$, i.e.
$J_0|q\rangle=q\,|q\rangle$, which is defined by \beq
\psi^{~}_r|q\rangle=\psi^*_s|q\rangle=0 ~~~~ \forall r>q\,,\,s>-q
\ . \label{vacchargeq}\eeq

Let us now introduce the free fermion Hamiltonian operator \beq
L_0:=\oint\limits_{\mathcal{C}_0}\NO\psi(z)~z\,\frac\dd{\dd z}
\psi^*(z)\NO=\sum_{r\in\zed+\frac12}r\,\NO\psi^{~}_r\,\psi^*_{-r}\NO
\label{FermiHamop}\eeq with $L_0|q\rangle=\frac{q^2}2\,|q\rangle$.
We are interested in the dynamics generated by this Hamiltonian in
a subspace $\mathcal{F}_{\rm
  F}^q$ of fixed $U(1)$ charge. According to (\ref{partbasis}), it is
in such subspaces that we can make contact with the Yang-Mills
vacuum amplitude (\ref{chiral1N}) through the representation of
the sum over Young diagrams in terms of partitions. For simplicity
we will work in the eigenspace of charge $q=0$, i.e.
$\mathcal{F}_{\rm F}^0=\ker J_0$. This essentially constrains us
to representations of $SU(N)$ and it describes the low energy
excitations (compared to $N$) of the fermion system around the
Fermi level (\ref{Fermisurface}).

We will study the correlation functions of the scaled charge
operators \beq J_0(w):=\oint\limits_{\mathcal{C}_0}
\NO\psi\left(w^{-1/2}\,z\right)\,\psi^*\left(w^{1/2}\,z
\right)\NO=\sum_{r\in\zed+\frac12}\NO\psi^{~}_r\,
\psi^*_{-r}\NO~w^r \ , \label{J0wdef}\eeq with $J_0(1)=J_0$, for
$w\in\complex$ in the subspace $\ker J_0$. Consider the amplitude,
defined for every non-negative integer $m$, given as a trace over
states of vanishing $U(1)$ charge by \beq
\left\langle\,\prod_{i=1}^mJ_0(w_i)\right\rangle_0:= \Tr^{~}_{\ker
J_0}\left(\e^{-\lambda\,L_0}\, \prod_{i=1}^mJ_0(w_i)\right) \ .
\label{nptfndef}\eeq By using the partition basis
(\ref{partbasis}) at $q=0$ this correlator may be written
explicitly as \beq
\left\langle\,\prod_{i=1}^mJ_0(w_i)\right\rangle_0=
\sum_{n=1}^\infty\e^{-n\,\lambda}\,\sum_{\stackrel{
\scriptstyle\mbf n\in\nat_0^n\,,\,\sum_ln_l=n}{\scriptstyle
n_1\geq n_2\geq\cdots\geq n_n}}~\prod_{k=1}^m\left(\,
\sum_{i=1}^nw_k^{n_i-i+\frac12}\right) \ . \label{nptfnexpl}\eeq

The correlation function (\ref{nptfnexpl}) admits a nice physical
interpretation. Let us write the complex coordinates in the
parametrization $z=\e^{t+\ii\sigma}$, where
$(t,\sigma)\in\real\times\mathbb{S}^1$ are coordinates on the
cylinder in radial quantization of the two-dimensional fermion
system. The partitions $\mbf n\in\nat_0^n$ occuring in the second
sum in (\ref{partbasis}), with $\sum_ln_l:=\sum_jj\,\nu_j=n$, are
in one-to-one correspondence with (unramified) $n$-sheeted
oriented coverings of the spatial circle $\mathbb{S}^1$~\cite{gt1,gt2}. Each
basis state in (\ref{partbasis}) may then be identified with a
state in the Fock space of closed strings defined by $\nu_j$
strings with winding number $j$, which we may interpret as
electric flux around the circle. The worldsheet swept out by each
such string in periodic time defines a $j$-fold cover of a torus
by a torus. Since the Hamiltonian (\ref{FermiHamop}) is not
diagonal in the partition basis, the interactions of strings
create branch cuts on the base torus $\torus^2$ and glue copies of
it into higher genus Riemann surfaces. The correlator
(\ref{nptfnexpl}) is the generating function for the (topological
classes of) covers of $\torus^2$ constructed from $m$ branch
points, i.e. for arbitrary (not necessarily simple) Hurwitz
numbers, with $w_k$ weighting the branch points of various orders.
The branch points themselves correspond to $U(1)$ charge
insertions at $w_k$ of the fermions defined by the field
(\ref{J0wdef}). The $k^{\rm th}$ Laurent coefficient of
(\ref{nptfnexpl}) near the point $w_1=\dots=w_m=1$ determines the
number of covers with branch points of order $k$. Extracting this
coefficient is tantamount to integrating over the positions of the
corresponding branch points.

In the case of chiral Yang-Mills theory on $\torus^2$, the
subspace $\mathcal{F}_{\rm F}^0$ may be identified with the space
of ``representation states'' contributing to the partition
function (\ref{chiralHurwitz}). In this way, the double scaling limit
precisely factors out the chiral sector from the weakly-coupled
chiral-antichiral string expansion of two-dimensional Yang-Mills
theory. In light of this discussion, the genus $h$ contribution
$Z_h^+(\lambda)$ to the Yang-Mills partition function
(\ref{chiralHurwitz}) may be extracted from the second Laurent
coefficient of (\ref{nptfnexpl}) with $m=2h-2$ (corresponding to the
case of covers of $\torus^2$ with only simple ramification). For this,
we associate to each non-increasing partition $\mbf n\in\nat_0^n$
occuring in the second sum in (\ref{nptfnexpl}) a Young diagram
$Y\in\mathcal{Y}_n$ and corresponding $U(N)$ representation $R(Y)$
as described in Section~\ref{TG-TET}, and evaluate the central
character $\widetilde{C}_2(\mbf n):=\widetilde{C}_2(R(Y))$ as in
(\ref{tildeC2expl}) to compute \beq
Z_h^+(\lambda)=\lambda^{2h-2}\,\sum_{n=1}^\infty
\e^{-n\,\lambda}\,\sum_{\stackrel{ \scriptstyle\mbf
n\in\nat_0^n\,,\,\sum_ln_l=n}{\scriptstyle n_1\geq
n_2\geq\cdots\geq n_n}}\widetilde{C}_2(\mbf n)^{2h-2} \ .
\label{ZhC2series}\eeq By using the identity \beq
\widetilde{C}_2(\mbf n)=\left.\left(w\,\frac\dd{\dd w}\right)^2
\right|_{w=1}\,\sum_{i=1}^nw^{n_i-i+\frac12}-\frac{n\left(4n^2-1\right)}{12}
\label{C2derivid}\eeq we may then write the partition function
(\ref{ZhC2series}) in terms of the correlator (\ref{nptfnexpl}) as
\beq Z_h^+(\lambda)=\lambda^{2h-2}\,
\prod_{i=1}^{2h-2}\left.\left(w_i\,\frac\dd{\dd w_i}
\right)^2\right|_{w_i=1}\left\langle\,\prod_{j=1}^{2h-2}
J_0(w_j)\right\rangle_0 \ . \label{Zhnptfn}\eeq

Once (\ref{Zhnptfn}) is determined, we may use the following
combinatorial trick to compute the genus~$h$ free energy
(\ref{FglambdaA})~\cite{eskin1}. Let $\mbf\Pi_h^+$ be the set of all partitions
of the set $\{1,\dots,2h-2\}$ into unordered disjoint unions of
non-empty subsets of even cardinality.
Given $\mbf\sigma:=\coprod_{k=1}^L\sigma_k\in\mbf\Pi_h^+$, the
number $L=L(\mbf\sigma)$ is the length of the partition
$\mbf\sigma$. By definition, since $Z_h^+(\lambda)$ is the
generating function of {\it all} genus $h$ simple covers of
$\torus^2$, and $F_h^+(\lambda)$ that of the connected coverings,
we have the identity \beq
Z_h^+(\lambda)=Z_1^+(\lambda)\,\sum_{\mbf\sigma\in\mbf\Pi_h^+}~
\prod_{k=1}^{L(\mbf\sigma)}F^+_{|\sigma_k|}(\lambda)
\label{ZhFhid}\eeq where
$Z_1^+(\tau)=\e^{-\pi\ii\tau/12}\,\eta(\tau)$ is the generating
function for coverings without unramified connected components.
Applying a standard M\"obius inversion to (\ref{ZhFhid})~\cite{Mob1}, we may
then determine the free energy contributions through the
combinatorial formula \beq
F_h^+(\lambda)=\sum_{\mbf\sigma\in\mbf\Pi_h^+}(-1)^{L(\mbf\sigma)}
\,\bigl(L(\mbf\sigma)-1\bigr)!\,\prod_{k=1}^{L(\mbf\sigma)}\,
\frac{Z^+_{|\sigma_k|}(\lambda)}{Z_1^+(\lambda)} \ .
\label{FhZhcomb}\eeq

The main advantage of this rewriting is that the amplitude
(\ref{nptfnexpl}) can be computed explicitly. It defines a
meromorphic function which is absolutely convergent in the domain
of the complex plane defined by the conditions
$\e^{-\lambda}<|w_{i_1}\cdots w_{i_k}|<\e^{\lambda}$ for any
subset $\{i_1,\dots,i_k\}\subset\{1,\dots,m\}$. Its only
singularities are simple poles along the divisors $w_i=1$,
$i=1,\dots,m$. This uniquely determines the correlation functions
up to an additive constant in terms of the single odd
characteristic elliptic Jacobi theta-function \beq
\vartheta(w,\tau):=\sum_{n=-\infty}^\infty(-1)^n~w^{n+\frac12}~
\e^{\pi\ii(n+\frac12)^2\,\tau} \label{oddthetafn}\eeq and its
derivatives \beq \vartheta^{(k)}(w,\tau):=\left(w\,\frac\dd{\dd
w}\right)^k \vartheta(w,\tau) \label{thetaderiv}\eeq as~\cite{bloch1} \beq
\left\langle\,\prod_{i=1}^mJ_0(w_i)\right\rangle_0= \sum_{\pi\in
S_m}\frac{\det\left[\frac{\vartheta^{(j-i+1)}\left(
w_{\pi(1)}\cdots w_{\pi(m-j)}\,,\,\tau\right)}{(j-i+1)!}
\right]_{1\leq i,j\leq
m}}{\vartheta\left(w_{\pi(1)}\,,\,\tau\right)\,
\vartheta\left(w_{\pi(1)}\,w_{\pi(2)}\,,\,\tau\right)\cdots
\vartheta\left(w_{\pi(1)}\cdots w_{\pi(m)}\,,\,\tau\right)}
\label{nptfntheta}\eeq with $\frac1{(-k)!}:=0~~\forall k\geq1$.
The free energy (\ref{FhZhcomb}) is then computed from
(\ref{nptfntheta}) using (\ref{Zhnptfn}) along with the
identity~\cite{bloch1}
\beq
\vartheta^{(2m+1)}(1,\tau)=(2m+1)!\,\sum_{\stackrel{\scriptstyle\mbf\nu
\in\nat_0^m}{\scriptstyle\sum_ll\,\nu_l=m}}~\prod_{k=1}^m
\frac1{\nu_k!}\,\left[\frac{B_{2k}\,E_{2k}(\tau)}{k\,(2k)!}
\right]^{\nu_k} \label{thetaderiv1}\eeq and
$\vartheta^{(2m)}(1,\tau)=0~~\forall m\geq1$. These formulas
demonstrate explicitly the quasi-modularity of the free energy
that we described in Section~\ref{TG-TET}~\cite{ochi1,eskin1,bloch1},
and moreover give, at least in principle, the explicit expansion of
the free energy into the Eisenstein series basis for the ring
$\mathcal{M}$ of quasi-modular forms.

In the present context the principal characteristic of the formula
(\ref{nptfntheta}) is that its form in the double scaling limit
can be made very explicit. By using the Poisson resummation
formula (\ref{Poissonresum}) one may derive the weak-coupling
behaviour of the Jacobi theta-function (\ref{oddthetafn}) to be
\beq
\frac{\vartheta(w,\tau)}{\vartheta^{(1)}(1,\tau)}=\frac{\lambda}
\pi\,\sin\left(\frac{\pi\,w}{\lambda}\right)~\e^{w^2/2\lambda}
+O\left(\e^{-1/\lambda}\right) \ , \label{weakthetafn}\eeq and use
the generating function \beq
\frac{\pi\,w}{\sin\pi\,w}=2\,\sum_{k=0}^\infty\left(1-2^{1-2k}\right)
\,\zeta(2k)~w^{2k} \ . \label{zetagenfn}\eeq One can now proceed
to derive the asymptotics of the formula (\ref{nptfntheta}), and
substitute into (\ref{Zhnptfn}) and (\ref{FhZhcomb}) to compute
finally the double scaling free energy at genus~$h$. We will not
enter into the details of this tedious combinatorial calculation,
but refer to~\cite{eskin1} for the technical details. The final
result agrees with Theorem \ref{xifinal}.

More generally, the two limits of two-dimensional Yang-Mills theory
that we have considered in this paper seem to have natural
realizations in the fermionic picture. The standard 't~Hooft limit is
the one in which the length of the (dual) circle where the fermions
move is kept finite, while their number goes to infinity. The
low-lying spectrum becomes universal and equidistant (with $\frac1N$
corrections), and this is the conformal field theory limit described
by a $c=1$ theory~\cite{douglas,MiPo}. The spatial density of
fermions, however, goes to infinity. On the other hand, the double
scaling limit appears to be more akin to the thermodynamic limit of
the free fermion system~\cite{LSA1}--\cite{MP2}, in
which the size of the circle scales like $N$ leading to an infinite
volume system with a finite particle density. In this limit the
spectrum becomes continuous and an infinity of states must be taken
into account to calculate correlation functions. It would be
interesting to make this connection more precise, and to extend it to
a fermionic formulation of noncommutative gauge theory in two
dimensions.

\end{document}